\PassOptionsToPackage{pdfpagelabels=false}{hyperref}
\documentclass[useAMS,usenatbib]{mnras}

\usepackage[pdftex]{graphicx}
\usepackage{color}
\usepackage{amsmath}  
\usepackage{amssymb}  
\usepackage{afterpage}  
\usepackage{times}  
\usepackage{paralist}  
\usepackage{pbox}  
\usepackage{booktabs}  
\usepackage[flushleft]{threeparttable}  

\usepackage{grffile}

\newcommand{\Msun}{{\rm M}_{\odot}}

\newcommand{\krot}{\kappa_{\rm rot}}
\newcommand{\PLate}{P({\rm Late})}
\newcommand{\PEarly}{P({\rm Early})}

\title[Galactic angular momentum in IllustrisTNG]{Galactic angular momentum in the IllustrisTNG simulation -- I. \\ Connection to morphology, halo spin, and black hole mass}

\author[V. Rodriguez-Gomez et al.]
{
	\parbox{18cm}{
		Vicente Rodriguez-Gomez,$^{1,2}$\thanks{E-mail: vrodgom.astro@gmail.com}
		Shy Genel,$^{3,4}$
		S. Michael Fall,$^{5}$
		Annalisa Pillepich,$^{6}$ \\
    Marc Huertas-Company,$^{7,8}$
		Dylan Nelson,$^{9}$
    Luis Enrique P\'erez-Monta\~no,$^{2}$ \\
		Federico Marinacci,$^{10}$
		R\"{u}diger Pakmor,$^{11}$
		Volker Springel,$^{11}$
		Mark Vogelsberger$^{12}$ \\
		and Lars Hernquist$^{13}$
	}
	\vspace{0.3cm} \\ 
	$^{1}$ Department of Physics \& Astronomy, Johns Hopkins University, 3400 N. Charles Street, Baltimore, MD 21218, USA \\
	$^{2}$ Instituto de Radioastronom\'ia y Astrof\'isica, Universidad Nacional Aut\'onoma de M\'exico, Apdo. Postal 72-3, 58089 Morelia, Mexico \\
	$^{3}$ Center for Computational Astrophysics, Flatiron Institute, 162 5th Avenue, New York, NY 10010, USA \\
	$^{4}$ Columbia Astrophysics Laboratory, Columbia University, 550 West 120th Street, New York, NY 10027, USA \\
	$^{5}$ Space Telescope Science Institute, 3700 San Martin Drive, Baltimore, MD 21218, USA \\
	$^{6}$ Max-Planck-Institut f\"{u}r Astronomie, K\"{o}nigstuhl 17, D-69117 Heidelberg, Germany \\
  $^{7}$ Departamento de Astrof\'isica, Instituto de Astrof\'isica de Canarias, Universidad de La Laguna, E-38200 La Laguna, Spain \\
  $^{8}$ LERMA, Observatoire de Paris, CNRS, PSL, Universit\'e Paris Diderot F-75013 Paris, France \\
  $^{9}$ Universit\"at Heidelberg, Zentrum f\"ur Astronomie, Institut f\"ur theoretische Astrophysik, Albert-Ueberle-Stra\ss{}e 2, D-69120 Heidelberg, Germany \\
  $^{10}$ Department of Physics \& Astronomy ``Augusto Righi'', University of Bologna, via Gobetti 93/2, I-40129 Bologna, Italy \\
	$^{11}$ Max-Planck-Institut f\"{u}r Astrophysik, Karl-Schwarzschild-Stra\ss{}e 1, D-85741 Garching bei M\"{u}nchen, Germany \\
	$^{12}$ Department of Physics, Kavli Institute for Astrophysics and Space Research, Massachusetts Institute of Technology, Cambridge, MA 02139, USA \\
	$^{13}$ Harvard-Smithsonian Center for Astrophysics, 60 Garden Street, Cambridge, MA 02138, USA
}

\begin{document}


\maketitle
\begin{abstract}
We use the TNG100 simulation of the IllustrisTNG project to investigate the stellar specific angular momenta ($j_{\ast}$) of $\sim$12~000 central galaxies at $z=0$ in a full cosmological context, with stellar masses ($M_{\ast}$) ranging from $10^{9}$ to $10^{12} \, \Msun$. We find that the $j_{\ast}$--$M_{\ast}$ relations for early-type and late-type galaxies in IllustrisTNG are in good overall agreement with observations, and that these galaxy types typically `retain' $\sim$10--20 and $\sim$50--60 per cent of their host haloes' specific angular momenta, respectively, with some dependence on the methodology used to measure galaxy morphology. We present results for \textit{kinematic} as well as \textit{visual-like} morphological measurements of the simulated galaxies. Next, we explore the scatter in the $j_{\ast}$--$M_{\ast}$ relation with respect to the spin of the dark matter halo and the mass of the supermassive black hole (BH) at the galactic centre. We find that galaxies residing in faster spinning haloes, as well as those hosting less massive BHs, tend to have a higher specific angular momentum. We also find that, at fixed galaxy or halo mass, halo spin and BH mass are anticorrelated with each other, probably as a consequence of more efficient gas flow towards the galactic centre in slowly rotating systems. Finally, we show that halo spin plays an important role in determining galaxy sizes -- larger discs form at the centres of faster rotating haloes -- although the trend breaks down for massive galaxies with $M_{\ast} \gtrsim 10^{11} \, \Msun$, roughly the mass scale at which a galaxy's stellar mass becomes dominated by accreted stars.

\end{abstract}


\section{Introduction}\label{sec:intro}
\renewcommand{\thefootnote}{\fnsymbol{footnote}}

\renewcommand{\thefootnote}{\arabic{footnote}}

The angular momentum ($J$) of a galaxy is one of its most fundamental properties, along with its mass ($M$) and energy. However, while energy is largely dissipated in the form of radiation during the formation of a galaxy, its mass and momentum are approximately conserved \citep[][]{Fall1980}. A related quantity, the \textit{specific} angular momentum $(j = J/M)$, can be used to normalize out the amount of material and thus help to understand the interplay between the various components of a galaxy, such as its gas and stars, as well as to establish a connection to the spin of the dark matter (DM) halo, which is acquired via tidal torques from neighbouring structures in the early Universe \citep[][]{Peebles1969, Doroshkevich1970, White1984}.

In this context, \cite{Fall1983} presented observational estimates of the stellar specific angular momenta ($j_{\ast}$) of a sample of nearby spiral and elliptical galaxies, and plotted them as a function of their stellar masses ($M_{\ast}$). He found that spiral and elliptical galaxies occupy approximately parallel tracks on the $j_{\ast}$--$M_{\ast}$ diagram (on a log--log scale) with a logarithmic slope of $\sim$0.7 but with a large offset, such that ellipticals had lower $j_{\ast}$ at fixed $M_{\ast}$ than spirals by a factor of $\sim$6. This subject has been revisited using updated observations in a series of papers \citep{Romanowsky2012, Fall2013, Fall2018} and the original findings from \cite{Fall1983} have been approximately maintained, with the latest iteration reporting a logarithmic slope of $0.67 \pm 0.07$ and an offset between `pure' discs and bulges of a factor of $8 \pm 2$ \citep{Fall2018}.

Other observational works have also studied the $j_{\ast}$--$M_{\ast}$ diagram in the local Universe, although these are generally limited to late-type morphologies \citep[e.g.][]{Obreschkow2014, Lapi2018, Posti2018a, DiTeodoro2021, ManceraPina2021, ManceraPina2021a, Hardwick2022} or consider measurements of $j_{\ast}$ confined to relatively small apertures, of about one effective radius \citep{Cortese2016, Tabor2019}. Measuring the total $j_{\ast}$ for elliptical galaxies remains difficult due to their more extended angular momentum profiles, and would benefit from more kinematic data out to $\sim$5 effective radii and beyond \citep{Romanowsky2012}.

On the numerical side, early hydrodynamic cosmological simulations suffered from the so-called `angular momentum catastrophe' \citep[][]{Navarro1995, Navarro1997}, in which baryons condensed too efficiently into their nearest gravitational potential well, and most of the baryonic angular momentum was transferred to the DM halo via the strong merging activity of these baryonic clumps. It soon became clear that the key to solving this problem was to include efficient stellar feedback at early times \citep[][]{SommerLarsen1999, Thacker2001}, which reheats gas into an extended reservoir that can then cool more gradually into the central galaxy, a phenomenon sometimes referred to as a galactic fountain. The numerical treatment of hydrodynamics has also been shown to make a difference in the angular momentum content of cosmologically simulated galaxies \citep[][]{Torrey2012}.

Using hydrodynamic cosmological simulations, only recently has it been possible to produce a statistically significant galaxy \textit{population} with realistic amounts of angular momentum \citep{Genel2015, Pedrosa2015, Teklu2015, Zavala2016, DeFelippis2017, Lagos2017, Lagos2018}, due to enormous improvements in computational capability and galaxy formation modelling \citep[see][for reviews]{Somerville2015, Naab2017, Vogelsberger2020}. The $j_{\ast}$--$M_{\ast}$ diagram has also been addressed with zoom-in hydrodynamic simulations, which model relatively small galaxy samples at higher resolution \citep{Obreja2016, Obreja2019, Grand2017, Grand2019, Sokolowska2017, El-Badry2018}, as well as with analytic, semi-analytic, and semi-empirical models \citep{Dutton2012, Romanowsky2012, Stevens2016, Shi2017, Lapi2018a, Posti2018, Zoldan2018, Irodotou2019}.

In order to understand the role of angular momentum in galaxy formation, it is often useful to define the specific angular momentum `retention fraction,' usually denoted by $f_j$, which quantifies the ratio between the \textit{specific} angular momentum of the galaxy and that of its parent halo. Traditionally, some analytic and semi-analytic models \citep[e.g.][]{Fall1980, Dalcanton1997, AvilaReese1998, Mo1998, Firmani2000, Firmani2009} have assumed that $f_j = 1$ for spiral galaxies, which is close to currently accepted values. Recent observational studies indicate $f_j \approx 0.7$--$0.8$ for spirals \citep{Fall2013, Posti2019a, DiTeodoro2021} and $f_j \approx 0.1$ for ellipticals \citep{Fall2013, Fall2018}.

Some possible reasons for a `loss' in angular momentum include gas stripping (in dense environments), gas ejection by galactic outflows (if the expelled gas comes from the outer, more angular momentum-rich regions of the galaxy), as well as so-called \textit{biased collapse} scenarios \citep[][]{VandenBosch1998, Dutton2012, Kassin2012}, in which the galaxy forms preferentially from material in the inner, more angular momentum-poor regions of the halo. On the other hand, gains in angular momentum can be induced by galactic fountains, as discussed above. Other processes can result in either a loss or gain of angular momentum, such as mergers \citep[][]{Fall1979, Barnes1988, Hernquist1995, Springel2005} and misaligned gas accretion \citep[][]{Sales2012}.

In particular, \cite{Genel2015} investigated the stellar angular momenta of galaxies in the Illustris simulation \citep{Genel2014a, Vogelsberger2014a, Vogelsberger2014} and found that, at fixed stellar mass, late-type galaxies have higher $j_{\ast}$ than early-type galaxies, with retention fractions of approximately 1 and 0.3, respectively. In addition, \cite{Genel2015} explored variations in galactic angular momentum with respect to modifications to the fiducial galaxy formation model, finding that stronger feedback from galactic winds results in higher galactic angular momentum, while stronger feedback from supermassive BHs acts in the opposite direction.

In this paper, we revisit the topic of galactic angular momentum from the perspective of hydrodynamic cosmological simulations, specifically using the state-of-the-art TNG100 simulation of the IllustrisTNG project \citep{Marinacci2018, Naiman2018, Nelson2018, Pillepich2018a, Springel2018}. Previous works have shown that the IllustrisTNG model returns observationally consistent correlations among galaxy stellar morphology, galaxy mass, and star formation rate at low redshift \citep{Huertas-Company2019, Rodriguez-Gomez2019, Tacchella2019, Donnari2021, Donnari2021a}, as well as the redshift evolution of the morphological fractions and of the degree of rotational versus dispersion-supported motions \citep{Pillepich2019, Varma2022}. Here, we focus on how $j_{\ast}$ relates to galaxy stellar morphology, to the spin of the DM halo, and to the mass of the central black hole (BH) over a wide range of stellar masses ($M_{\ast} = 10^9$--$10^{12} \, \Msun$) at $z=0$. We also discuss the retention fraction $f_j$ in the context of galaxy morphology, and determine whether the relation between halo spin and galaxy size predicted by some theoretical models arises naturally in IllustrisTNG.

This paper is structured as follows. In Section \ref{sec:methodology}, we describe the simulation used for this work, provide details about the main calculations, and define the galaxy sample. Section \ref{sec:results} contains our main results, which comprise a study of the interplay between $j_{\ast}$, $f_j$ and galaxy morphology (Section \ref{subsec:connection_to_morphology}), an analysis of the role of halo spin and BH mass in establishing galactic angular momentum (Section \ref{subsec:connection_to_halo_spin_and_bh_mass}), and an exploration of the link between halo spin and galaxy size (Section \ref{subsec:halo_spin_and_galaxy_size}). Finally, in Section \ref{sec:discussion_and_conclusions} we discuss these results and present our conclusions.

\section{Methodology}\label{sec:methodology}

\subsection{The IllustrisTNG simulation suite}
\label{subsec:tng_simulations}

The IllustrisTNG project \citep{Marinacci2018, Naiman2018, Nelson2018, Nelson2019a, Pillepich2018a, Pillepich2019, Springel2018} is a suite of magnetohydrodynamic cosmological simulations carried out with the moving-mesh code \textsc{arepo} \citep{Springel2010, Pakmor2013, Pakmor2016}, featuring a galaxy formation model that includes prescriptions for radiative cooling, star formation and evolution, metal enrichment, supernova feedback, and supermassive BH growth and feedback \citep{Weinberger2017, Pillepich2018}. We provide more details about the numerical implementation of these physical processes in Section \ref{subsec:tng_model}.

In this work we use the highest resolution version of the TNG100 simulation (TNG100-1), which covers a periodic volume of $(75h^{-1} \, {\rm Mpc})^3 \approx (110.7 \; {\rm Mpc})^3$ and follows the evolution of $1820^3$ DM particles and approximately $1820^3$ baryonic resolution elements (gas cells and stellar particles), which have masses of $7.47 \times 10^6$ and $1.39 \times 10^6 \, \Msun$ (on average), respectively. The gravitational softening length for both DM and stellar particles is $0.5 h^{-1} \approx 0.74$ kpc at $z=0$, while for gas cells it is tied to their radius and in principle can be as low as $0.19$ kpc.

Haloes and subhaloes are identified with the friends-of-friends \citep[FoF,][]{Davis1985} and \textsc{subfind} \citep{Springel2001, Dolag2009a} algorithms, respectively. We define galaxies as being composed of the stellar and star-forming gas components of subhaloes. Unless otherwise noted, we measure all properties of a galaxy (e.g. stellar mass or angular momentum) for the entire \textsc{subfind} object, i.e. without removing the particles found beyond some fiducial aperture (e.g. twice the stellar half-mass radius).

The initial conditions of the simulation have been set at $z = 127$ using the \textsc{n-genic} code \citep{Springel2005a}, based on a power spectrum generated by \textsc{camb} \citep{Lewis2000}. The cosmological parameters adopted in IllustrisTNG are $\Omega_{\rm m} = 0.3089$, $\Omega_{\rm b} = 0.0486$, $\Omega_{\Lambda} = 0.6911$, $\sigma_{8} = 0.8159$, $n_{\rm s} = 0.9667$, and $h = 0.6774$, in accordance with Planck measurements \citep{Planck2016}.

\subsection{The galaxy formation model}
\label{subsec:tng_model}

Here we summarize some of the most salient features of the IllustrisTNG galaxy formation model, which is based on the original Illustris model \citep{Vogelsberger2013, Torrey2014}. For brevity, we omit details about the gravitational and magnetohydrodynamic calculations, and instead give an overview of the physical prescriptions for radiative cooling, star formation and evolution, metal enrichment, galactic wind feedback, supermassive BH growth, and feedback from active galactic nuclei (AGNs). For a complete description of the IllustrisTNG model, we refer the reader to \cite{Weinberger2017} and \cite{Pillepich2018}.

Gas in the simulation is allowed to cool through primordial and metal-line cooling in the presence of a redshift-dependent, spatially uniform, ionizing UV background \citep{Faucher-Giguere2009} that is switched on at $z=6$, taking self-shielding corrections into account \citep{Vogelsberger2013}. Gas cells with densities above $\rho_{\rm thres} = 0.13$ cm$^{-3}$ (in units of hydrogen number density) are considered to be star-forming, and become candidates to be stochastically converted into stellar particles according to the subresolution model of \cite{Springel2003}, with some modifications such as using a \cite{Chabrier2003} initial mass function (IMF) instead of a \cite{Salpeter1955} IMF, as detailed in \cite{Vogelsberger2013}. This star-forming gas is modelled with a two-phase, \textit{effective} equation of state that describes the formation and evaporation of unresolved cold clouds embedded in a hot ambient medium, including the effects of supernovae that inject metal-enriched gas and thermal energy to the ambient phase. As shown in \cite{Springel2003}, this model quickly leads to a self-regulated, `quiescent' mode of star formation.

Stellar particles in the simulation represent coeval stellar populations with a \cite{Chabrier2003} IMF. These are allowed to evolve in time while depositing mass and metals -- originating from asymptotic giant branch (AGB) stars and supernovae of Types Ia (SNIa) and II (SNII) -- into the surrounding gas, keeping track of the production and evolution of nine elements (H, He, C, N, O, Ne, Mg, Si, Fe). The minimum mass for a star to end its life as a core-collapse supernova (SNII) was increased from $6 \, \Msun$ in Illustris to $8 \, \Msun$ in IllustrisTNG \citep{Pillepich2018}. Accordingly, stars less massive than $8 \, \Msun$ are assumed to enter an AGB phase. Finally, given the uncertainty about the progenitors of SNIa events, the SNIa rate is obtained from a simple delay-time distribution, regardless of the metallicity and IMF of the parent stellar population.

Galactic winds are launched (in the form of wind particles) from star-forming gas cells in a random direction with a speed that is proportional to $\sigma_{\rm DM} H(z)^{-1/3}$, where $\sigma_{\rm DM}$ is the local, one-dimensional DM velocity dispersion and $H(z)$ is the Hubble parameter at redshift $z$. A velocity floor of $350 \, {\rm km \, s^{-1}}$ is also imposed, as described in \cite{Pillepich2018}. When launched, the wind particles are hydrodynamically decoupled until they leave their local interstellar medium, which typically happens when the density of the gas cell where they are currently located falls below $0.05 \rho_{\rm thres}$ (5 per cent of the density threshold for star formation), transferring their mass, momentum, metals, and energy to that gas cell. The wind mass loading factor, $\eta_{\rm w}$,\footnote{The wind mass loading factor is defined as $\eta_{\rm w} \equiv \dot{M}_{\rm w} / \dot{M}_{\rm SFR}$, where $\dot{M}_{\rm w}$ is the rate of gas mass ejected in the form of wind particles and $\dot{M}_{\rm SFR}$ is the local star formation rate.} depends on the available energy for wind generation from core-collapse supernovae (SNII), as in the original Illustris implementation \citep{Vogelsberger2013}, but now also depends on gas metallicity, such that $\eta_{\rm w}$ becomes smaller in higher metallicity environments. In addition, the winds are now allowed to carry a small fraction of thermal energy.

Supermassive BHs are `seeded' at the centres of haloes that reach a FoF group mass of $5 \times 10^{10} h^{-1} \, \Msun$. The BH seed mass is $8 \times 10^5 h^{-1} \, \Msun$, a value that is approximately eight times larger than the seed mass in the original Illustris model. The supermassive BHs can then grow by accreting material from their surroundings, as well as by merging with other BHs. The accretion rate on to supermassive BHs is assumed to be given by the Bondi accretion rate \citep{Bondi1952}, $\dot{M}_{\rm Bondi}$, limited by the Eddington accretion rate, $\dot{M}_{\rm Edd}$:

\begin{flalign}
\dot{M}_{\rm BH} &= \min(\dot{M}_{\rm Bondi}, \dot{M}_{\rm Edd}), &&
\label{eq:bh_accretion_rate}
\end{flalign}
where
\begin{flalign}
\dot{M}_{\rm Bondi} = \frac{4 \pi G^2 M_{\rm BH}^2 \rho}{c_{\rm s}^3}, &&
\label{eq:bondi_rate}
\end{flalign}
\begin{flalign}
\dot{M}_{\rm Edd} = \frac{4 \pi G M_{\rm BH} m_{\rm p}}{\epsilon_{\rm r} \sigma_{\rm T} c}. &&
\label{eq:eddington_rate}
\end{flalign}
Here, $M_{\rm BH}$ denotes the BH mass, $G$ the gravitational constant, $c$ the speed of light in vacuum, $m_{\rm p}$ the proton mass, and $\sigma_{\rm T}$ the Thomson cross-section. The factor $\epsilon_{\rm r}$ is the radiative efficiency, which quantifies the radiated luminosity in terms of the accreted rest-mass energy ($L_{\rm r} = \epsilon_{\rm r} \dot{M}_{\rm BH} c^2 $), and in this model is assumed to have a fixed value of 0.2. Finally, $\rho$ and $c_{\rm s}$ are the density and sound speed of the gas around the supermassive BH, which are calculated in a kernel-weighted fashion over a spherical region containing approximately 256 gas cells (at TNG100 resolution).

It is worth noting that current cosmological and galaxy-scale hydrodynamic simulations cannot resolve the region actually associated with the accretion disc around the supermassive BH, by several orders of magnitude. This led some early models of AGN feedback in hydrodynamic simulations \citep[][]{Springel2005b, Sijacki2007} to introduce a dimensionless `boost factor' $\alpha$ when calculating $\dot{M}_{\rm Bondi}$, which was also included in the original Illustris model \citep{Vogelsberger2013} with a value of $\alpha =$ 100. This factor has been eliminated in IllustrisTNG, being effectively replaced by the larger BH seed mass mentioned above, which achieves a similar effect by promoting early BH growth \citep{Weinberger2017}.

Once the BH accretion rate has been determined, AGN feedback can operate in two different modes depending on the accretion state: (i) a high-accretion mode, sometimes referred to as the `quasar' mode, which corresponds to a classic accretion disc \citep{Shakura1973}, and (ii) a low-accretion mode, which operates for BHs accreting below some threshold value of the Eddington ratio, $\dot{M}_{\rm Bondi} / \dot{M}_{\rm Edd}$. In IllustrisTNG, the feedback associated with the latter mode is described with the `kinetic wind' model of \cite{Weinberger2017}, which was inspired by observational findings \citep[e.g. the prevalence of `red geyser' galaxies,][]{Cheung2016} as well as theoretical considerations \citep[e.g.][]{Yuan2014}, and replaces the `radio bubble' model of \cite{Sijacki2007} that was used in Illustris.

The transition between the two AGN feedback modes takes place when the Eddington ratio ($\dot{M}_{\rm Bondi} / \dot{M}_{\rm Edd}$) reaches a threshold $\chi$ given by
\begin{flalign}
\chi = \min\left[ \chi_0 \left( \frac{M_{\rm BH}}{10^8 \, \Msun} \right)^{\beta}, 0.1 \right], &&
\label{eq:eddington_ratio}
\end{flalign}
where $\chi_0 =$ 0.002 and $\beta =$ 2. For BHs with $\dot{M}_{\rm Bondi} / \dot{M}_{\rm Edd} \geq \chi$, \textit{thermal} energy is injected in a kernel-weighted fashion over the same region used to calculate the accretion rate (i.e. a sphere enclosing approximately 256 gas cells) at a rate $\dot{E}_{\rm high} = 0.02 \dot{M}_{\rm BH} c^2$. Conversely, BHs with $\dot{M}_{\rm Bondi} / \dot{M}_{\rm Edd} < \chi$ impart \textit{kinetic} energy to neighbouring gas cells (also in a kernel-weighted manner over the same spherical region) at a rate $\dot{E}_{\rm low} = \epsilon_{\rm f, kin} \dot{M}_{\rm BH} c^2$, where the efficiency $\epsilon_{\rm f, kin}$ is typically equal to its maximum value of 0.2, but decreases proportionally to $\rho$ at sufficiently low densities ($\rho < 0.01 \rho_{\rm thres}$) in order to prevent the kinetic feedback mode from pushing the densities to ever lower values. The kinetic energy is injected in the form of a momentum kick in a random direction for each injection event \citep{Weinberger2017}. This approach violates strict momentum conservation for a single injection event, but ultimately conserves momentum by averaging over many injection events. As with galactic winds, this represents an inherently isotropic feedback model.

Overall, the IllustrisTNG galaxy formation model has numerous free parameters (some of them better constrained than others) that have been calibrated to match several observables at $z=$ 0 -- namely, the galaxy stellar mass function, the stellar-to-halo and BH-to-halo mass relations, the halo gas fraction as a function of halo mass, and the galaxy size versus stellar mass relation -- as well as the global star formation rate density at $z=$ 0--8 \citep{Pillepich2018}. We note, however, that the model was not tuned to match the angular momentum content of galaxies, nor to enforce any correlation between galactic angular momentum and other galaxy or halo properties.

\subsection{Measuring angular momentum}
\label{subsec:measuring_am}

The galaxies' stellar angular momenta are measured following \cite{Genel2015}. For each subhalo, the calculation frame is centred on the particle (of any type) with the lowest gravitational potential. It is important not to set the origin at the centre of mass, which can be sensitive to structure at large radii. However, the velocity of the calculation frame coincides with that of the stellar centre of mass, which is not sensitive to the velocity of any individual particle. As mentioned in Section \ref{subsec:tng_simulations}, these measurements are not restricted to any fiducial radius, so they are carried out for the entire \textsc{subfind} object (accounting only for the stellar particles).

The angular momenta of the parent haloes are obtained by means of a spherical overdensity calculation equivalent to that of \cite{Zjupa2017}. For each FoF group, the calculation frame is centred on the particle with the lowest gravitational potential, while velocities are measured with respect to the centre-of-mass velocity, which is calculated by including all resolution elements (i.e. DM, gas, stars, supermassive BHs) within $R_{200} \equiv R_{\rm 200, crit}$, i.e. the radius enclosing an average density equal to 200 times the critical density of the Universe. The magnitude of the angular momentum, $J_{200} \equiv J_{\rm 200, crit}$, and the mass of the halo, $M_{200} \equiv M_{\rm 200, crit}$, are also measured within $R_{\rm 200}$. The \textit{specific} angular momentum of the halo is defined as $j_{200} \equiv J_{200} / M_{200}$.

Once these halo properties are known, the halo spin parameter $\lambda$ \citep[][]{Peebles1969} can be quantified using the closely related definition by \cite{Bullock2001},
\begin{flalign}
\lambda^{\prime} \equiv \frac{J_{\rm 200}}{\sqrt{2} M_{\rm 200} V_{\rm 200} R_{\rm 200}} = \frac{j_{\rm 200}}{\sqrt{2} V_{\rm 200} R_{\rm 200}}, &&
\label{eq:spin_parameter}
\end{flalign}
where $V_{\rm 200} = \sqrt{G M_{\rm 200} / R_{200}}$ is the circular velocity at $R_{200}$. The halo spin parameter is predicted to follow a lognormal distribution that is relatively insensitive to halo mass, environment, or redshift. We note that these calculations include all particles contained within $R_{\rm 200}$, regardless of whether they belong to the FoF group or not \citep[see][for more details]{Zjupa2017}.

In previous work \citep{Rodriguez-Gomez2017}, we additionally matched each halo from the hydrodynamic simulation to its counterpart from a corresponding DM-only (DMO) run and used the spin measurement from the latter, $\lambda^{\prime}_{\rm DMO}$, in order to remove any possible influence of the galaxy itself on the halo spin. Here, we do not take this extra step, having verified that none of our results changes appreciably if we replace $\lambda^{\prime}$ with $\lambda^{\prime}_{\rm DMO}$, and instead follow the simpler approach of taking both the galaxy and halo measurements from the hydrodynamic run. This choice is also supported by \cite{Zjupa2017}, who found that the spin parameter of DM haloes is affected in a minimal way by the dissipative collapse of baryons within them.

\subsection{Quantifying galaxy morphology}
\label{subsec:quantifying_morphology}

We employ two different measures of galaxy morphology throughout this paper. The first one is the so-called kappa parameter, $\krot$, which quantifies the fraction of the stellar kinetic energy that is invested into ordered circular motion \citep[][]{Sales2010, Rodriguez-Gomez2017}. More precisely, it is defined as 

\begin{flalign}
  \kappa_{\rm rot} = \frac{K_{\rm rot}}{K} = \frac{1}{K}\sum_{i}\frac{1}{2} m_{i}\left(\frac{j_{z,i}}{R_{i}}\right)^{2}, &&
  \label{eq:kappa}
\end{flalign}
where $K$ is the total kinetic energy of the stellar component, $m_i$ is the mass of the $i$-th particle, $j_{z,i}$ is the $z$-component of the particle's specific angular momentum, $R_i$ is the projected radius, and the sum is carried out over all the stellar particles in the galaxy. The $z$-axis is set to coincide with the stellar angular momentum of the galaxy, as defined in Section \ref{subsec:measuring_am}.

Like other \textit{kinematic} measures of galaxy morphology \citep[e.g.][]{Abadi2003, Scannapieco2009, Scannapieco2012, Aumer2013, Dubois2016}, $\krot$ is a good proxy for the amount of rotational support in a galaxy \citep[see][for a discussion]{Rodriguez-Gomez2017}. For example, \cite{Tacchella2019} quantified the spheroid-to-total ratio ($S/T$) and the concentration of the stellar mass density profile ($C_{82}$) of TNG100 galaxies, while \cite{Pillepich2019} quantified both structural morphologies via 3D stellar shapes as well as $v/\sigma$. Some recent works \citep[e.g.][]{Correa2017} use an alternative definition of $\krot$ where the sum in equation (\ref{eq:kappa}) is only carried out for corotating particles, thus excluding counterrotating stars that are probably part of the bulge (or a counterrotating disc, e.g. \citealt{Starkenburg2019}). For simplicity, here we maintain the original definition and do not distinguish between corotating and counterrotating particles in equation (\ref{eq:kappa}). A comparison between both definitions can be found in appendix A from \cite{Rodriguez-Gomez2017}.

While such measures of kinematic morphology are very useful in hydrodynamic simulations, they generally cannot be directly compared to observations. Therefore, we also employ the \textit{visual-like} morphologies presented in \cite{Huertas-Company2019}, which were obtained by applying a neural network trained on visual morphologies from the Sloan Digital Sky Survey (SDSS) to synthetic images from the TNG100 simulation \citep{Rodriguez-Gomez2019} designed to match such SDSS observations. Here, we use the first level of the binary classification hierarchy described in \cite{Huertas-Company2019}, which returns the probability for a galaxy to be late-type, $\PLate = 1 - \PEarly$.

As discussed in \cite{Huertas-Company2019}, galaxies with $\PLate < 0.5$ mostly consist of elliptical and lenticular galaxies, while those with $\PLate \geq 0.5$ include all types of spirals. Although in Section \ref{subsec:connection_to_morphology} we will explore the variation in some quantities for galaxies with $\PLate$ values ranging continuously from 0 to 1, we note that this quantity was not designed to be used as a precise mapping for the Hubble sequence. Instead, our purpose is simply to complement our results with an alternative morphological measurement that is more observationally meaningful than $\krot$.

Finally, we note that the DM particle mass in the TNG100 simulation, $m_{\rm DM} \approx 7.5 \times 10^6 \, \Msun$, exceeds the critical value of $\sim \! 10^6 \, \Msun$ for significant heating of stellar motions in galactic discs over a Hubble time \citep{Lacey1985, Ludlow2021}. This spurious heating causes simulated discs to become more spheroidal (earlier morphology) and to rotate less rapidly (lower $j_{\ast}$). According to idealized simulations of isolated stellar discs embedded in DM haloes with particle masses similar to that of TNG100, the loss of $j_{\ast}$ over $10^{10}$ yr increases from $\sim$1 per cent for $M_{\ast} \! \sim \! 10^{11} \, \Msun$ to $\sim$30 per cent for $M_{\ast} \! \sim \! 10^{9} \, \Msun$ (Wilkinson et al., in preparation). The importance of this heating effect for the evolution of galactic discs in hydrodynamic cosmological simulations such as IllustrisTNG is now under active investigation (Ludlow et al., in preparation).

\subsection{The galaxy sample}

\begin{figure*}
  \centerline{\vbox{
    \hbox{
	    \includegraphics[width=8cm]{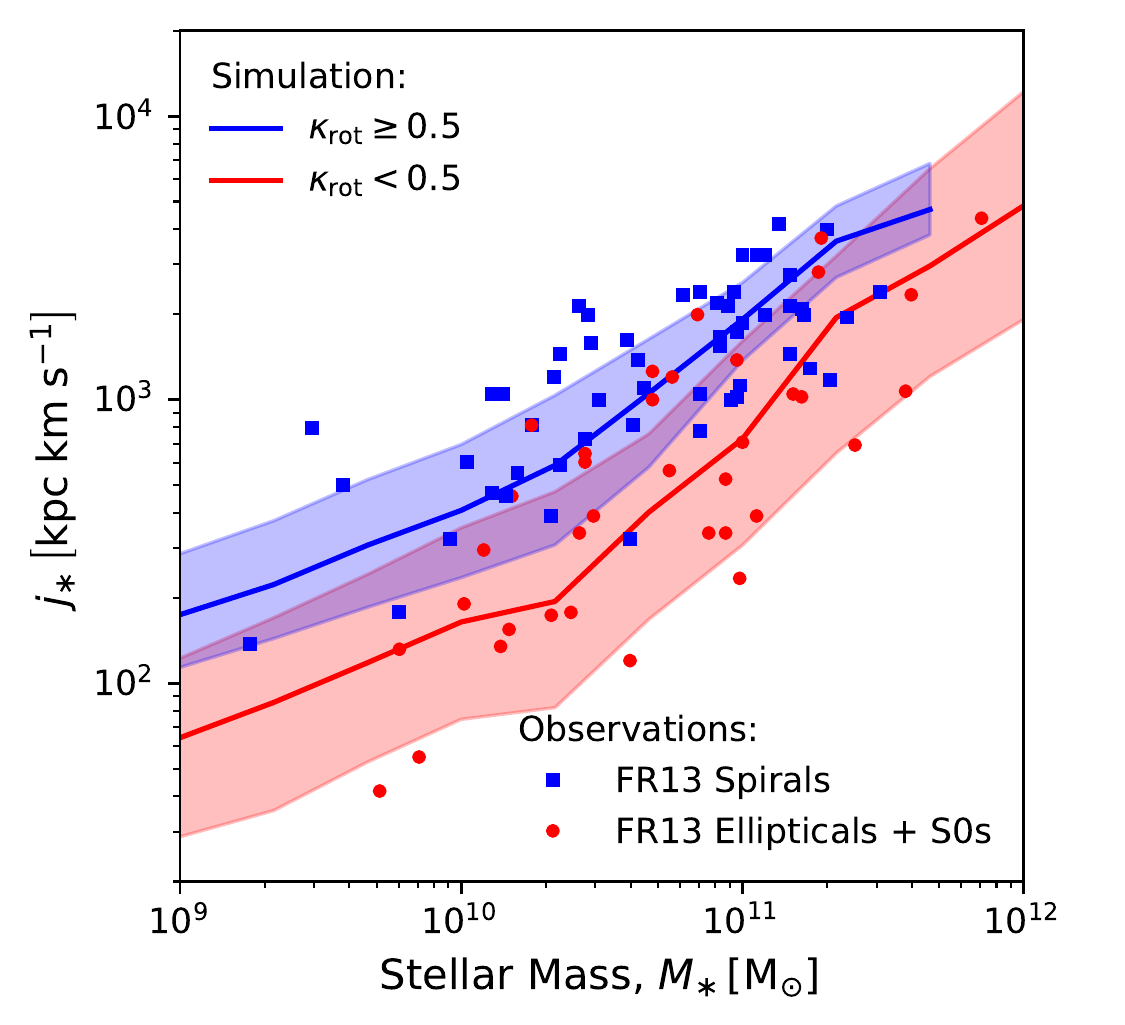}
	    \includegraphics[width=8cm]{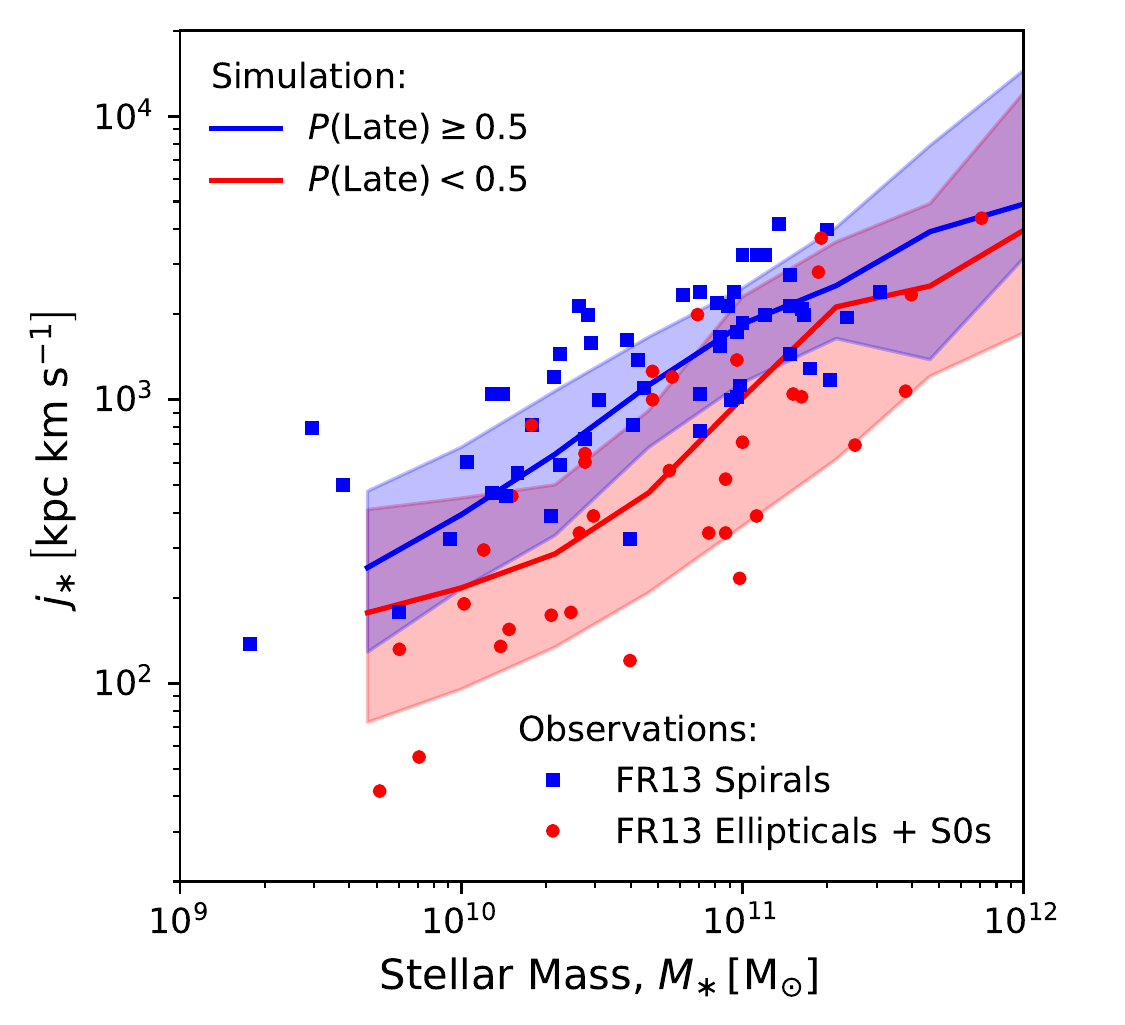}
     }
    \hbox{
	    \includegraphics[width=8cm]{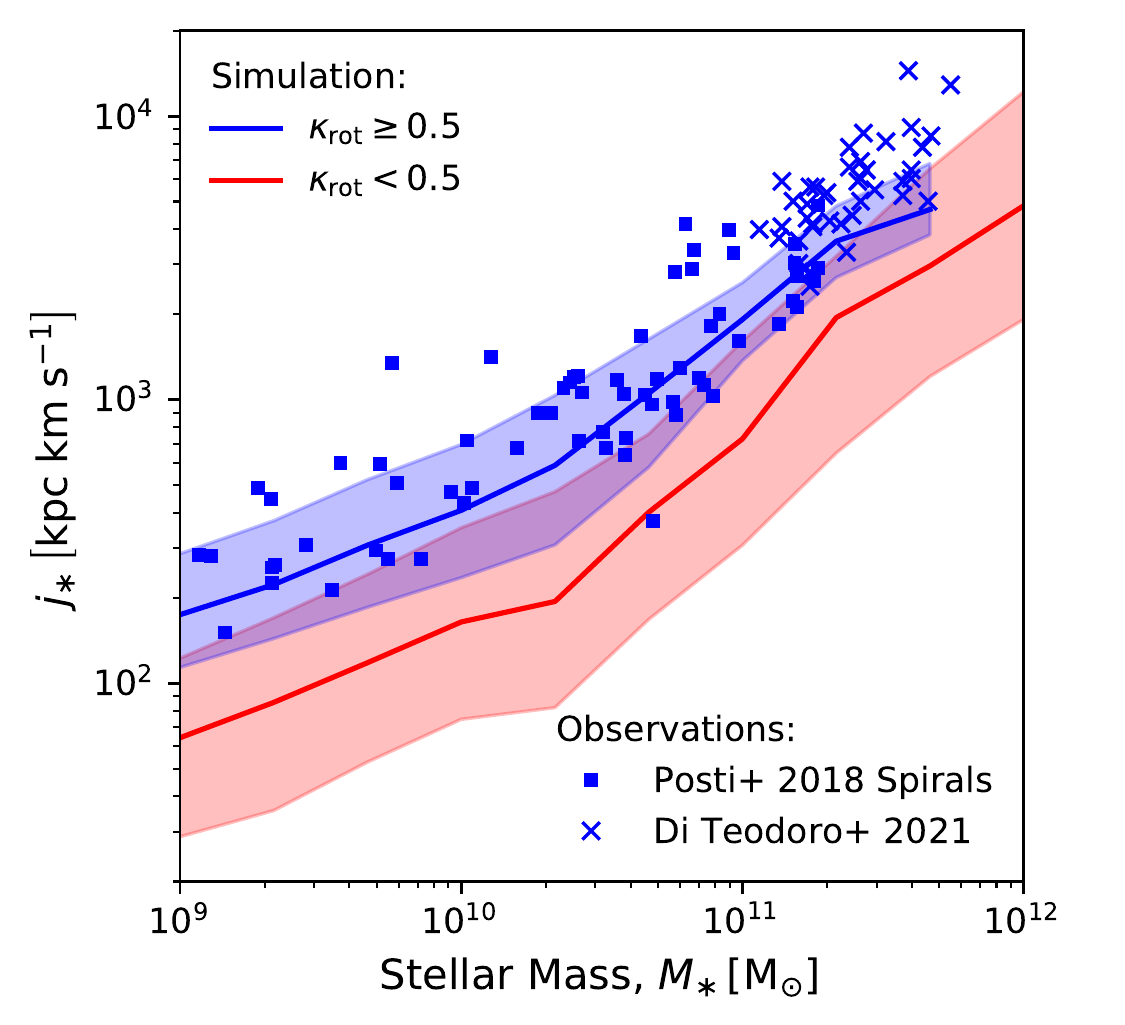}
	    \includegraphics[width=8cm]{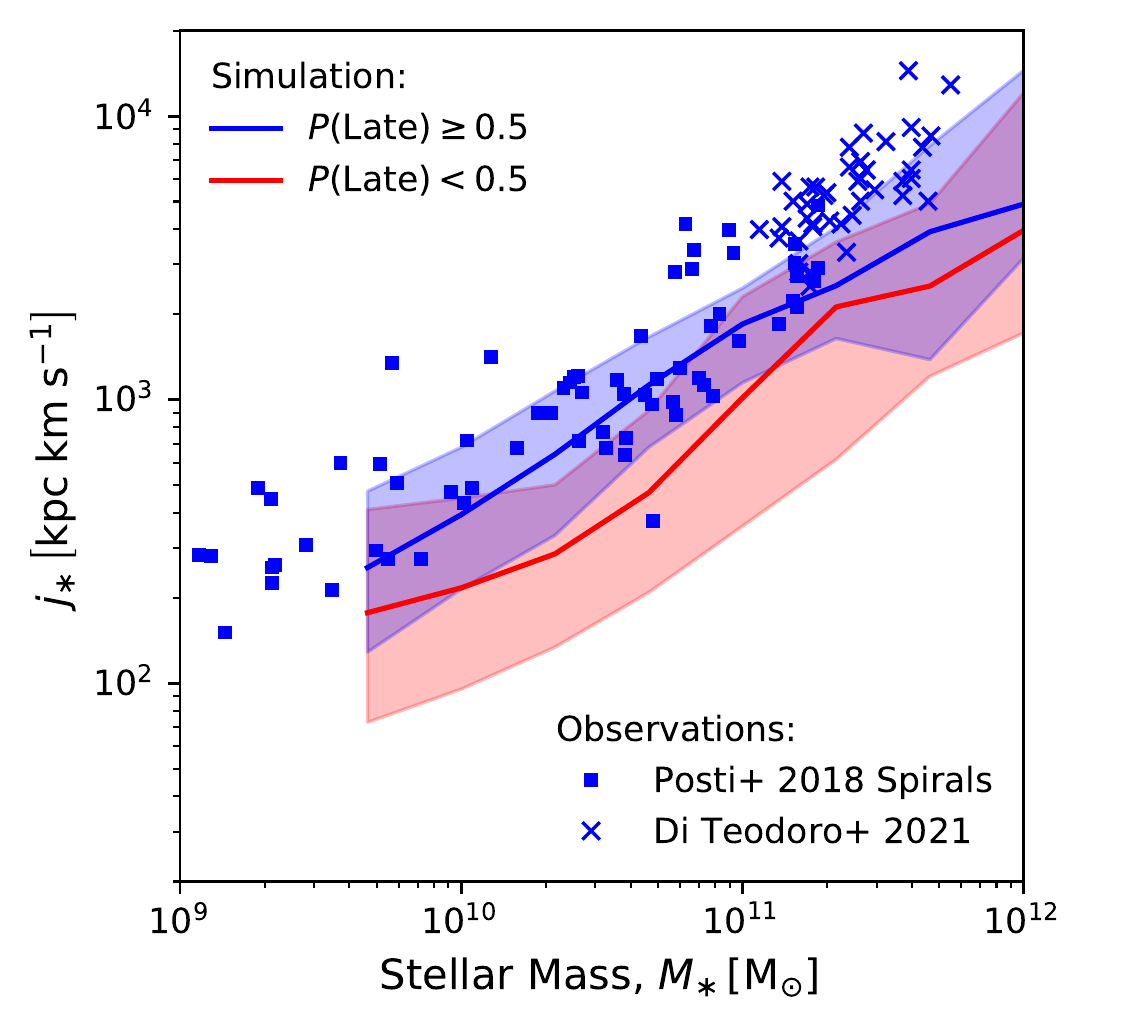}
     }

   }}
	\caption{Stellar specific angular momentum ($j_{\ast}$) as a function of stellar mass ($M_{\ast}$) for central galaxies from the IllustrisTNG simulation at $z=0$. The red and blue lines show median trends for early-type and late-type galaxies, respectively, while the shaded regions represent the corresponding 16th to 84th percentile ranges at fixed stellar mass. On the left-hand panels, galaxies are classified according to \textit{kinematic} morphology, using the $\krot$ parameter \citep{Sales2010}, while on the right-hand panels they are separated according to \textit{visual-like} morphology, using a deep-learning algorithm \citep{Huertas-Company2019}. On the top panels, the blue squares show observational $j_{\ast}$ estimates for spiral galaxies by \protect\cite{Fall2013}, while the red circles show the corresponding measurements for elliptical and lenticular (S0) galaxies. On the bottom panels, the blue symbols represent observational estimates by \protect\cite{Posti2018a} and \protect\cite{DiTeodoro2021} for spiral galaxies.}
	\label{fig:jstar_vs_mstar}
\end{figure*}

\begin{figure*}
  \centerline{\hbox{
	\includegraphics[width=8cm]{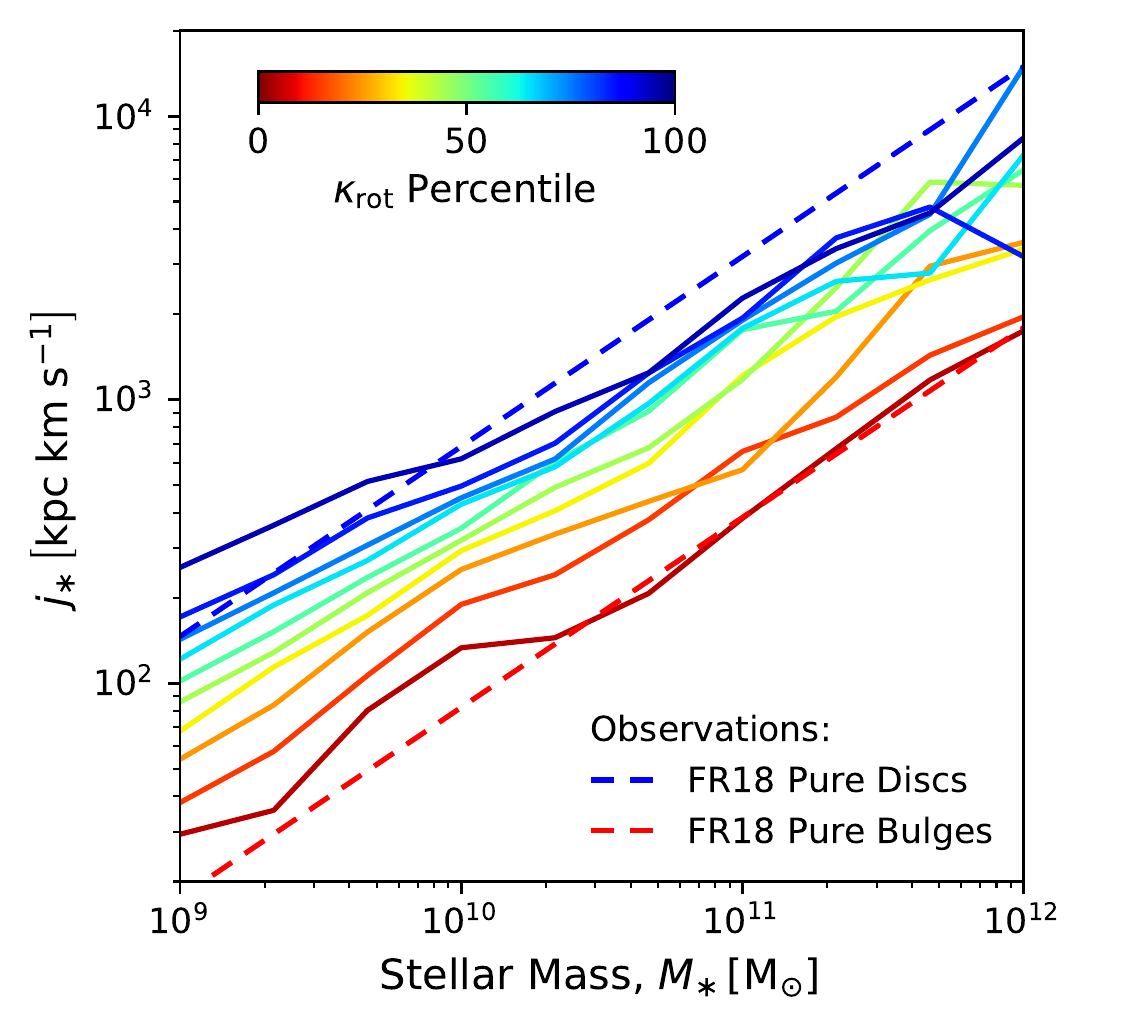}
	\includegraphics[width=8cm]{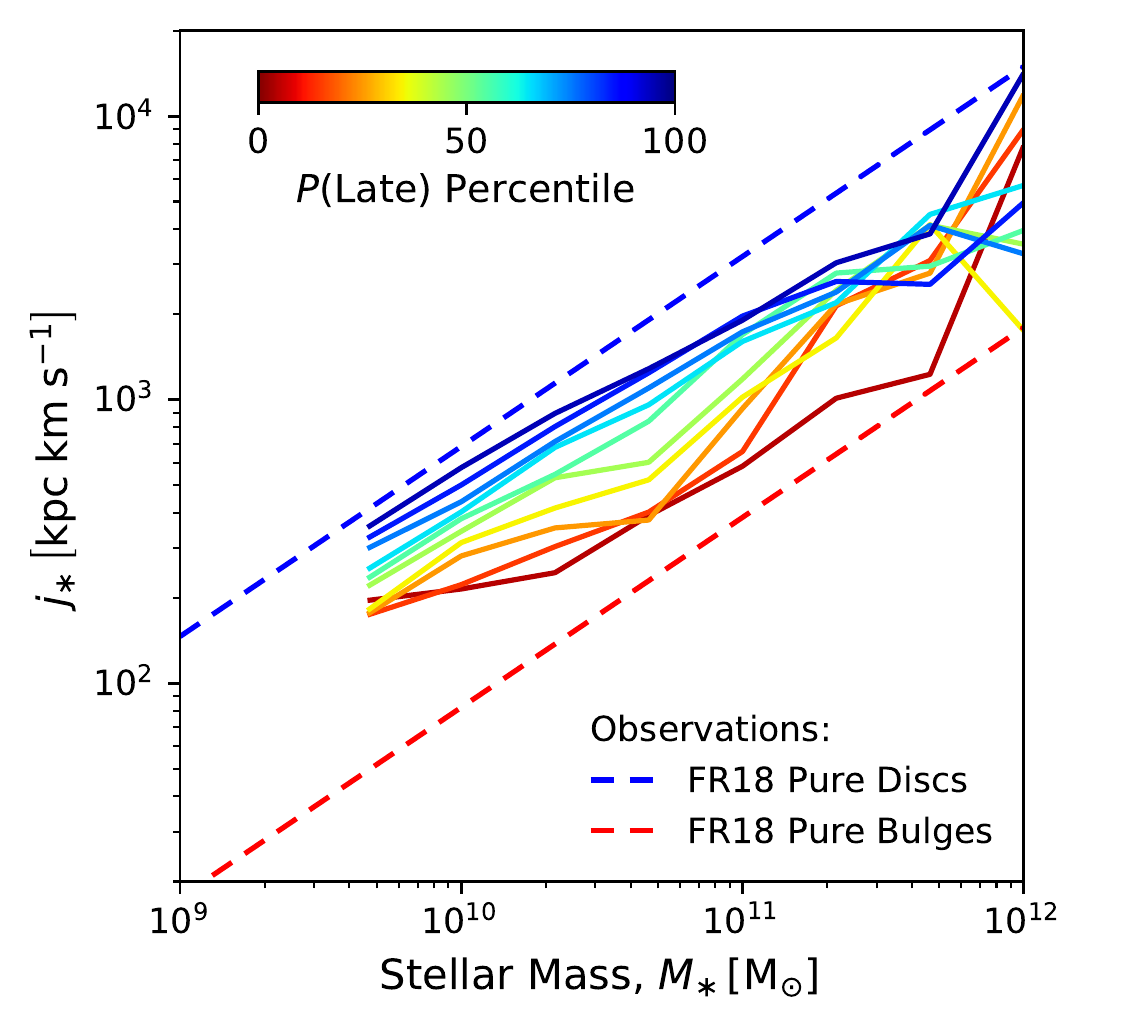}
  }}
	\caption{The $j_{\ast}$--$M_{\ast}$ relation for central galaxies at $z=0$, separating them according to the \textit{percentile at a fixed stellar mass} of $\kappa_{\rm rot}$ (left) and $P({\rm Late})$ (right), as indicated by the colour scale. Galaxies are thus classified in a way that is less sensitive to the details of the morphological measurement, instead depending only on the relative ordering of their morphologies at a fixed stellar mass. This approach results in approximately parallel tracks on the $j_{\ast}$--$M_{\ast}$ diagram for the different morphological `types'. The blue and red dashed lines show the trends inferred from the 3D fits from \protect\cite{Fall2018} for `pure' discs and bulges, respectively, which have a logarithmic slope $\alpha = 0.67 \pm 0.07 \approx 2/3$.}
	\label{fig:jstar_vs_mstar_by_percentile}
\end{figure*}

We consider all central galaxies at $z=0$ with stellar masses $M_{\ast}$ above $10^9 \, \Msun$, which represents a total of 12~235 objects. We limit this work to central galaxies because we are interested in the connection to properties of their host haloes, which would be difficult to interpret in the case of satellites. However, we note that none of our other results (i.e. the ones not involving halo quantities) would change significantly if we had included satellite galaxies as well.

When considering the visual-like morphologies from \cite{Huertas-Company2019}, only galaxies with $M_{\ast} \geq 10^{9.5} \, \Msun$ are shown, which corresponds to the stellar mass range of the galaxies with synthetic images in \cite{Rodriguez-Gomez2019}. Smaller galaxies were excluded due to the computational cost of running the \textsc{skirt} radiative transfer code \citep{Baes2011, Camps2015} on a larger galaxy sample. We also note that the analyses in \cite{Rodriguez-Gomez2019} and \cite{Huertas-Company2019}, originally carried out for galaxies at $z \approx 0.05$ (snapshot 95 in IllustrisTNG), were afterwards repeated for galaxies at $z=0$ (snapshot 99), but `mock-observing' them at $z \approx 0.05$ for the sake of consistency. This extension of previous work resulted in the visual-like morphologies used for this paper.

\section{Results: Galactic angular momentum}
\label{sec:results}

This section contains our main results, which comprise an exploration of galactic angular momentum at $z=0$ and its relation to morphology (Section \ref{subsec:connection_to_morphology}), as well as to halo spin and BH mass (Section \ref{subsec:connection_to_halo_spin_and_bh_mass}). We also test whether there is a correlation between halo spin and galaxy size in IllustrisTNG (Section \ref{subsec:halo_spin_and_galaxy_size}).

\subsection{Connection to morphology}
\label{subsec:connection_to_morphology}

Observations reveal that galaxies of different morphologies (or bulge fractions, B/T) follow roughly parallel scaling relations (in log--log space) between stellar specific angular momentum and stellar mass of the form $j_{\ast} \propto M_{\ast}^{\alpha}$, with similar indices $\alpha \approx 2/3$ but different normalizations, by a factor of $\sim$8 between late-type spirals and ellipticals \citep{Fall1983, Romanowsky2012, Fall2013, Fall2018}. These scaling relations for the stellar components of galaxies are reminiscent of the one for DM haloes, $j_{200} \propto M_{200}^{2/3} \lambda$ (since the halo spin $\lambda$ is essentially independent of $M_{200}$). Connecting the $j_{\ast}$--$M_{\ast}$ and $j_{200}$--$M_{200}$ relations by means of the intervening $j_{\ast}$--$j_{200}$ and $M_{\ast}$--$M_{200}$ relations provides some important clues about galaxy formation, as we explore in detail later in this section.

Fig. \ref{fig:jstar_vs_mstar} shows how early-type and late-type galaxies distribute on the $j_{\ast}$--$M_{\ast}$ plane, often referred to as the \textit{Fall diagram}, comparing simulated galaxies from IllustrisTNG at $z=0$ to observational measurements from \cite{Fall2013} in the upper panels,\footnote{\cite{Fall2013} revised the original $j_{\ast}$--$M_{\ast}$ data of \cite{Romanowsky2012} using more accurate mass-to-light ratios and listed these revisions in table 1 of \cite{Fall2018}. The last of these papers also presents the power-law fits to the revised $j_{\ast}$--$M_{\ast}$ data that we use here.} and from \cite{Posti2018a} and \cite{DiTeodoro2021} in the lower panels.\footnote{Three lenticular (S0) galaxies were removed from the sample by \cite{Posti2018a}, so that it consists only of late-type galaxies.} We quantify the morphologies of the simulated galaxies according to the two different methods described in Section \ref{subsec:quantifying_morphology}. On the left-hand panels, IllustrisTNG galaxies are separated according to the $\krot$ parameter \citep{Sales2010, Rodriguez-Gomez2017}, i.e. the fraction of kinetic energy contributed by the azimuthal component of the stellar velocities, while galaxies on the right-hand panels are separated by $\PLate$, the probability of having a late-type morphology according to the deep learning classifier from \cite{Huertas-Company2019}. For both $\krot$ and $\PLate$, early types are distinguished from late types by imposing a cut at a value of 0.5 \citep{Rodriguez-Gomez2017, Huertas-Company2019}, as indicated by the figure labels.

The quantities $\krot$ and $\PLate$ are examples of \textit{kinematic} and \textit{visual-like} morphologies, respectively, and we will consider both types of measurement throughout the rest of this paper in order to reduce the dependence of our results on any particular type of morphological measurement, as well as to reflect the fact that galaxy morphologies in hydrodynamic cosmological simulations are not yet in perfect agreement with observations \citep[e.g.][]{Snyder2015, Bottrell2017, Rodriguez-Gomez2019}.

Overall, Fig. \ref{fig:jstar_vs_mstar} shows good agreement between simulations and observations, except perhaps for the sample of extremely massive spirals from \cite{DiTeodoro2021}, which have somewhat higher $j_{\ast}$ values than our simulated galaxies. Other hydrodynamic cosmological simulations have also been able to reproduce the observed $j_{\ast}$--$M_{\ast}$ relation to a reasonable degree (within a factor of $\sim$2). For example, \cite{Teklu2015} and \cite{Zavala2016} report broad agreement with observational trends in the $j_{\ast}$--$M_{\ast}$ diagram for simulated galaxies from the Magneticum Pathfinder \citep{Hirschmann2014} and EAGLE \citep{Crain2015, Schaye2015} simulations, respectively, although in both cases they find that their simulated galaxies have slightly lower $j_{\ast}$ values than observed galaxies. Similarly, using the original Illustris simulation \citep{Genel2014a, Vogelsberger2014a, Vogelsberger2014}, \cite{Genel2015} find $j_{\ast}$--$M_{\ast}$ trends in reasonable agreement with observations, although with a somewhat shallower logarithmic slope for the late types. By comparison, the $j_{\ast}$--$M_{\ast}$ trends in IllustrisTNG display even better agreement with observations in both normalization and slope.

\begin{figure*}
  \centerline{\hbox{
	\includegraphics[width=8cm]{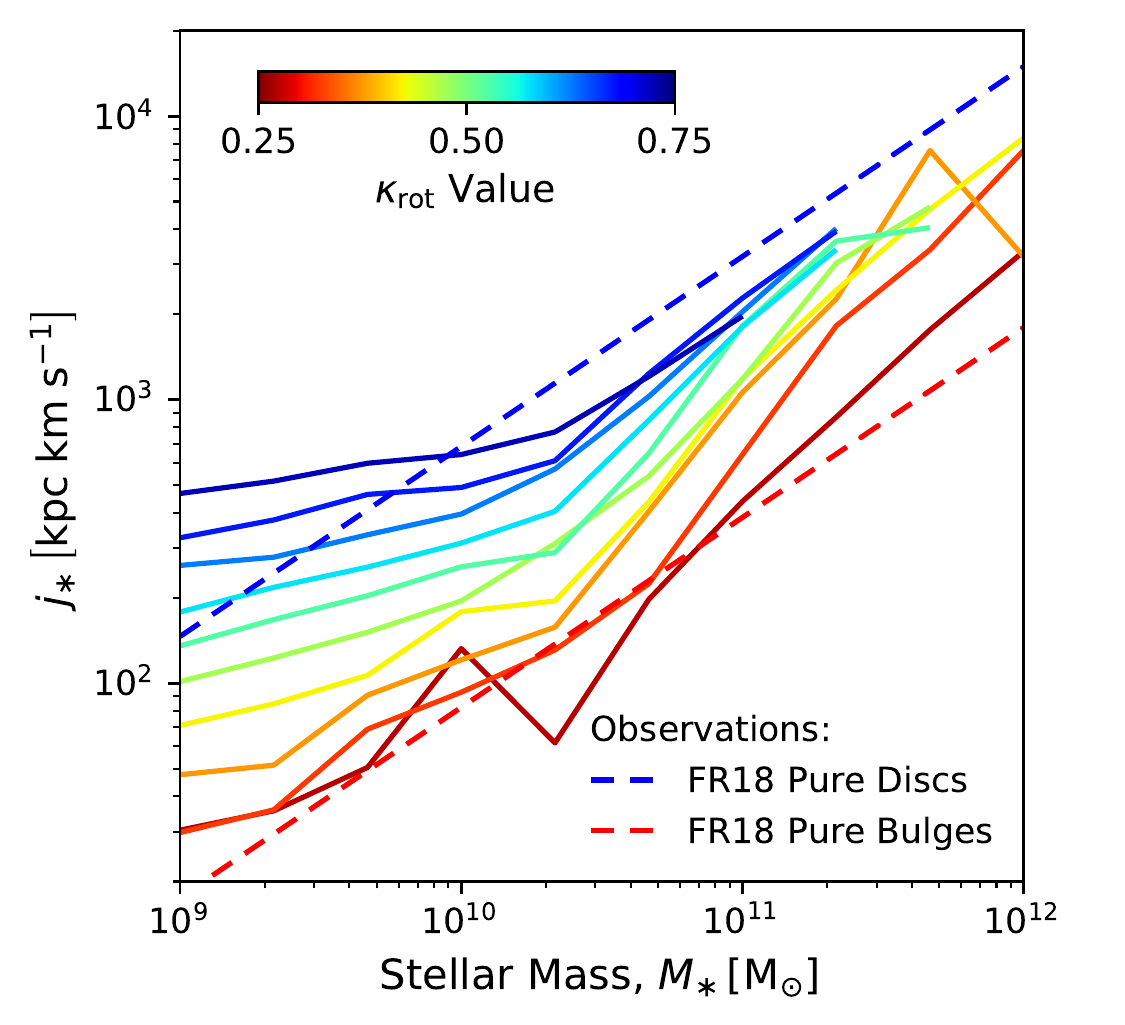}
	\includegraphics[width=8cm]{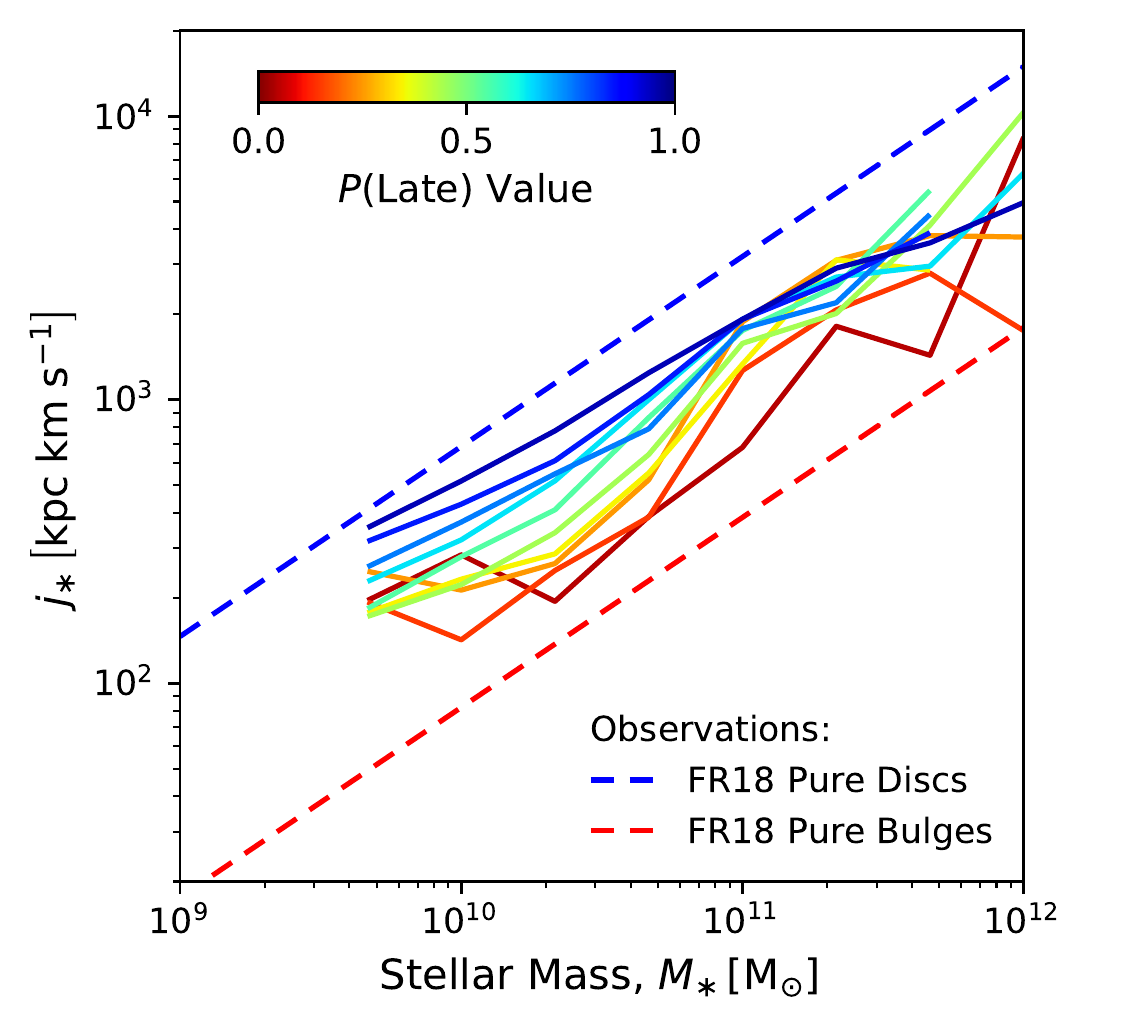}
  }}
	\caption{Same as Fig. \ref{fig:jstar_vs_mstar_by_percentile}, but separating galaxies according to the actual values of $\kappa_{\rm rot}$ (left) and $P({\rm Late})$ (right), as indicated by the colour scale, instead of using percentiles at a fixed stellar mass. While the results for $P({\rm Late})$ are relatively unchanged compared to Fig. \ref{fig:jstar_vs_mstar_by_percentile}, the $j_{\ast}$--$M_{\ast}$ trends for different $\kappa_{\rm rot}$ values now show a bend near $M_{\ast} \! \sim \! 2$--$3 \times 10^{10} \, \Msun$, such that the logarithmic slope becomes shallower than 2/3 at low masses and slightly steeper than 2/3 at the massive end.}
	\label{fig:jstar_vs_mstar_by_value}
\end{figure*}

Fig. \ref{fig:jstar_vs_mstar_by_percentile} also shows the $j_{\ast}$--$M_{\ast}$ plane in IllustrisTNG at $z=0$, but in this case the simulated galaxies are separated according to their \textit{percentile at a fixed stellar mass} in $\krot$ (left) or $\PLate$ (right), where the different coloured lines correspond to 10 equally spaced percentile ranges between 0 and 100. The motivation for doing this is to remove the dependence on the actual value of each morphological classification, and instead show the variation in $j_{\ast}$ due to the `ranking' of galaxy morphologies in each stellar mass bin. In principle, this procedure could be mimicked in observations, effectively constituting a `morphological matching' at fixed stellar mass. However, this approach would require $j_{\ast}$ measurements for complete, volume-limited galaxy samples.

The blue and red dashed lines in Fig. \ref{fig:jstar_vs_mstar_by_percentile} represent the 3D fits from \cite{Fall2018} for `pure' discs and bulges, respectively, which have a logarithmic slope $\alpha = 0.67 \pm 0.07 \approx 2/3$. These lines are remarkably close to the trends for IllustrisTNG galaxies in the highest (90--100) and lowest (0--10) percentile ranges, indicating that the most extreme morphological types in the simulation have similar $j_{\ast}$ values to those of pure discs and spheroids in the observations. Close inspection of Fig. \ref{fig:jstar_vs_mstar_by_percentile} also shows that the $j_{\ast}$ trends are slightly shallower than $M_{\ast}^{2/3}$, with logarithmic slopes of $\sim$0.5--0.6 for the various percentile intervals.

Fig. \ref{fig:jstar_vs_mstar_by_value} is analogous to Fig. \ref{fig:jstar_vs_mstar_by_percentile} except that galaxies are now split according to the actual values of $\krot$ and $\PLate$ instead of their percentiles at a fixed stellar mass. While the trends for different $\PLate$ values (right-hand panel) are relatively unchanged, the trends for different $\krot$ values (left-hand panel) now exhibit a bend near $M_{\ast} \sim 2$--$3 \times 10^{10} \, \Msun$, with logarithmic slopes steeper (shallower) than $2/3$ for galaxies with stellar masses above (below) this transition.

The main reason for the qualitative differences between Figs \ref{fig:jstar_vs_mstar_by_percentile} and \ref{fig:jstar_vs_mstar_by_value}, which are more noticeable in the case of the $\krot$ parameter (left-hand panels), is that galaxy morphology is mass-dependent: there are more disc-dominated galaxies at $\log_{10}(M_{\ast}/\Msun) \! \sim \! 10.5$ and more spheroid-dominated galaxies at lower and higher masses (this trend is reproduced in IllustrisTNG, e.g. \citealt{Tacchella2019}). However, the fact that IllustrisTNG predicts bends for the individual galaxy types (Fig. \ref{fig:jstar_vs_mstar_by_value}) is in tension with the observations by \cite{Fall2013}, which show unbent power laws. We note that \cite{DiTeodoro2021} found a slightly steeper logarithmic slope for their observational sample of `super spirals,' such that a very gradual bend appears when combining their measurements with those of \cite{Posti2018a} (see Fig. \ref{fig:jstar_vs_mstar}), but this variation in slope is much weaker than in IllustrisTNG.

\begin{figure*}
  \centerline{\hbox{
	\includegraphics[width=8cm]{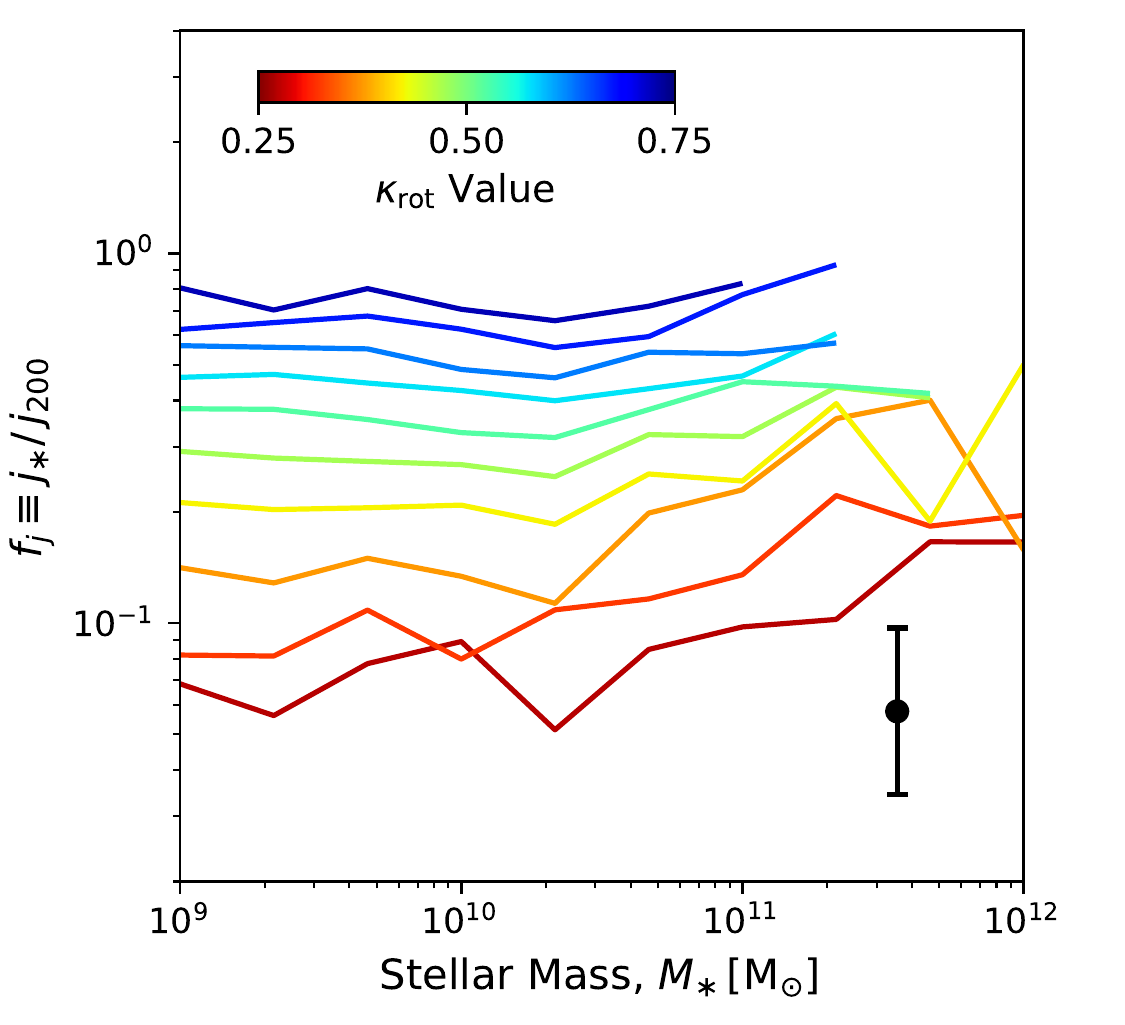}
	\includegraphics[width=8cm]{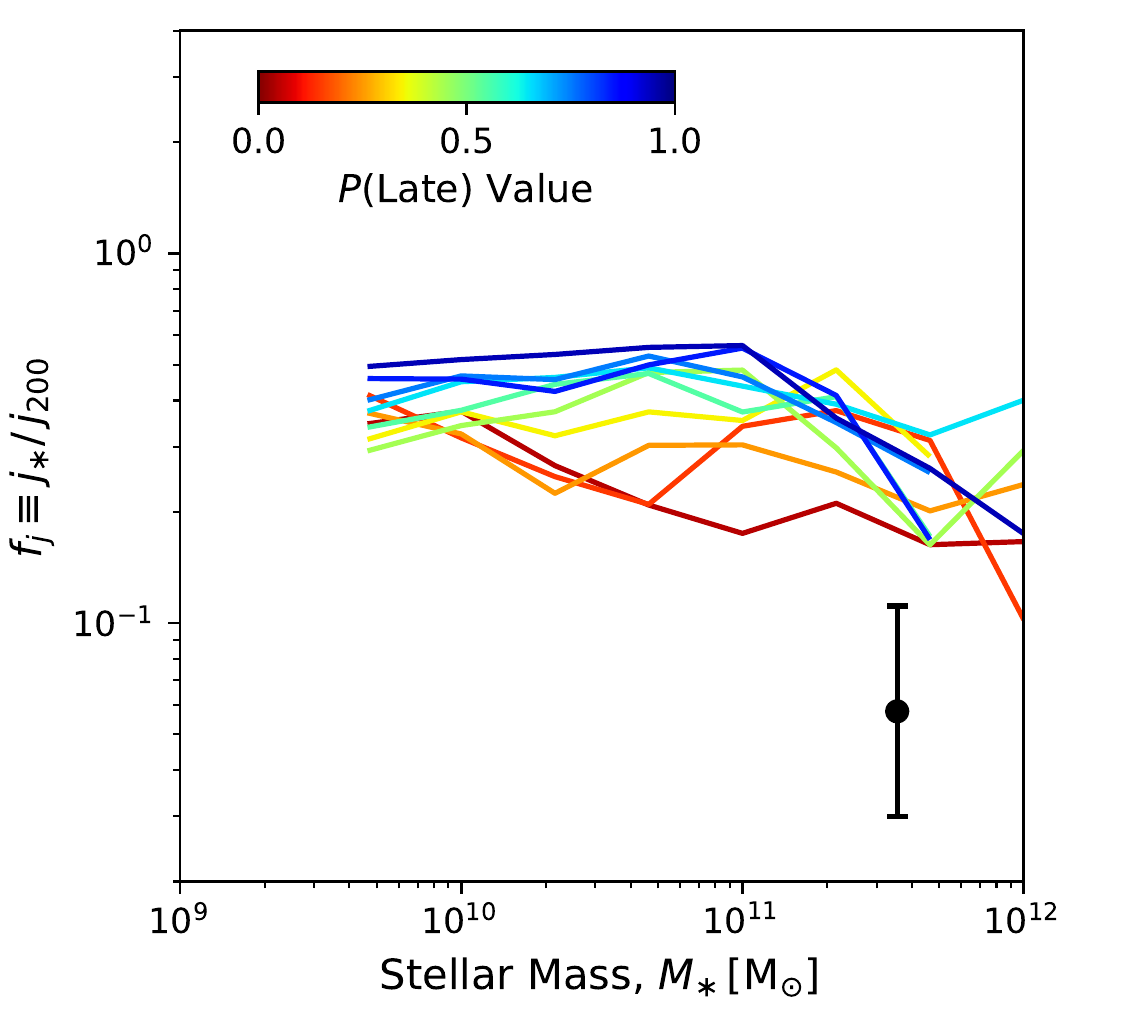}
  }}
	\caption{The specific angular momentum `retention fraction' ($f_j \equiv j_{\ast} / j_{200}$) plotted against stellar mass ($M_{\ast}$) for IllustrisTNG central galaxies at $z=0$. The different coloured lines correspond to different actual values of $\kappa_{\rm rot}$ (left) and $P({\rm Late})$ (right), as indicated by the colour scale. The error bar in the lower right-hand corner of each panel shows the typical scatter, defined as the median of the 16th to 84th percentile range (in logarithmic units) of all the data points in the figure. Note that when measuring galaxy morphology with the $\kappa_{\rm rot}$ parameter (left-hand panel), the $f_j$ distributions become remarkably constant across a wide range of stellar masses, with median values $f_j \approx 0.5$--$0.6$ for spiral galaxies ($\kappa_{\rm rot} \gtrsim 0.6$) and $f_j \approx 0.1$--$0.2$ for ellipticals ($\kappa_{\rm rot} \approx 1/3$).}
	\label{fig:retention_factor_vs_mstar}
\end{figure*}

In Fig. \ref{fig:retention_factor_vs_mstar}, we plot the specific angular momentum `retention fraction', defined as the ratio between the specific angular momentum of the galaxy and that of its host halo ($f_j \equiv j_{\ast} / j_{200}$), as a function of stellar mass for central galaxies at $z=0$. As in Fig. \ref{fig:jstar_vs_mstar_by_value}, the various lines correspond to different $\krot$ (left) and $\PLate$ (right) values, as indicated by the colour scale. Clearly, for both types of morphological measurement, late-type galaxies typically retain a higher fraction of their host haloes' specific angular momenta than early-type galaxies.

The precise behaviour of the $f_j$--$M_{\ast}$ curves in Fig. \ref{fig:retention_factor_vs_mstar} is somewhat dependent on the morphological parameter used. In the case of kinematic morphology (left-hand panel), the retained fraction of angular momentum is remarkably constant for each galaxy type across a wide range of stellar masses, with median values $f_j \approx 0.5$--$0.6$ for spirals ($\krot \gtrsim 0.6$) and $f_j \approx 0.1$--$0.2$ for ellipticals ($\krot \approx 1/3$). These values are consistent with observations showing $f_j \approx 0.7$--$0.8$ for highly disc-dominated galaxies \citep{Fall2013, Posti2019a, DiTeodoro2021} and $f_j \approx 0.1$ for highly bulge-dominated galaxies \citep{Fall2013, Fall2018}. Our $f_j$ measurements from IllustrisTNG are also broadly consistent with results from other hydrodynamic simulations for both spiral and elliptical galaxies \citep[][]{Genel2015, Pedrosa2015, Teklu2015, Zavala2016, Sokolowska2017, El-Badry2018}.

\begin{figure*}
  \centerline{\hbox{
	\includegraphics[width=8cm]{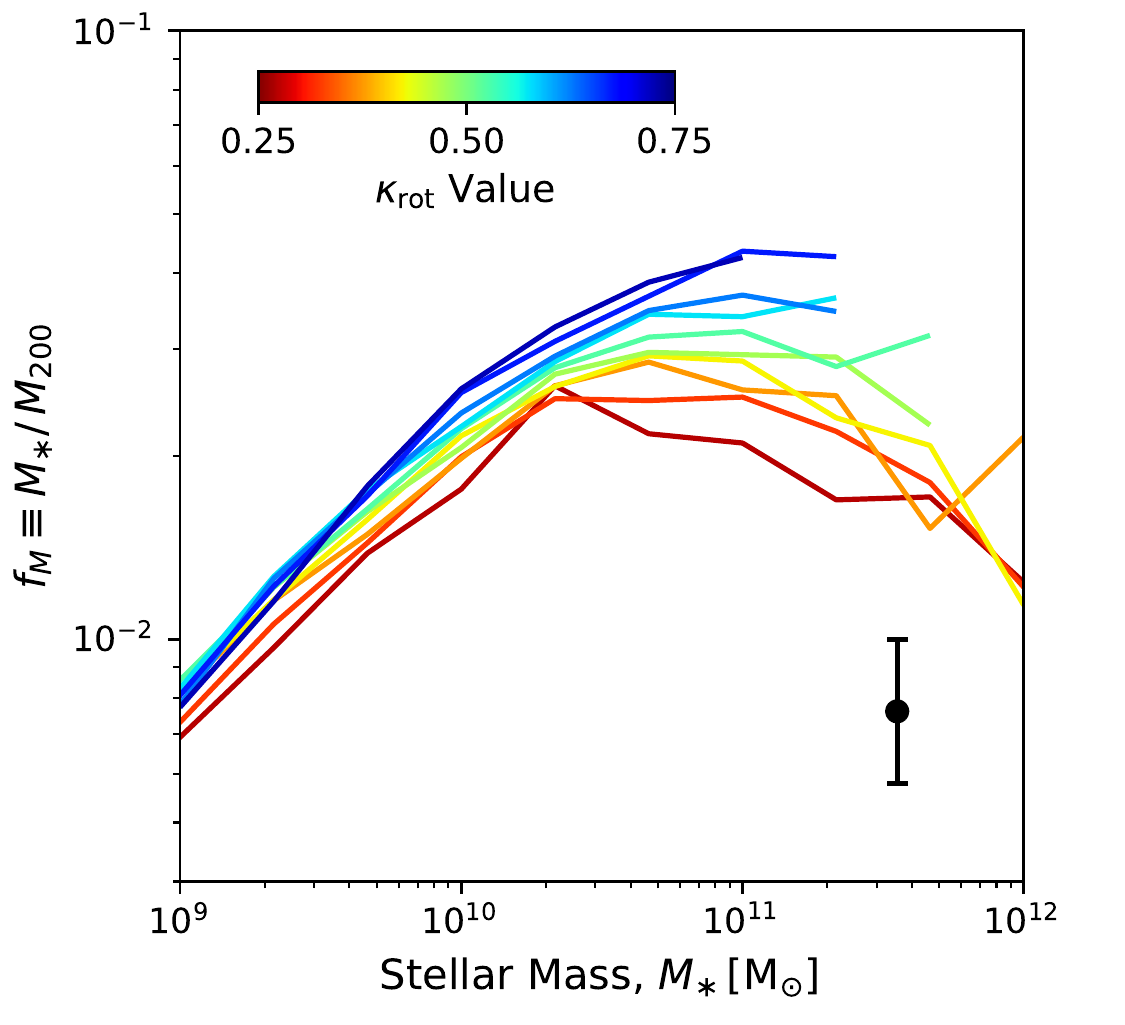}
	\includegraphics[width=8cm]{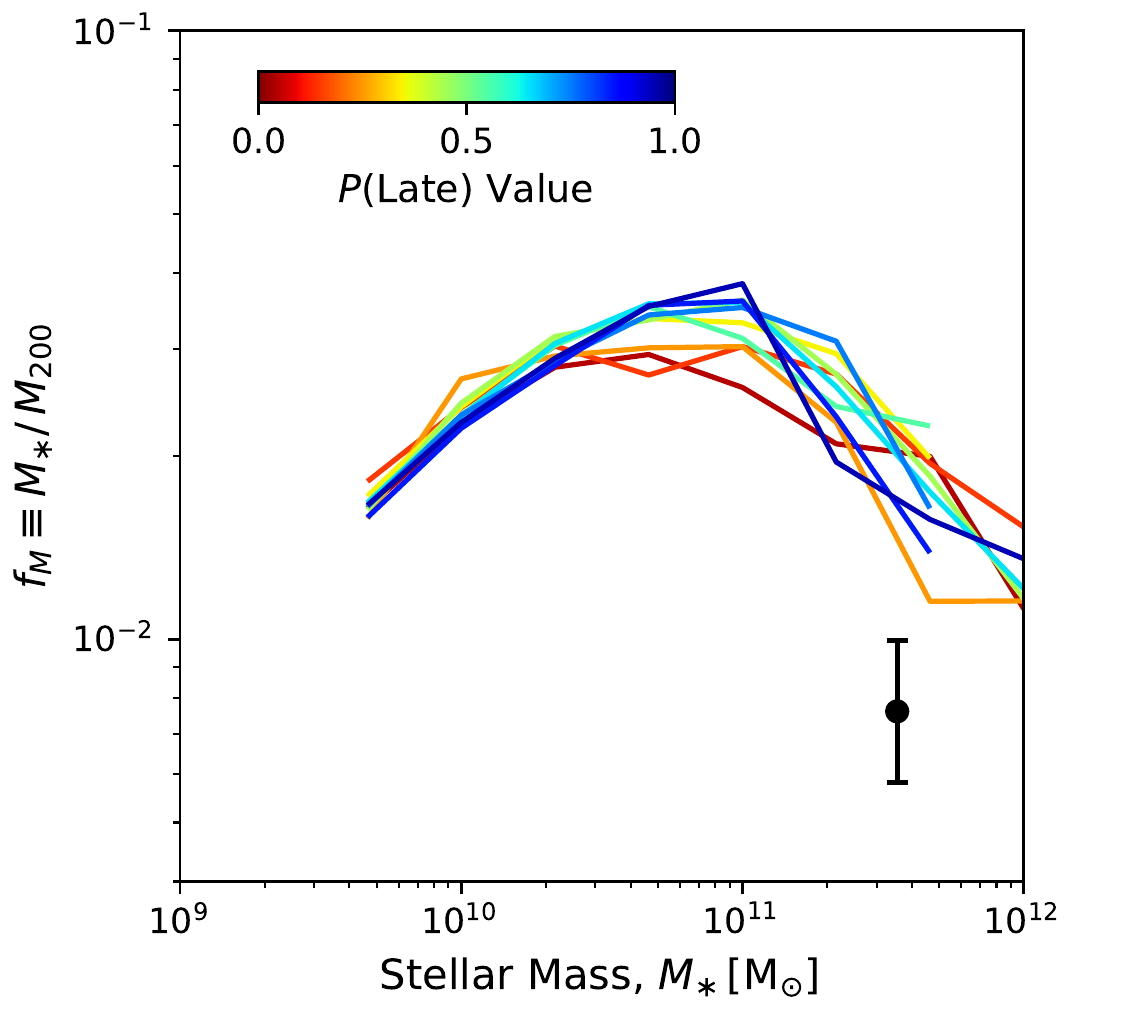}
  }}
	\caption{The stellar-to-halo mass ratio ($f_{M} \equiv M_{\ast} / M_{200}$) as a function of stellar mass ($M_{\ast}$) for IllustrisTNG central galaxies at $z=0$. As before, the different coloured lines represent different actual values of $\kappa_{\rm rot}$ (left) and $P({\rm Late})$ (right), as indicated by the colour scale, while the error bar in the lower right-hand corner shows the typical scatter. Note that when separating galaxies according to the $\krot$ parameter (left-hand panel), the $f_{M}$--$M_{\ast}$ relation is monotonically increasing for spirals ($\kappa_{\rm rot} \gtrsim 0.6$) and inverted-U-shaped for ellipticals ($\kappa_{\rm rot} \approx 1/3$), in qualitative agreement with recent works by \protect\cite{Posti2019} and \protect\cite{Posti2021}.}
	\label{fig:mstar_to_m200_vs_mstar}
\end{figure*}

When splitting galaxies according to visual-like morphology (right-hand panel of Fig. \ref{fig:retention_factor_vs_mstar}), some differences arise with respect to kinematic morphology. One of them is that the range of $f_j$ values becomes smaller, especially at low masses. This effect, which is also visible on the right-hand panel of Fig. \ref{fig:jstar_vs_mstar_by_value}, could result from the neural network having some difficulty in identifying early types at low masses. In addition, it is to some degree expected that $j_{\ast}$ should correlate more strongly with $\krot$ than with any image-based statistic, since both $j_{\ast}$ and $\krot$ are kinematic measurements based on the 3D velocities of the stellar particles -- information that is unavailable to the neural network.

Another difference between the two panels in Fig. \ref{fig:retention_factor_vs_mstar} is that the galaxies classified as late types according to $\PLate$ show a decrease in $f_j$ at high masses. In this case, one should note that this happens at \textit{very} high stellar masses, $\log_{10}(M_{\ast}/\Msun) \gtrsim 11.5$, where spirals would be exceedingly rare in the real Universe (or even in the simulation, according to the $\krot$ parameter). Therefore, any results based on $\PLate$ classifications should be taken with caution at such high masses. Additionally, these $f_j$ trends become approximately constant at $M_{\ast} \lesssim 2 \times 10^{11} \, \Msun$, which makes them qualitatively consistent with the $f_j$ trends for $\krot$-selected galaxies, as long as the spirals are limited to a realistic range of stellar masses.

The flatness of the $f_j$--$M_{\ast}$ trends in Fig. \ref{fig:retention_factor_vs_mstar} has interesting implications for the galaxy--halo connection. While \cite{Romanowsky2012} favoured a nearly constant value of $f_j$ across all stellar masses, \cite{Posti2018} argued that the $f_j$--$M_{\ast}$ relation should show an `inverted U' shape with a peak around $\log_{10}(M_{\ast}/\Msun) \! \sim \! 10.5$, which we do not see in Fig. \ref{fig:retention_factor_vs_mstar}. This prediction was obtained by enforcing agreement both with the Fall relation ($j_{\ast} \propto M_{\ast}^{2/3}$) and with an inverted-U-shaped stellar-to-halo mass ratio, $f_{M} \equiv M_{\ast} / M_{200}$, for galaxies of all morphological types from \cite{Rodriguez-Puebla2015}.

In a subsequent work, \cite{Posti2019} found that $f_{M}$ is, in fact, a monotonically increasing function of $M_{\ast}$ for spiral galaxies, with no sign of decline at high masses. This finding solves the inconsistency previously pointed out by \cite{Posti2018}, as discussed in \cite{Posti2019a} and \cite{Posti2021}. According to their statistically favoured model (a single power law), \cite{Posti2019a} found a nearly constant value of $f_j \approx 0.7$ for their entire sample of spiral galaxies, with a very weak dependence on stellar mass. This result has been confirmed with a larger sample of galaxies by \cite{DiTeodoro2021}. Our Fig. \ref{fig:retention_factor_vs_mstar} provides strong support to this scenario, in which spiral galaxies `inherit' a fixed fraction, on average, of their parent haloes' specific angular momenta, with similar findings for ellipticals.

For completeness, in Fig. \ref{fig:mstar_to_m200_vs_mstar} we plot the stellar-to-halo mass ratio ($f_{M} \equiv M_{\ast} / M_{200}$) as a function of stellar mass for central galaxies at $z=0$. As before, the left-hand and right-hand panels separate galaxies according to the $\krot$ and $\PLate$ morphological measurements, respectively. We can see that when separating galaxies according to $\PLate$ (right-hand panel), the $f_{M}$--$M_{\ast}$ relation has approximately the same shape for all morphological types, in qualitative agreement with the semi-empirical model of \cite{Rodriguez-Puebla2015}. On the other hand, when classifying galaxies according to $\krot$ (left-hand panel), $f_{M} (M_{\ast})$ is monotonically increasing for spirals and inverted-U-shaped for ellipticals, in qualitative agreement with recent works by \cite{Posti2019} and \cite{Posti2021}, respectively. A closer inspection of the $f_{M}$--$M_{\ast}$ trend for spirals, however, shows a slight bend at $M_{\ast} \gtrsim 10^{10} \, \Msun$, coming close but not quite matching the single power law favoured by \cite{Posti2019}. This slight bend is the main reason for the corresponding bend in the $j_{\ast}$--$M_{\ast}$ trend for spirals previously seen in Fig. \ref{fig:jstar_vs_mstar_by_value}, which follows from the basic relation\footnote{For the coefficient of proportionality in equation (\ref{eq:jstar_vs_mstar}), see \cite{Romanowsky2012} or \cite{Posti2018}.}
\begin{flalign}
  j_{\ast} \propto f_j f_{M}^{-2/3} M_{\ast}^{2/3} \lambda &&
  \label{eq:jstar_vs_mstar}
\end{flalign}
by noting that $f_j$ is approximately constant for the different morphological types and that $\lambda$ is independent of mass.

\begin{figure}
  \centerline{\hbox{
	\includegraphics[width=8cm]{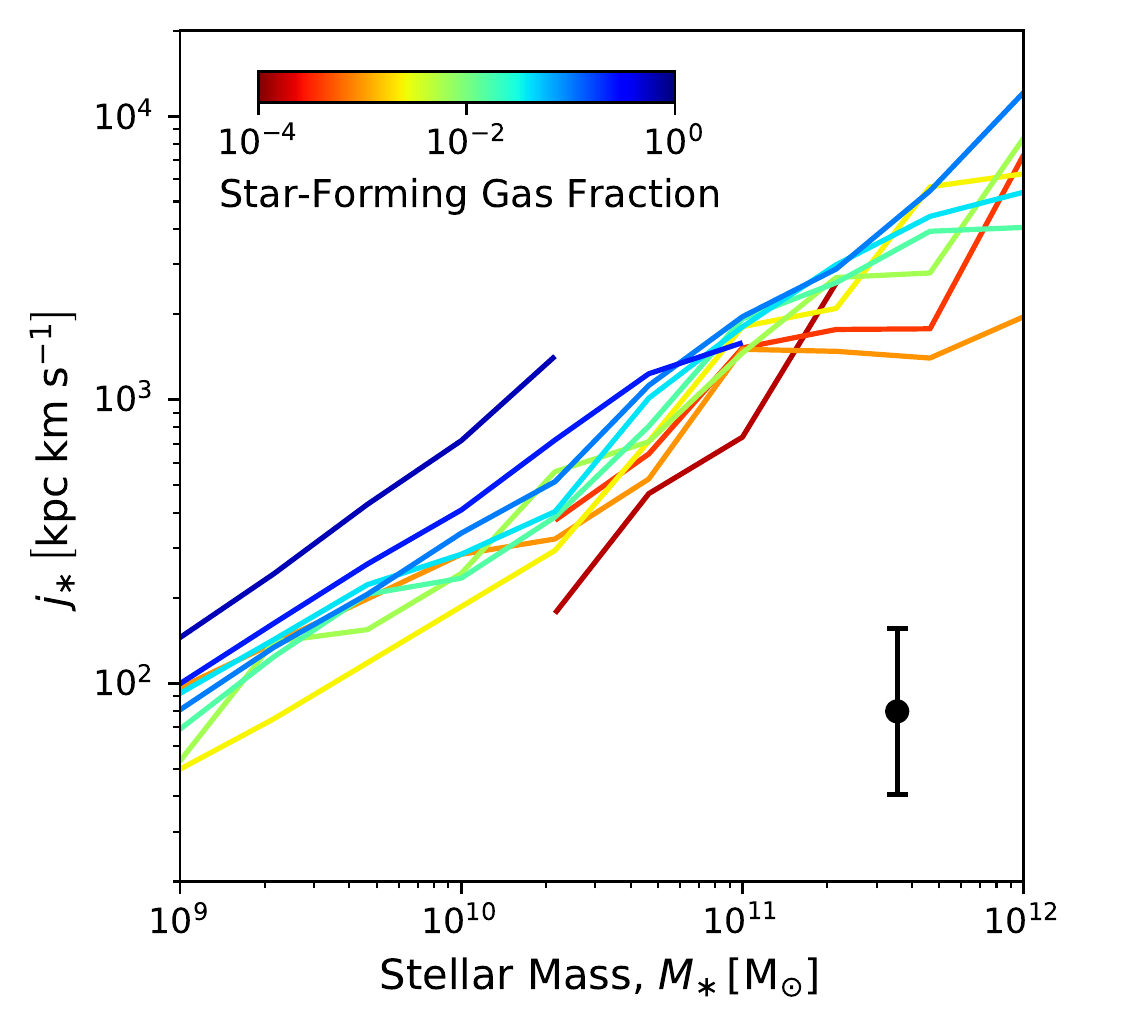}
  }}
	\caption{Stellar specific angular momentum ($j_{\ast}$) as a function of stellar mass ($M_{\ast}$) for IllustrisTNG central galaxies at $z=0$, separating them according to their star-forming gas fraction, $f_{\rm gas, sf} \equiv M_{\rm gas, sf} / (M_{\rm gas, sf} + M_{\ast})$. The error bar in the lower right corner indicates the typical scatter. This figure confirms that, at fixed stellar mass, gas-rich galaxies tend to have higher stellar angular momentum content than gas-poor galaxies, in qualitative agreement with recent works by \protect\cite{ManceraPina2021a} and \protect\cite{Hardwick2022}.}
	\label{fig:jstar_vs_mstar_by_fgas_sf}
\end{figure}

Finally, motivated by recent observations by \cite{ManceraPina2021a} and \cite{Hardwick2022}, in Fig. \ref{fig:jstar_vs_mstar_by_fgas_sf} we investigate whether the scatter in the $j_{\ast}$--$M_{\ast}$ plane is related to the cold gas fraction. As a proxy for cold gas, we consider the mass contributed by all the star-forming gas cells, $M_{\rm gas, sf}$ (see Section \ref{subsec:tng_model}), and then define the star-forming gas fraction as $f_{\rm gas, sf} \equiv M_{\rm gas, sf} / (M_{\rm gas, sf} + M_{\ast})$. Thus, Fig. \ref{fig:jstar_vs_mstar_by_fgas_sf} shows the  $j_{\ast}$--$M_{\ast}$ trends for galaxies with different values of $f_{\rm gas, sf}$, as indicated by the colour scale. We conclude from this figure that the observed correlation between stellar specific angular momentum and cold gas fraction (at fixed stellar mass) is qualitatively reproduced by the IllustrisTNG model. We note, however, that the dependence of the $j_{\ast}$--$M_{\ast}$ relation on cold gas fraction, while significant, is weaker than the dependence on galactic morphology (compare Figs \ref{fig:jstar_vs_mstar_by_percentile}, \ref{fig:jstar_vs_mstar_by_value}, and \ref{fig:jstar_vs_mstar_by_fgas_sf}).

\subsection{Connection to halo spin and BH mass}
\label{subsec:connection_to_halo_spin_and_bh_mass}

\begin{figure*}
  \centerline{\hbox{
	\includegraphics[width=8cm]{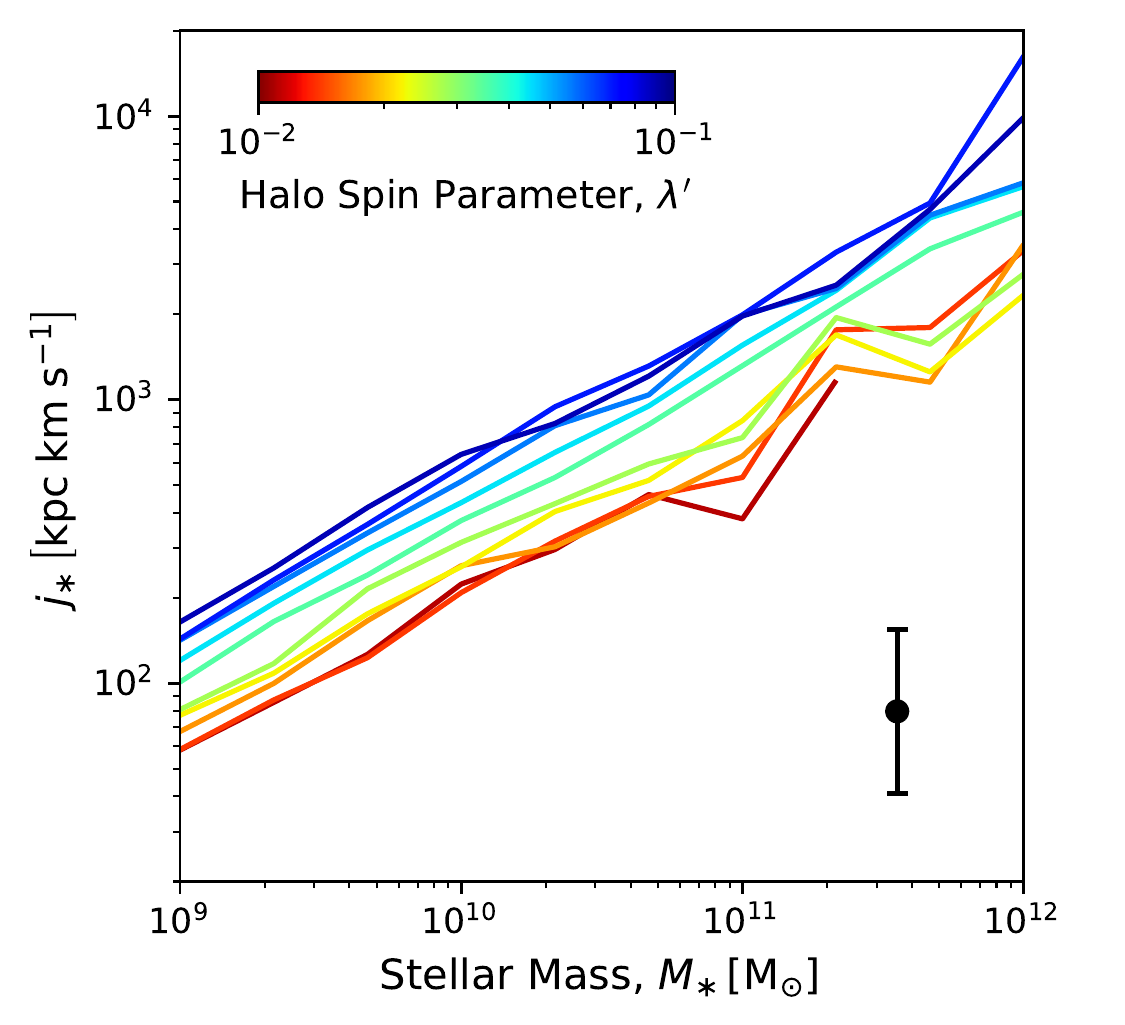}
	\includegraphics[width=8cm]{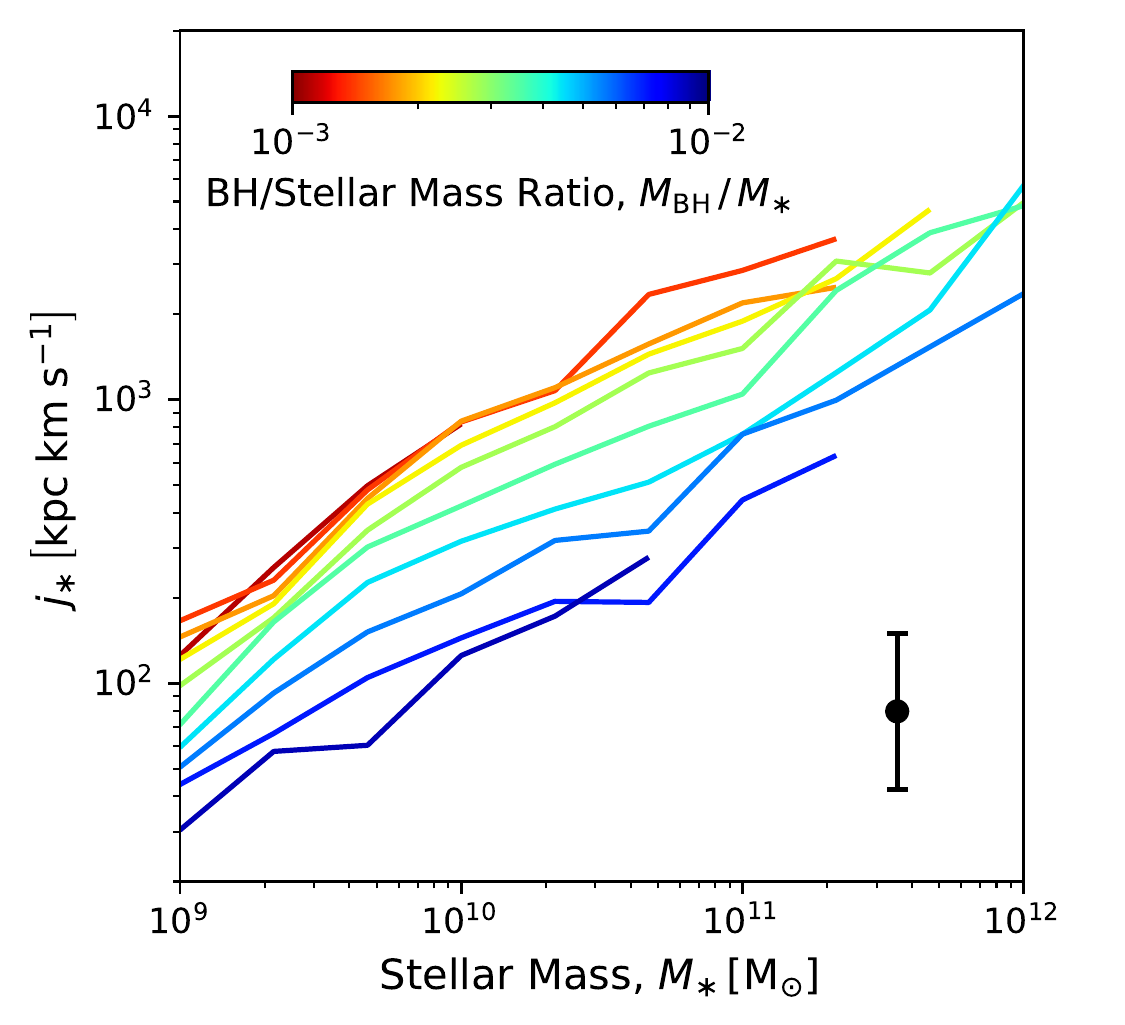}
  }}
	\caption{Stellar specific angular momentum ($j_{\ast}$) plotted against stellar mass ($M_{\ast}$) for central galaxies at $z=0$, separating galaxies according to the spin parameter of their host haloes (left) and the mass of their central BHs (right). Each coloured line shows the median $j_{\ast}$--$M_{\ast}$ trend for galaxies with the value indicated by the colour scale. The error bar in the lower right-hand corner of each panel shows the typical scatter. This figure shows that galaxies found at the centres of faster rotating haloes, as well as galaxies hosting less massive BHs, tend to have a higher angular momentum. In addition, the effects of halo spin and BH mass have approximately the same importance at all stellar masses, separating galaxies into roughly parallel tracks on the $j_{\ast}$--$M_{\ast}$ diagram.}
	\label{fig:jstar_vs_mstar_by_lambda_and_mbh_to_mstar}
\end{figure*}

\begin{figure*}
  \centerline{\hbox{
	\includegraphics[width=17.5cm]{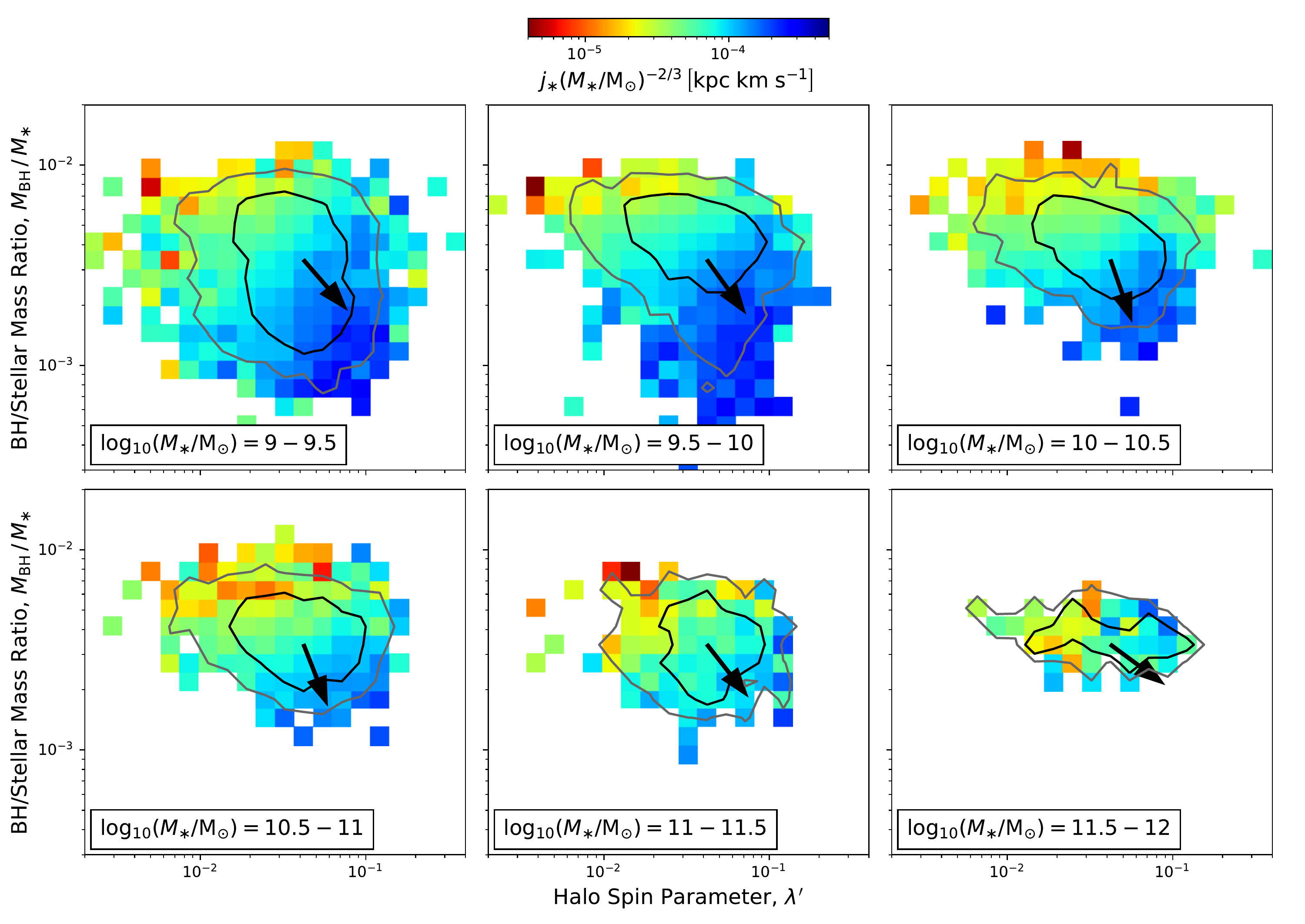}
  }}
	\caption{Variation of the stellar specific angular momentum (colour scale) with respect to the halo spin parameter ($x$-axis) and the BH-to-stellar mass ratio ($y$-axis), shown for central galaxies at $z=0$ in different stellar mass bins (different panels). The stellar specific angular momentum $j_{\ast}$ has been divided by $M_{\ast}^{2/3}$, its approximate scaling on the $j_{\ast}$--$M_{\ast}$ diagram, so that each panel covers a similar range of values. Thus, each two-dimensional bin is coloured according to the median value of $j_{\ast} / M_{\ast}^{2/3}$ of the galaxies that fall into that bin. The black and grey contours contain 68 and 95 per cent of the galaxy population in each panel, respectively, while the black arrows represent the direction of increasing $j_{\ast} / M_{\ast}^{2/3}$. Clearly, the colour gradient is approximately diagonal in each panel, which indicates that both halo spin and BH mass are important in establishing galactic angular momentum, regardless of stellar mass.}
	\label{fig:compare_three_quantities_manymasses}
\end{figure*}

In this section we explore what drives the scatter in the $j_{\ast}$--$M_{\ast}$ relation, focusing on two important physical quantities: the spin of the host halo and the mass of the supermassive BH at the galactic centre (which we use as a proxy for the amount of energy injected by AGN feedback into the gas over a galaxy's history). We also explore whether there is a correlation between these two quantities, independently of $j_{\ast}$, and whether they play a role in shaping galaxy morphology.

Fig. \ref{fig:jstar_vs_mstar_by_lambda_and_mbh_to_mstar} once again shows how IllustrisTNG galaxies distribute on the $j_{\ast}$--$M_{\ast}$ plane at $z=0$, but in this case the different coloured lines correspond to different values of the halo spin parameter (left) and the BH mass (right). As mentioned in Section \ref{subsec:measuring_am}, instead of using the original spin parameter definition by \cite{Peebles1969}, we use a very similar one by \cite{Bullock2001}, which is usually denoted by $\lambda^{\prime}$ instead of $\lambda$. The differences between these two quantities have been discussed in various works \citep[e.g.][]{Zjupa2017} and are usually of the order of 10 per cent. In the case of the BH mass, we normalize it by the galaxy's stellar mass ($M_{\rm BH} / M_{\ast}$) in order to create a more stable, dimensionless quantity that only spans $\sim$1 order of magnitude across $\sim$3 orders of magnitude in stellar mass. Fig. \ref{fig:jstar_vs_mstar_by_lambda_and_mbh_to_mstar} clearly shows that both halo spin and BH mass are correlated with galactic angular momentum: galaxies that form inside haloes with higher (lower) spin, as well as galaxies hosting less (more) massive BHs at their centres, tend to have higher (lower) values of $j_{\ast}$.

The relation between $j_{\ast}$ and halo spin parameter ($\lambda^{\prime}$) shown in Fig. \ref{fig:jstar_vs_mstar_by_lambda_and_mbh_to_mstar} has important implications for galaxy formation \citep{Fall1980, Mo1998}. It means that galaxy rotation (measured by $j_{\ast}$) depends on halo rotation (measured by $\lambda^{\prime}$), which implies that galaxy sizes should correlate with halo spin, as we will show in Section \ref{subsec:halo_spin_and_galaxy_size}. We note that a correlation between $j_{\ast}$ and halo spin had been previously found in other hydrodynamic cosmological simulations, in particular by \cite{Genel2015} and \cite{Teklu2015} for the Illustris and Magneticum Pathfinder simulations, respectively.

The relation between $j_{\ast}$ and BH mass ($M_{\rm BH}$) shown in Fig. \ref{fig:jstar_vs_mstar_by_lambda_and_mbh_to_mstar} is less straightforward to interpret. First of all, we note that such a relation is expected on purely empirical grounds, regardless of the underlying physical process: $M_{\rm BH}$ is known to correlate with bulge mass \citep[e.g.][]{Kormendy2013}, which anticorrelates with $j_{\ast}$ at fixed $M_{\ast}$ \citep[e.g.][]{Romanowsky2012}. The physical link between $M_{\rm BH}$ and bulge mass is believed to be AGN feedback, mergers, or some combination of both processes. However, while galaxy mergers (especially `dry' mergers) are known to contribute to bulge mass growth \citep[e.g.][and references therein]{Rodriguez-Gomez2017}, a preliminary analysis suggests that mergers play a less significant role than AGN feedback in establishing the $j_{\ast}$--$M_{\ast}$ relation (Rodriguez-Gomez et al., in preparation).

A plausible explanation for the effect of AGN feedback on galactic angular momentum is that AGN feedback acts mostly at late times, suppressing late-time gas accretion in galaxies with more massive BHs. Since the gas accreted at late times has higher specific angular momentum, the lack of it would explain why galaxies with higher $M_{\rm BH}$ have lower $j_{\ast}$. In this sense, the effect of AGN feedback on galactic angular momentum is opposite to that of stellar feedback \citep[e.g.][]{Ubler2014}, which favours late-time, favourably oriented gas accretion via galactic fountains. The complementary roles of AGN and stellar feedback in establishing galactic angular momentum are also evident in \cite{Genel2015}, who plotted the $j_{\ast}$--$M_{\ast}$ relation for several variations of the original Illustris model \citep{Vogelsberger2013, Torrey2014} and found that the inclusion of galactic winds increased $j_{\ast}$ at fixed $M_{\ast}$, while a model with stronger AGN feedback produced galaxies with lower $j_{\ast}$. Furthermore, using hydrodynamic zoom-in simulations of Milky Way-like galaxies, \cite{Grand2017} found an anticorrelation between disc size and BH mass growth since $z=1$, which is fundamentally the same effect.

At this point, one might wonder whether the variations in $j_{\ast}$ (at fixed $M_{\ast}$) with respect to $\lambda^{\prime}$ and $M_{\rm BH}$ seen in Fig. \ref{fig:jstar_vs_mstar_by_lambda_and_mbh_to_mstar} are independent of each other. To check this, in Fig. \ref{fig:compare_three_quantities_manymasses} we plot the joint distribution of the halo spin parameter (horizontal axis) and the BH-to-stellar mass ratio (vertical axis), while colouring each 2D bin according to the median value of $j_{\ast} / M_{\ast}^{2/3}$. This last quantity is a proxy for $j_{\ast}$ that captures its approximate scaling with $M_{\ast}$, essentially quantifying the normalization of the corresponding $j_{\ast} \propto M_{\ast}^{2/3}$ track (similar to the `$b$-values' from \citealt{Teklu2015}). The different panels show galaxies from different stellar mass ranges, as indicated by the text labels, and the black and grey contours contain 68 and 95 per cent of the galaxy population in each panel.

Examining the direction of the colour gradient (indicated by the black arrows) in each panel of Fig. \ref{fig:compare_three_quantities_manymasses}, in combination with the overall shape of the galaxy distribution (grey and black contours), provides information about the relative importance of halo spin and AGN feedback in establishing galactic angular momentum. The fact that the colour gradients are predominantly diagonal (rather than horizontal or vertical) in all panels indicates that both halo spin and AGN feedback play an important role in determining the stellar specific angular momentum of a galaxy, regardless of stellar mass. In other words, any panel from Fig. \ref{fig:compare_three_quantities_manymasses} shows that even at fixed $\lambda^{\prime}$, more (less) massive BHs are associated with lower (higher) $j_{\ast}$ values. Similarly, at fixed $M_{\rm BH} / M_{\ast}$, faster (slower) rotating haloes host galaxies with higher (lower) $j_{\ast}$ at their centres.\footnote{We also checked the halo concentration, but found that it does not correlate significantly with $j_{\ast}$ or $M_{\rm BH}$. However, we found that the halo concentration anticorrelates slightly with halo spin, an effect usually attributed to haloes out of equilibrium \citep{Maccio2007}.}

\begin{figure}
  \centerline{\hbox{
	\includegraphics[width=8cm]{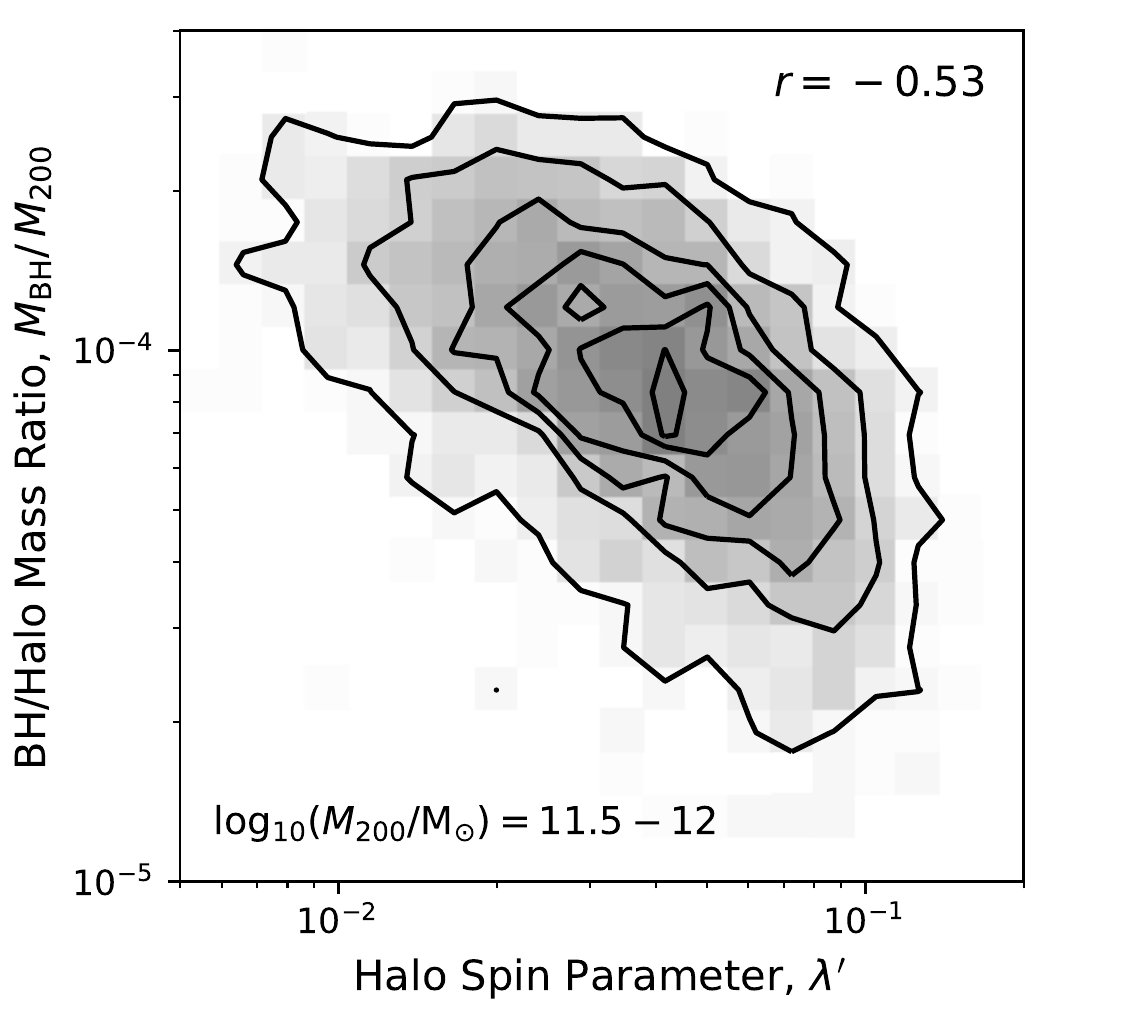}
  }}
	\caption{The correlation between BH mass ($y$-axis, normalized by $M_{200}$) and halo spin ($x$-axis) at $z=0$. This plot is limited to central galaxies within a relatively narrow range of host halo masses ($M_{200} = 10^{11.5}$--$10^{12} \, \Msun$), which is approximately where we found the correlation to be strongest, although such a connection to the BH mass exists for all sufficiently massive haloes ($M_{200} \gtrsim 10^{11} \, \Msun$) in IllustrisTNG. The contours contain 5, 20, 40, 60, 80 and 95 per cent of the galaxies, while the grey-scale shows their overall distribution. The Pearson correlation coefficient $r$ of the data (in logarithmic units) is indicated in the upper right corner of the figure.}
	\label{fig:mbh_to_m200_vs_lambda_halo}
\end{figure}

Close inspection of the contours in Fig. \ref{fig:compare_three_quantities_manymasses} reveals that there is also a connection between halo spin and BH mass, which deserves further attention. This correlation becomes more evident if we essentially bypass the galaxy and replace the ratio $M_{\rm BH} / M_{\ast}$ with $M_{\rm BH} / M_{200}$, i.e. if we consider only halo- and BH-related properties. This is done in Fig. \ref{fig:mbh_to_m200_vs_lambda_halo}, which shows the joint distribution of $M_{\rm BH} / M_{200}$ and $\lambda^{\prime}$ for a range of halo masses $M_{200} = 10^{11.5}$--$10^{12} \, \Msun$. This is roughly the mass range where we find the correlation to be strongest, with a Pearson correlation coefficient of $r = -0.53$ (for the logarithms of $M_{\rm BH} / M_{200}$ and $\lambda^{\prime}$), although we verified that such a correlation exists for all sufficiently massive haloes, with $M_{200} \gtrsim 10^{11} \, \Msun$. We find that such a relation is also present in the original Illustris simulation, although it is somewhat weaker, with a Pearson correlation coefficient of $r = -0.33$ for the same halo mass range.

This inverse relation between halo spin and BH mass is likely a consequence of the higher centrifugal acceleration in faster rotating haloes that hinders gas flow to the galactic centre, thus inhibiting the growth of the central BH. Recently, also using the TNG100 simulation, \cite{Lu2022} found decreased star formation activity and lower BH accretion rates in galaxies at the centres of faster spinning haloes, which they also attributed to less efficient gas inflow in these systems. These findings lead to a picture in which the supermassive BHs to some extent `amplify' the effect of halo spin on galactic angular momentum: galaxies with lower $j_{\ast}$ form at the centres of haloes with lower $\lambda^{\prime}$, but the lower centrifugal acceleration in such haloes results in stronger gas accretion towards the central BH, which contributes to further decreasing $j_{\ast}$ via AGN feedback.

When interpreting these results, it is important to note that the process of accretion on to supermassive BHs takes place at scales much smaller than the spatial resolution of current cosmological and galaxy-scale simulations. For example, the `radius of influence' of supermassive BHs, $r_{\rm infl} \equiv G M_{\rm BH} / \sigma^2$ (where $\sigma$ is the velocity dispersion of the galactic bulge), has typical values between 1 and 100 pc, much below the $\sim$kpc resolution of TNG100. Furthermore, at even smaller scales, the angular momentum transport mechanisms that funnel gas from the interstellar medium into the BH accretion disc (of size $\lesssim 0.01$ pc) are not fully understood \citep[e.g.][and references therein]{Hopkins2022}. Despite these issues, the fact that the BH accretion model adopted in IllustrisTNG is coarse (see Section \ref{subsec:tng_model}) does not mean that it is unphysical. In particular, it still seems plausible that low-angular-momentum gas can flow more efficiently towards the central BH than high-angular-momentum gas, regardless of the details of how this gas is transported at subresolution scales.

\begin{figure*}
  \centerline{\hbox{
	\includegraphics[width=8cm]{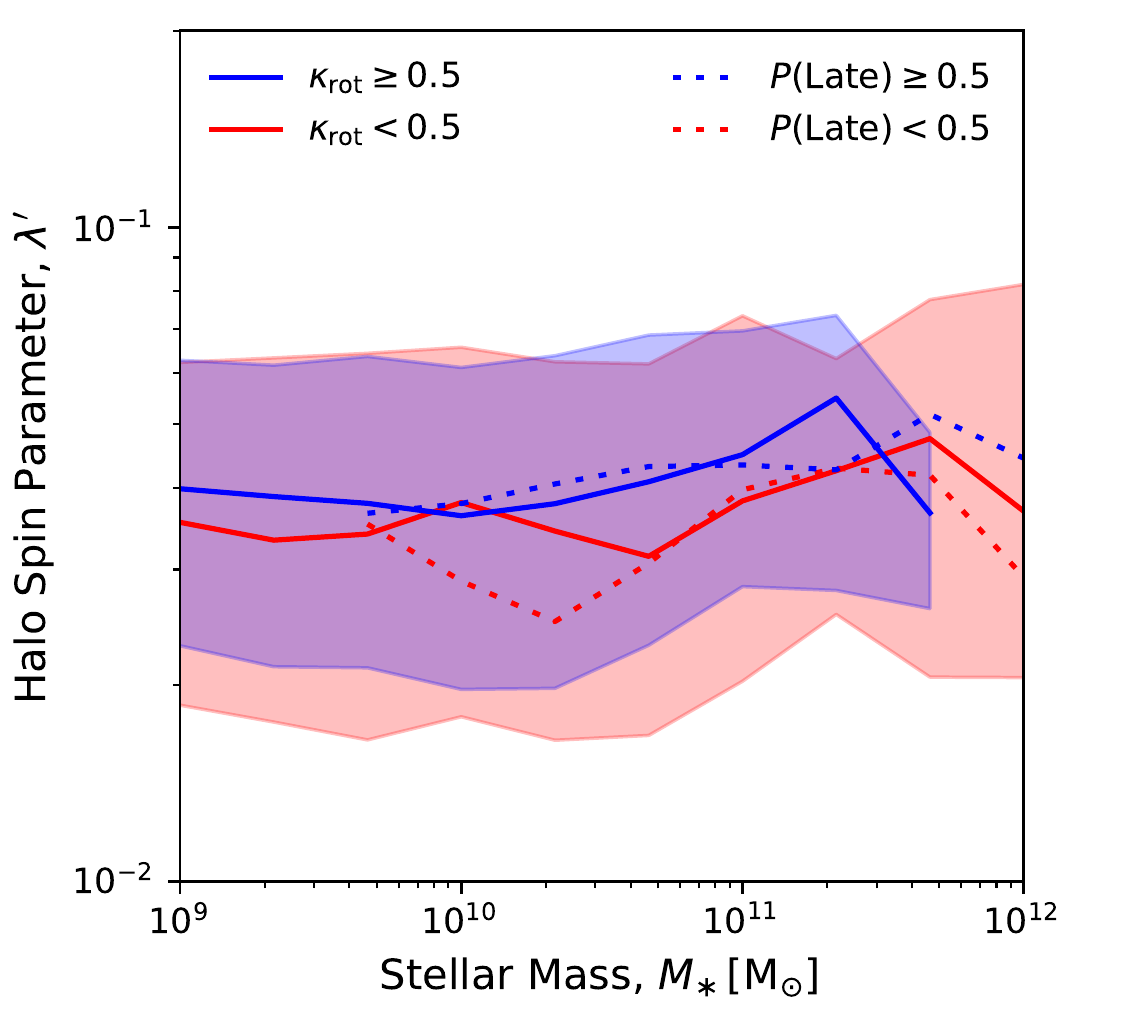}
	\includegraphics[width=8cm]{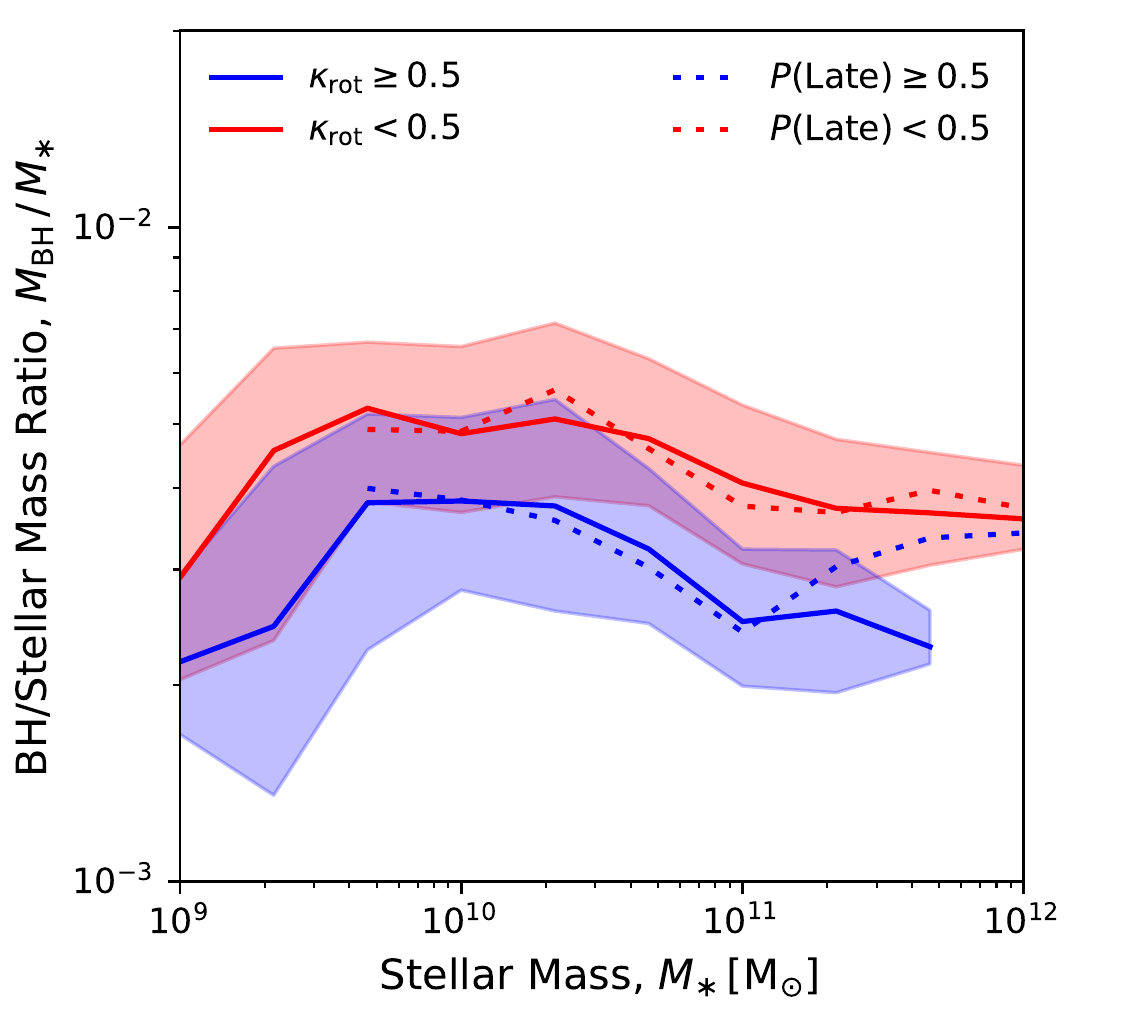}
  }}
	\caption{The effect of halo spin (left) and BH mass (right) on galaxy morphology as a function of stellar mass, again shown for central galaxies at $z=0$. The red and blue solid lines show the median trends for early-type and late-type galaxies, respectively, classified according to \textit{kinematic} morphology ($\kappa_{\rm rot}$), while the shaded regions indicate the corresponding 16th to 84th percentile ranges. Similarly, the red and blue dotted lines show the median trends for early-type and late-type galaxies classified according to \textit{visual-like} ($P({\rm Late})$) morphology. This figure shows that, in the IllustrisTNG simulation, halo spin has a marginal effect on galaxy morphology, while BH mass plays a more important role.}
	\label{fig:lambda_halo_and_mbh_to_mstar_vs_mstar}
\end{figure*}

Having seen that both halo spin and AGN feedback have an effect on galactic angular momentum, one might wonder whether they also have an effect on galaxy morphology. This is tested in Fig. \ref{fig:lambda_halo_and_mbh_to_mstar_vs_mstar}, which shows the halo spin parameter (left) and the BH-to-stellar mass ratio (right) as a function of stellar mass, separating galaxies into early types and late types according to the two morphological quantities described in Section \ref{subsec:quantifying_morphology}. The solid lines and shaded regions show the median and scatter of $\krot$-selected early types (red) and late types (blue), as indicated by the figure labels, while the dotted lines show the median trends for late-type and early-type galaxies selected according to $\PLate$.

The right-hand panel of Fig. \ref{fig:lambda_halo_and_mbh_to_mstar_vs_mstar} clearly shows that BH mass is correlated with galaxy morphology at all stellar masses, with early-type (late-type) galaxies hosting more (less) massive BHs. This is an expected result of the AGN feedback implementation in IllustrisTNG, which heats the gas surrounding a galaxy, suppressing late-time gas accretion and hence the formation of galactic discs, although galaxy mergers are also expected to play a role at the high-mass end \citep[][]{Rodriguez-Gomez2017}. A similar correlation between galaxy morphology and BH mass in IllustrisTNG was found by \cite{Li2020}. These findings are qualitatively consistent with the well-known scaling relation between BH mass and bulge mass \citep[e.g.][]{Kormendy2013}.

On the other hand, the left-hand panel of Fig. \ref{fig:lambda_halo_and_mbh_to_mstar_vs_mstar} shows a negligible correlation between halo spin and galaxy morphology at all inspected masses, which might seem puzzling considering that we find a strong correlation between morphology and $j_{\ast}$ (Figs \ref{fig:jstar_vs_mstar}--\ref{fig:jstar_vs_mstar_by_value}), as well as between $j_{\ast}$ and $\lambda^{\prime}$ (Figs \ref{fig:jstar_vs_mstar_by_lambda_and_mbh_to_mstar}--\ref{fig:compare_three_quantities_manymasses}). The reason for this apparent contradiction is that kinematic morphological measurements such as $\krot$ are more strongly correlated with the \textit{inner} $j_{\ast}$ \citep[][]{Rodriguez-Gomez2017}, while the halo spin parameter is more strongly correlated with the \textit{outer} $j_{\ast}$. This can lead to a situation where $j_{\ast}$ is strongly correlated with both $\krot$ and $\lambda^{\prime}$, while these two quantities are very weakly correlated with each other.

Some previous works have reported a correlation between halo spin and morphology in hydrodynamic cosmological simulations, such as \cite{Teklu2015} for the Magneticum Pathfinder simulation and \cite{Rodriguez-Gomez2017} for the original Illustris simulation (at low masses). Regarding the latter case, one possible explanation for why we do not see this effect in IllustrisTNG could be the fact that Illustris galaxies were larger by a factor of $\sim$2, especially at the low-mass end \citep[][]{Snyder2015, Pillepich2018, Rodriguez-Gomez2019}, which could make them more `dynamically connected' to their parent DM haloes by having an approximately eight times larger `effective volume' in common. In addition, as a consequence of weaker stellar feedback, the stellar masses in Illustris were higher by a factor of $\sim$2 at the low-mass end, suggesting that stellar angular momentum could also have been affected.

\subsection{Halo spin and galaxy size}
\label{subsec:halo_spin_and_galaxy_size}

\begin{figure*}
  \centerline{\hbox{
	\includegraphics[width=8cm]{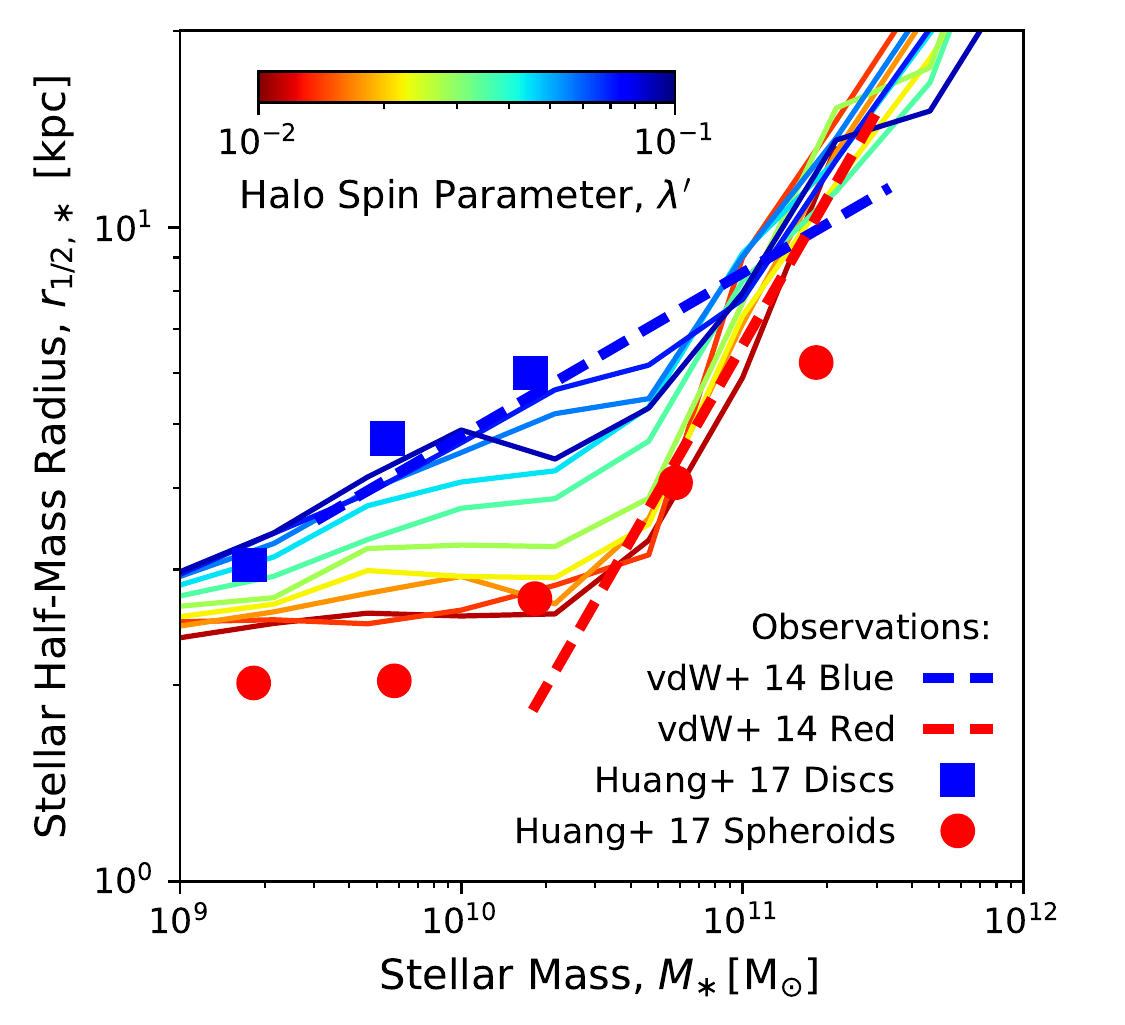}
	\includegraphics[width=8cm]{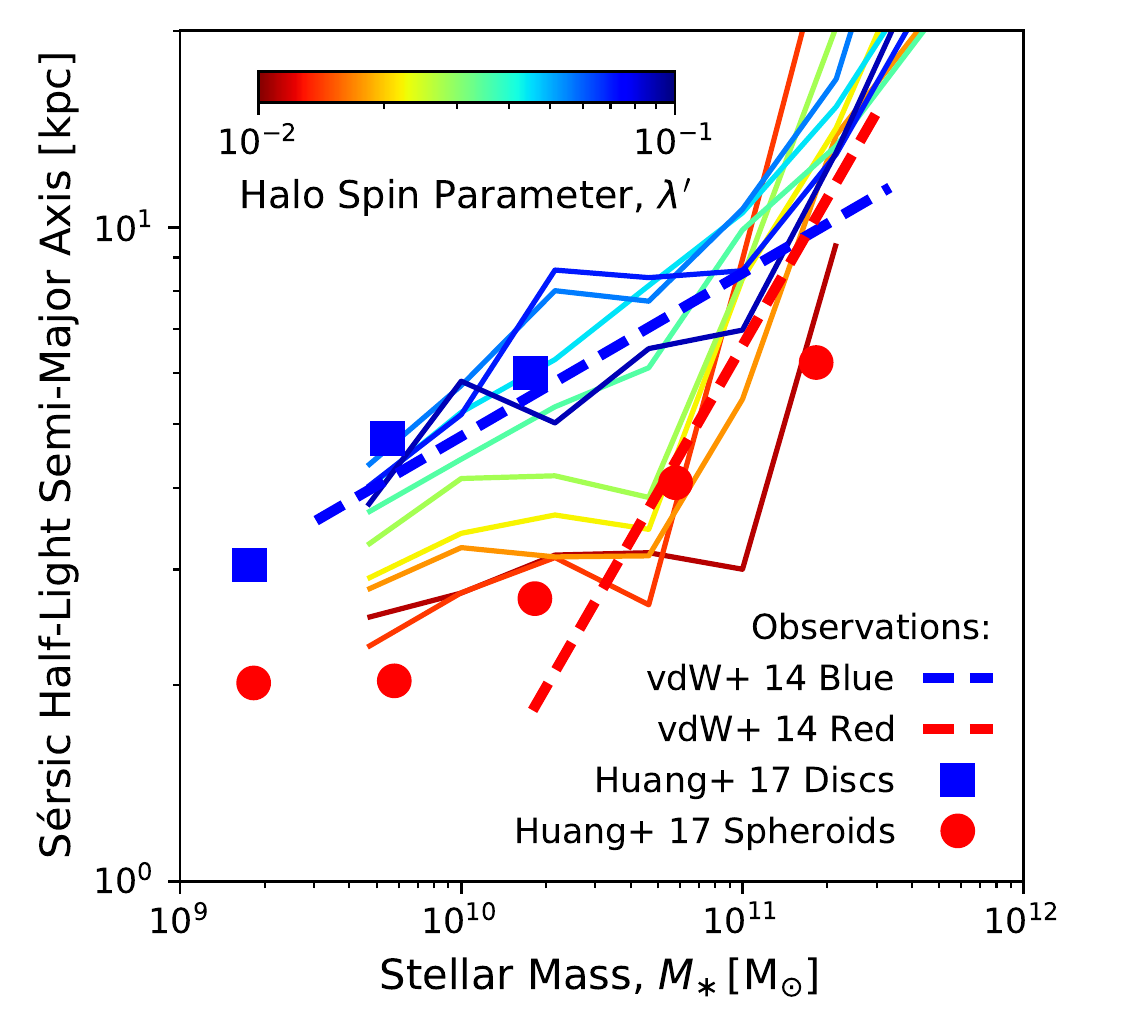}
  }}
	\caption{Galaxy size, parametrized by the (3D) stellar half-mass radius (left) and the (2D) S\'ersic half-light semimajor axis (right), plotted against stellar mass for central galaxies at $z=0$. The different solid coloured lines show median trends for galaxies found at the centres of haloes with different spin parameters, as indicated by the colour scale. The blue squares and red circles show the median observational trends from \protect\cite{Huang2017} for disc-dominated and spheroid-dominated galaxies, respectively, while the dashed lines represent power-law fits by \protect\cite{vanderWel2014} for blue and red galaxies. This figure shows that larger galaxies formed at the centres of faster rotating haloes, in agreement with simple analytical arguments, and that the variations in halo spin span the observed range of morphologies and colours. However, the trend breaks down at $M_{\ast} \gtrsim 10^{11} \, \Msun$ in both observations and simulations, presumably as a result of the increased importance of dry mergers in the formation of massive galaxies and the associated destruction of galactic discs.}
	\label{fig:rhalf_vs_mstar_by_lambda_halo}
\end{figure*}

A well-known prediction of galaxy formation models is that the size of a galactic disc should be correlated to the spin of its parent halo \citep[][]{Fall1980, Mo1998}. In Fig. \ref{fig:rhalf_vs_mstar_by_lambda_halo}, we investigate whether such a relation exists in IllustrisTNG by plotting galaxy size as a function of stellar mass for central galaxies at $z=0$. The left-hand panel quantifies galaxy size in a very straightforward manner, using the (3D) stellar half-mass radius, while the right-hand panel follows a more observationally motivated approach, showing the half-light semimajor axis obtained from 2D S\'ersic fits to the galaxy light distribution in the Pan-STARRS $i$-band \citep{Rodriguez-Gomez2019}. The coloured solid lines represent different values of the halo spin parameter in logarithmically spaced bins, as indicated by the colour scale. Clearly, a correlation between galaxy size and halo spin exists in IllustrisTNG, at least at stellar masses $M_{\ast} \lesssim 10^{11} \, \Msun$, with larger discs forming at the centres of faster rotating haloes. Such a relation had been previously found using hydrodynamic zoom-in simulations of Milky Way-like galaxies \citep{Grand2017}.\footnote{Some hydrodynamic zoom-in simulations produce a weaker correlation between galaxy size and halo spin \citep{Jiang2019}. The origin of this discrepancy will be clarified in future work.} However, this is the first time, to our knowledge, that this correlation has been shown to appear naturally in a fully cosmological volume.

We complement Fig. \ref{fig:rhalf_vs_mstar_by_lambda_halo} by also showing observational determinations of the effective size versus stellar mass relation for different galaxy types. The blue squares and red circles represent median observational trends obtained by \cite{Huang2017} for disc-dominated and spheroid-dominated galaxies, respectively, using data from the CANDELS survey \citep[][]{Grogin2011, Koekemoer2011} at $z < 0.5$. Similarly, the blue and red dashed lines show power-law fits by \cite{vanderWel2014} for blue and red galaxies (selected according to their rest-frame UVJ colours), using the same CANDELS measurements at $z < 0.5$. Interestingly, this comparison shows that the variations in the spin of the DM halo span the observed range of galaxy morphologies and colours, again suggesting that halo spin plays a significant role in galaxy formation, as discussed in Section \ref{subsec:connection_to_halo_spin_and_bh_mass}.

The connection between halo spin and galaxy size is closely related to the idea that low surface brightness galaxies (LSBGs), as well as ultradiffuse galaxies (UDGs), might form at the centres of faster spinning haloes \citep[][]{Amorisco2016, Perez-Montano2019, Salinas2021}. These tendencies for LSBGs and (isolated) UDGs have been recently verified using hydrodynamic simulations (\citealt{DiCintio2019}; \citealt{Kulier2020}; \citealt{Benavides2021}; Pérez-Montaño et al., in preparation).

As also shown in Fig. \ref{fig:rhalf_vs_mstar_by_lambda_halo}, the correlation between galaxy size and halo spin breaks down at $M_{\ast} \gtrsim 10^{11} \, \Msun$ in both observations and simulations, which is perhaps not surprising considering that the theoretical prediction is for galactic \textit{discs}, while the high-mass end is dominated by elliptical galaxies. This transition at $M_{\ast} \! \sim \! 10^{11} \, \Msun$ roughly corresponds to the mass scale at which dry mergers become the dominant growth mechanism of galaxies \citep[][]{Lee2013, Rodriguez-Gomez2016}, driven mostly by the effectiveness of AGN feedback in massive galaxies and their consequent quenching, leading to the formation of spheroid-dominated systems at the massive end \citep{Rodriguez-Gomez2017}. \cite{Genel2018} also noted a different size evolution for galaxies above and below this mass scale, such that the sizes of galaxies with $M_{\ast} \gtrsim 10^{11} \, \Msun$ at $z=0$ showed very slow growth after quenching. In future work we will explore the role that galaxy mergers play in establishing galactic angular momentum at different masses.

\section{Discussion and conclusions}\label{sec:discussion_and_conclusions}

We have studied the stellar specific angular momenta ($j_{\ast}$) of central galaxies in the IllustrisTNG simulation at $z=0$, covering a stellar mass range from $10^9$ to $10^{12} \, \Msun$, with an emphasis on how $j_{\ast}$ relates to the morphology of the galaxy, the spin of the host halo, and the mass of the central BH. We also explored the fraction of the halo specific angular momentum `retained' by galaxies of different morphological types, and tested whether a connection between halo spin and galaxy size, which is predicted by simple analytical arguments, arises naturally in a cosmological context.

We quantified the morphology of our simulated galaxies using two different indicators: the $\kappa_{\rm rot}$ parameter \citep{Sales2010, Rodriguez-Gomez2017}, which measures the fraction of kinetic energy of the stars that is invested into ordered circular motion, and $\PLate$, which gives the probability that a galaxy has a late-type morphology according to a deep learning algorithm \citep{Huertas-Company2019}. The quantities $\kappa_{\rm rot}$ and $\PLate$ represent \textit{kinematic} and \textit{visual-like} morphological measurements, respectively, and the simulated galaxies can be roughly separated into early types and late types by adopting cuts at $\kappa_{\rm rot} = 0.5$ and $\PLate = 0.5$. For both types of morphological measurement, we found that the distribution of IllustrisTNG galaxies on the $j_{\ast}$--$M_{\ast}$ plane is in good overall agreement with observational estimates from \cite{Fall2013} for both early types and late types, and with \cite{Posti2018a} and \cite{DiTeodoro2021} for late types (Fig. \ref{fig:jstar_vs_mstar}), with the late types exhibiting higher $j_{\ast}$ at fixed galaxy stellar mass.

We further explored the relation between galactic angular momentum and morphology by showing several finely spaced median trends on the $j_{\ast}$--$M_{\ast}$ diagram, corresponding to galaxies classified according to the percentile at a fixed stellar mass (Fig. \ref{fig:jstar_vs_mstar_by_percentile}) and the actual value (Fig. \ref{fig:jstar_vs_mstar_by_value}) of the morphological parameters $\kappa_{\rm rot}$ and $\PLate$. This allowed us to probe the entire range of galaxy morphologies produced by the simulation in more detail. In both cases, we compared the obtained trends with the 3D fits provided by \cite{Fall2018} for the extreme cases of `pure' discs and spheroids, which have a logarithmic slope of $2/3$ but are separated by a factor of approximately eight in normalization.

We found that when separating morphological types according to the percentile at fixed mass (Fig. \ref{fig:jstar_vs_mstar_by_percentile}), the $j_{\ast}$--$M_{\ast}$ trends for both $\kappa_{\rm rot}$ and $\PLate$ behave very similarly to each other, and the range of $j_{\ast}$ values covered by these `morphological classes' is comparable to the separation between the pure discs and spheroids from \cite{Fall2018}, although with a somewhat smaller spread in the case of $\PLate$.

When separating galaxies according to the actual values of $\kappa_{\rm rot}$ and $\PLate$ (Fig. \ref{fig:jstar_vs_mstar_by_value}), the behaviour for $\PLate$ is relatively unchanged, but the $j_{\ast}$--$M_{\ast}$ trends for different $\kappa_{\rm rot}$ values now deviate from the simple $j_{\ast} \propto M_{\ast}^{2/3}$ relation favoured by \cite{Fall2018}, displaying instead a bend near $M_{\ast} \! \sim \! 2$--$3 \times 10^{10} \, \Msun$, such that the logarithmic slopes in the $j_{\ast}$--$M_{\ast}$ plane at stellar masses above (below) this transition point become steeper (shallower) than $2/3$. The main reason for these qualitative differences between Figs \ref{fig:jstar_vs_mstar_by_percentile} and \ref{fig:jstar_vs_mstar_by_value} is that galaxy morphology is mass-dependent, with more disc-dominated galaxies at $\log_{10}(M_{\ast}/\Msun) \! \sim \! 10.5$ and more spheroid-dominated galaxies at lower and higher masses.

We next considered the specific angular momentum `retention fraction', defined as the ratio between the specific angular momenta of the galaxy and its host halo, $f_j \equiv j_{\ast} / j_{200}$, and plotted it as a function of stellar mass (Fig. \ref{fig:retention_factor_vs_mstar}). When splitting the simulated galaxies by their $\krot$ values, we found that the resulting $f_j$ trends are remarkably constant as a function of stellar mass, with spiral galaxies ($\krot \gtrsim 0.6$) `retaining' 50--60 per cent of their host haloes' specific angular momenta and ellipticals ($\krot \approx 1/3$) retaining 10--20 per cent. These values are close to the best empirical estimates of $f_j \approx 0.7$--$0.8$ for highly disc-dominated galaxies \citep{Fall2013, Posti2019a, DiTeodoro2021} and $f_j \approx 0.1$ for spheroid-dominated galaxies \citep{Fall2013, Fall2018}.

For completeness, we also considered the stellar-to-halo mass ratio ($f_{M} \equiv M_{\ast} / M_{200}$), which is related to $j_{\ast}$ via equation (\ref{eq:jstar_vs_mstar}), and plotted it in Fig. \ref{fig:mstar_to_m200_vs_mstar} as a function of stellar mass. When separating galaxies according to their $\krot$ values, we found that spirals show a monotonically increasing $f_{M}(M_{\ast})$, while ellipticals show an `inverted U' shape, in qualitative agreement with \cite{Posti2019} and \cite{Posti2021}, respectively. Finally, in Fig. \ref{fig:jstar_vs_mstar_by_fgas_sf} we showed that IllustrisTNG galaxies also display a correlation between $j_{\ast}$ and cold gas fraction at fixed $M_{\ast}$, in qualitative agreement with recent observational works by \cite{ManceraPina2021a} and \cite{Hardwick2022}, although the dependence of the $j_{\ast}$--$M_{\ast}$ relation on cold gas fraction is weaker than its dependence on galactic morphology.

Having explored the interplay between galactic angular momentum and morphology, we proceeded to investigate the physical mechanisms that drive the scatter in the $j_{\ast}$--$M_{\ast}$ relation at fixed $M_{\ast}$. In particular, we considered the spin of the host halo ($\lambda^{\prime}$) and the mass of the central BH ($M_{\rm BH}$, which we used as a proxy for the amount of energy injected into the gas by AGN feedback over a galaxy's history). We found that galaxies that formed at the centres of faster (slower) rotating haloes, as well as those hosting less (more) massive BHs, tend to have higher (lower) stellar specific angular momenta (Fig. \ref{fig:jstar_vs_mstar_by_lambda_and_mbh_to_mstar}).

We also found that the correlations between galactic angular momentum, halo spin, and BH mass hold at fixed $\lambda^{\prime}$ or $M_{\rm BH}/M_{\ast}$ (Fig. \ref{fig:compare_three_quantities_manymasses}), indicating that halo spin and AGN feedback play distinct roles in driving the scatter of the $j_{\ast}$--$M_{\ast}$ relation. This paper supports a scenario where the rotation of a galaxy (measured by $j_{\ast}$) depends on that of its host halo (measured by $\lambda^{\prime}$), while AGN feedback (by suppressing late-time gas accretion, which has higher specific angular momentum) has an effect opposite to that of stellar feedback in establishing galactic angular momentum, in agreement with earlier findings \citep{Genel2015, Grand2017}.

Interestingly, we also found an anticorrelation between the spin of the halo and the mass of the supermassive BH at fixed galaxy or halo mass (Fig. \ref{fig:mbh_to_m200_vs_lambda_halo}). This could be explained by stronger (weaker) gas flows feeding the central BH in slower (faster) rotating haloes, where the centrifugal acceleration is easier (harder) to overcome. This would also mean that BHs to some extent `amplify' the effect of the halo spin: galaxies tend to have lower $j_{\ast}$ if they form inside slowly rotating haloes, but such slow rotation leads to increased BH growth and a stronger effect of AGN feedback, further decreasing $j_{\ast}$. It will be interesting and important to check with other simulations whether this prediction holds for models of gas accretion into supermassive BHs that account for the dynamics of the gas in the innermost regions of galaxies, e.g. in the case of the gravitational torque-limited accretion models of the Simba simulation \citep{Dave2019}.

We also studied the effects of halo spin and BH mass on galaxy morphology (Fig. \ref{fig:lambda_halo_and_mbh_to_mstar_vs_mstar}). As expected, we found that more massive BHs are associated with earlier-type morphologies, in agreement with observations \citep[][]{Kormendy2013} as well as with earlier findings using the IllustrisTNG simulation \citep[][]{Li2020}. On the other hand, we found a negligible correlation between halo spin and galaxy morphology, in disagreement with some recent works. In particular, using the original Illustris simulation, \cite{Rodriguez-Gomez2017} found a significant correlation between $\krot$ and $\lambda^{\prime}$ at low masses ($M_{\ast} \lesssim 10^{10} \, \Msun$), which we do not reproduce with the updated IllustrisTNG model. This could be explained by the significantly larger sizes of low-mass Illustris galaxies \citep[][]{Snyder2015, Pillepich2018, Rodriguez-Gomez2019}, which could lead to increased dynamical interaction with their host haloes, or to the fact that stellar feedback was directional in Illustris but isotropic in IllustrisTNG \citep{Pillepich2018}.

Finally, we revisited a well-known prediction of some analytical models of galaxy formation: that haloes with a higher spin parameter should form larger discs at their centres \citep[][]{Fall1980, Mo1998}. We tested this hypothesis by plotting the galaxy size as a function of stellar mass, showing the median trends for different $\lambda^{\prime}$ values (Fig. \ref{fig:rhalf_vs_mstar_by_lambda_halo}). We verified that there is indeed a correlation between galaxy size and halo spin in IllustrisTNG, which breaks down at the massive end ($M_{\ast} \gtrsim 10^{11} \, \Msun$), probably due to the increased importance of galaxy mergers in this regime \citep[][]{Lee2013, Rodriguez-Gomez2016} and the associated change in galaxy morphology, from discs to spheroids \citep{Rodriguez-Gomez2017}. Furthermore, we found that the variations in galaxy size with respect to halo spin span the observed range of galaxy morphologies and colours \citep{vanderWel2014, Huang2017}.

Overall, we have found that the properties of galactic angular momentum at $z=0$ in the IllustrisTNG simulation are in good agreement with observational constraints, despite the fact that the IllustrisTNG model was not tuned to match these observations (although it was designed to roughly match galaxy sizes at $z=0$). We have also investigated how galactic angular momentum relates to morphology, halo spin, and BH mass within this galaxy formation framework. All of this has been possible due to enormous theoretical and computational advances in the last few decades. Further observational data will be needed to test some of the predictions made in this paper, especially kinematic measurements for early-type galaxies at large radii.

On the theoretical side, further progress can be made with additional studies about the redshift evolution of galactic angular momentum, its detailed spatial distribution, its alignment to the gaseous and DM components, and the effect of mergers, which we will address in upcoming work. Furthermore, enhanced and more robust comparisons between theory and observations, such as with forward modelling of integral field spectroscopic observations \citep[][]{Ibarra-Medel2019, Bouche2021}, will be pivotal in order to constrain and improve future models of galaxy formation.

\section*{Acknowledgements}

We thank Vladimir Avila-Reese, Bernardo Cervantes-Sodi, and Jolanta Zjupa for useful comments and discussions. The IllustrisTNG flagship simulations were run on the HazelHen Cray XC40 supercomputer at the High Performance Computing Center Stuttgart (HLRS) as part of project GCS-ILLU of the Gauss Centre for Supercomputing (GCS). Ancillary and test runs of the project were also run on the compute cluster operated by HITS, on the Stampede supercomputer at TACC/XSEDE (allocation AST140063), at the Hydra and Draco supercomputers at the Max Planck Computing and Data Facility, and on the MIT/Harvard computing facilities supported by FAS and MIT MKI. The Flatiron Institute is supported by the Simons Foundation.

\section*{Data availability}

The data from the Illustris and IllustrisTNG simulations used in this work are publicly available at the websites \href{https://www.illustris-project.org}{https://www.illustris-project.org} and \href{https://www.tng-project.org}{https://www.tng-project.org}, respectively \citep{Nelson2015, Nelson2019}.

\bibliographystyle{mnras}

\bibliography{paper}

\begin{thebibliography}{}
\makeatletter
\relax
\def\mn@urlcharsother{\let\do\@makeother \do\$\do\&\do\#\do\^\do\_\do\%\do\~}
\def\mn@doi{\begingroup\mn@urlcharsother \@ifnextchar [ {\mn@doi@}
  {\mn@doi@[]}}
\def\mn@doi@[#1]#2{\def\@tempa{#1}\ifx\@tempa\@empty \href
  {http://dx.doi.org/#2} {doi:#2}\else \href {http://dx.doi.org/#2} {#1}\fi
  \endgroup}
\def\mn@eprint#1#2{\mn@eprint@#1:#2::\@nil}
\def\mn@eprint@arXiv#1{\href {http://arxiv.org/abs/#1} {{\tt arXiv:#1}}}
\def\mn@eprint@dblp#1{\href {http://dblp.uni-trier.de/rec/bibtex/#1.xml}
  {dblp:#1}}
\def\mn@eprint@#1:#2:#3:#4\@nil{\def\@tempa {#1}\def\@tempb {#2}\def\@tempc
  {#3}\ifx \@tempc \@empty \let \@tempc \@tempb \let \@tempb \@tempa \fi \ifx
  \@tempb \@empty \def\@tempb {arXiv}\fi \@ifundefined
  {mn@eprint@\@tempb}{\@tempb:\@tempc}{\expandafter \expandafter \csname
  mn@eprint@\@tempb\endcsname \expandafter{\@tempc}}}

\bibitem[\protect\citeauthoryear{Abadi, Navarro, Steinmetz  \& Eke}{Abadi
  et~al.}{2003}]{Abadi2003}
Abadi M.~G.,  Navarro J.~F.,  Steinmetz M.,   Eke V.~R.,  2003, \mn@doi [ApJ]
  {10.1086/378316}, 597, 21

\bibitem[\protect\citeauthoryear{Amorisco \& Loeb}{Amorisco \&
  Loeb}{2016}]{Amorisco2016}
Amorisco N.~C.,  Loeb A.,  2016, \mn@doi [MNRAS] {10.1093/mnrasl/slw055}, 459,
  L51

\bibitem[\protect\citeauthoryear{Aumer, White, Naab  \& Scannapieco}{Aumer
  et~al.}{2013}]{Aumer2013}
Aumer M.,  White S. D.~M.,  Naab T.,   Scannapieco C.,  2013, \mn@doi [MNRAS]
  {10.1093/mnras/stt1230}, 434, 3142

\bibitem[\protect\citeauthoryear{Avila‐Reese, Firmani  \&
  Hernandez}{Avila‐Reese et~al.}{1998}]{AvilaReese1998}
Avila‐Reese V.,  Firmani C.,   Hernandez X.,  1998, \mn@doi [ApJ]
  {10.1086/306136}, 505, 37

\bibitem[\protect\citeauthoryear{Baes, Verstappen, {De Looze}, Fritz, Saftly,
  {Vidal P{\'{e}}rez}, Stalevski  \& Valcke}{Baes et~al.}{2011}]{Baes2011}
Baes M.,  Verstappen J.,  {De Looze} I.,  Fritz J.,  Saftly W.,  {Vidal
  P{\'{e}}rez} E.,  Stalevski M.,   Valcke S.,  2011, \mn@doi [ApJS]
  {10.1088/0067-0049/196/2/22}, 196, 22

\bibitem[\protect\citeauthoryear{Barnes}{Barnes}{1988}]{Barnes1988}
Barnes J.~E.,  1988, \mn@doi [ApJ] {10.1086/166593}, 331, 699

\bibitem[\protect\citeauthoryear{Benavides et~al.,}{Benavides
  et~al.}{2021}]{Benavides2021}
Benavides J.~A.,  et~al., 2021, \mn@doi [Nat. Astron.]
  {10.1038/s41550-021-01458-1}, 5, 1255

\bibitem[\protect\citeauthoryear{Bondi}{Bondi}{1952}]{Bondi1952}
Bondi H.,  1952, \mn@doi [MNRAS] {10.1093/mnras/112.2.195}, 112, 195

\bibitem[\protect\citeauthoryear{Bottrell, Torrey, Simard  \& Ellison}{Bottrell
  et~al.}{2017}]{Bottrell2017}
Bottrell C.,  Torrey P.,  Simard L.,   Ellison S.~L.,  2017, \mn@doi [MNRAS]
  {10.1093/mnras/stx276}, 467, 2879

\bibitem[\protect\citeauthoryear{Bouch{\'{e}} et~al.,}{Bouch{\'{e}}
  et~al.}{2021}]{Bouche2021}
Bouch{\'{e}} N.~F.,  et~al., 2021, \mn@doi [A&A] {10.1051/0004-6361/202040225},
  654, A49

\bibitem[\protect\citeauthoryear{Bullock, Dekel, Kolatt, Kravtsov, Klypin,
  Porciani  \& Primack}{Bullock et~al.}{2001}]{Bullock2001}
Bullock J.~S.,  Dekel A.,  Kolatt T.~S.,  Kravtsov A.~V.,  Klypin A.~A.,
  Porciani C.,   Primack J.~R.,  2001, \mn@doi [ApJ] {10.1086/321477}, 555, 240

\bibitem[\protect\citeauthoryear{Camps \& Baes}{Camps \&
  Baes}{2015}]{Camps2015}
Camps P.,  Baes M.,  2015, \mn@doi [Astron. Comput.]
  {10.1016/j.ascom.2014.10.004}, 9, 20

\bibitem[\protect\citeauthoryear{Chabrier}{Chabrier}{2003}]{Chabrier2003}
Chabrier G.,  2003, \mn@doi [PASP] {10.1086/376392}, 115, 763

\bibitem[\protect\citeauthoryear{Cheung et~al.,}{Cheung
  et~al.}{2016}]{Cheung2016}
Cheung E.,  et~al., 2016, \mn@doi [Nature] {10.1038/nature18006}, 533, 504

\bibitem[\protect\citeauthoryear{Correa, Schaye, Clauwens, Bower, Crain,
  Schaller, Theuns  \& Thob}{Correa et~al.}{2017}]{Correa2017}
Correa C.~A.,  Schaye J.,  Clauwens B.,  Bower R.~G.,  Crain R.~A.,  Schaller
  M.,  Theuns T.,   Thob A. C.~R.,  2017, \mn@doi [MNRAS]
  {10.1093/mnrasl/slx133}, 472, L45

\bibitem[\protect\citeauthoryear{Cortese et~al.,}{Cortese
  et~al.}{2016}]{Cortese2016}
Cortese L.,  et~al., 2016, \mn@doi [MNRAS] {10.1093/mnras/stw1891}, 463, 170

\bibitem[\protect\citeauthoryear{Crain et~al.,}{Crain et~al.}{2015}]{Crain2015}
Crain R.~A.,  et~al., 2015, \mn@doi [MNRAS] {10.1093/mnras/stv725}, 450, 1937

\bibitem[\protect\citeauthoryear{Dalcanton, Spergel  \& Summers}{Dalcanton
  et~al.}{1997}]{Dalcanton1997}
Dalcanton J.~J.,  Spergel D.~N.,   Summers F.~J.,  1997, \mn@doi [ApJ]
  {10.1086/304182}, 482, 659

\bibitem[\protect\citeauthoryear{Dav{\'{e}}, Angl{\'{e}}s-Alc{\'{a}}zar,
  Narayanan, Li, Rafieferantsoa  \& Appleby}{Dav{\'{e}}
  et~al.}{2019}]{Dave2019}
Dav{\'{e}} R.,  Angl{\'{e}}s-Alc{\'{a}}zar D.,  Narayanan D.,  Li Q.,
  Rafieferantsoa M.~H.,   Appleby S.,  2019, \mn@doi [MNRAS]
  {10.1093/mnras/stz937}, 486, 2827

\bibitem[\protect\citeauthoryear{Davis, Efstathiou, Frenk  \& White}{Davis
  et~al.}{1985}]{Davis1985}
Davis M.,  Efstathiou G.,  Frenk C.~S.,   White S. D.~M.,  1985, \mn@doi [ApJ]
  {10.1086/163168}, 292, 371

\bibitem[\protect\citeauthoryear{DeFelippis, Genel, Bryan  \& Fall}{DeFelippis
  et~al.}{2017}]{DeFelippis2017}
DeFelippis D.,  Genel S.,  Bryan G.~L.,   Fall S.~M.,  2017, \mn@doi [ApJ]
  {10.3847/1538-4357/aa6dfc}, 841, 16

\bibitem[\protect\citeauthoryear{{Di Cintio}, Brook, Macci{\`{o}}, Dutton  \&
  Cardona-Barrero}{{Di Cintio} et~al.}{2019}]{DiCintio2019}
{Di Cintio} A.,  Brook C.~B.,  Macci{\`{o}} A.~V.,  Dutton A.~A.,
  Cardona-Barrero S.,  2019, \mn@doi [MNRAS] {10.1093/mnras/stz985}, 486, 2535

\bibitem[\protect\citeauthoryear{{Di Teodoro}, Posti, Ogle, Fall  \&
  Jarrett}{{Di Teodoro} et~al.}{2021}]{DiTeodoro2021}
{Di Teodoro} E.~M.,  Posti L.,  Ogle P.~M.,  Fall S.~M.,   Jarrett T.,  2021,
  \mn@doi [MNRAS] {10.1093/mnras/stab2549}, 507, 5820

\bibitem[\protect\citeauthoryear{Dolag, Borgani, Murante  \& Springel}{Dolag
  et~al.}{2009}]{Dolag2009a}
Dolag K.,  Borgani S.,  Murante G.,   Springel V.,  2009, \mn@doi [MNRAS]
  {10.1111/j.1365-2966.2009.15034.x}, 399, 497

\bibitem[\protect\citeauthoryear{Donnari et~al.,}{Donnari
  et~al.}{2021a}]{Donnari2021}
Donnari M.,  et~al., 2021a, \mn@doi [MNRAS] {10.1093/mnras/staa3006}, 500, 4004

\bibitem[\protect\citeauthoryear{Donnari, Pillepich, Nelson, Marinacci,
  Vogelsberger  \& Hernquist}{Donnari et~al.}{2021b}]{Donnari2021a}
Donnari M.,  Pillepich A.,  Nelson D.,  Marinacci F.,  Vogelsberger M.,
  Hernquist L.,  2021b, \mn@doi [MNRAS] {10.1093/mnras/stab1950}, 506, 4760

\bibitem[\protect\citeauthoryear{Doroshkevich}{Doroshkevich}{1970}]{Doroshkevich1970}
Doroshkevich A.~G.,  1970, \mn@doi [Astrofizika] {10.1007/BF01001625}, 6, 581

\bibitem[\protect\citeauthoryear{Dubois, Peirani, Pichon, Devriendt, Gavazzi,
  Welker  \& Volonteri}{Dubois et~al.}{2016}]{Dubois2016}
Dubois Y.,  Peirani S.,  Pichon C.,  Devriendt J.,  Gavazzi R.,  Welker C.,
  Volonteri M.,  2016, \mn@doi [MNRAS] {10.1093/mnras/stw2265}, 463, 3948

\bibitem[\protect\citeauthoryear{Dutton \& van~den Bosch}{Dutton \& van~den
  Bosch}{2012}]{Dutton2012}
Dutton A.~A.,  van~den Bosch F.~C.,  2012, \mn@doi [MNRAS]
  {10.1111/j.1365-2966.2011.20339.x}, 421, 608

\bibitem[\protect\citeauthoryear{El-Badry et~al.,}{El-Badry
  et~al.}{2018}]{El-Badry2018}
El-Badry K.,  et~al., 2018, \mn@doi [MNRAS] {10.1093/mnras/stx2482}, 473, 1930

\bibitem[\protect\citeauthoryear{Fall}{Fall}{1979}]{Fall1979}
Fall S.~M.,  1979, \mn@doi [Nature] {10.1038/281200a0}, 281, 200

\bibitem[\protect\citeauthoryear{Fall}{Fall}{1983}]{Fall1983}
Fall S.~M.,  1983, in IAU Symp. 100, Internal Kinematics and Dynamics of
  Galaxies, ed. E. Athanassoula (Cambridge: Cambridge Univ. Press), p.~391

\bibitem[\protect\citeauthoryear{Fall \& Efstathiou}{Fall \&
  Efstathiou}{1980}]{Fall1980}
Fall S.~M.,  Efstathiou G.,  1980, \mn@doi [MNRAS] {10.1093/mnras/193.2.189},
  193, 189

\bibitem[\protect\citeauthoryear{Fall \& Romanowsky}{Fall \&
  Romanowsky}{2013}]{Fall2013}
Fall S.~M.,  Romanowsky A.~J.,  2013, \mn@doi [ApJ]
  {10.1088/2041-8205/769/2/L26}, 769, L26

\bibitem[\protect\citeauthoryear{Fall \& Romanowsky}{Fall \&
  Romanowsky}{2018}]{Fall2018}
Fall S.~M.,  Romanowsky A.~J.,  2018, \mn@doi [ApJ] {10.3847/1538-4357/aaeb27},
  868, 133

\bibitem[\protect\citeauthoryear{Faucher-Gigu{\`{e}}re, Lidz, Zaldarriaga  \&
  Hernquist}{Faucher-Gigu{\`{e}}re et~al.}{2009}]{Faucher-Giguere2009}
Faucher-Gigu{\`{e}}re C.-A.,  Lidz A.,  Zaldarriaga M.,   Hernquist L.,  2009,
  \mn@doi [ApJ] {10.1088/0004-637X/703/2/1416}, 703, 1416

\bibitem[\protect\citeauthoryear{Firmani \& Avila-Reese}{Firmani \&
  Avila-Reese}{2000}]{Firmani2000}
Firmani C.,  Avila-Reese V.,  2000, \mn@doi [MNRAS]
  {10.1046/j.1365-8711.2000.03338.x}, 315, 457

\bibitem[\protect\citeauthoryear{Firmani \& Avila-Reese}{Firmani \&
  Avila-Reese}{2009}]{Firmani2009}
Firmani C.,  Avila-Reese V.,  2009, \mn@doi [MNRAS]
  {10.1111/j.1365-2966.2009.14844.x}, 396, 1675

\bibitem[\protect\citeauthoryear{Genel et~al.,}{Genel
  et~al.}{2014}]{Genel2014a}
Genel S.,  et~al., 2014, \mn@doi [MNRAS] {10.1093/mnras/stu1654}, 445, 175

\bibitem[\protect\citeauthoryear{Genel, Fall, Hernquist, Vogelsberger, Snyder,
  Rodriguez-Gomez, Sijacki  \& Springel}{Genel et~al.}{2015}]{Genel2015}
Genel S.,  Fall S.~M.,  Hernquist L.,  Vogelsberger M.,  Snyder G.~F.,
  Rodriguez-Gomez V.,  Sijacki D.,   Springel V.,  2015, \mn@doi [ApJ]
  {10.1088/2041-8205/804/2/L40}, 804, L40

\bibitem[\protect\citeauthoryear{Genel et~al.,}{Genel et~al.}{2018}]{Genel2018}
Genel S.,  et~al., 2018, \mn@doi [MNRAS] {10.1093/mnras/stx3078}, 474, 3976

\bibitem[\protect\citeauthoryear{Grand et~al.,}{Grand et~al.}{2017}]{Grand2017}
Grand R. J.~J.,  et~al., 2017, \mn@doi [MNRAS] {10.1093/mnras/stx071}, 467,
  stx071

\bibitem[\protect\citeauthoryear{Grand et~al.,}{Grand et~al.}{2019}]{Grand2019}
Grand R. J.~J.,  et~al., 2019, \mn@doi [MNRAS] {10.1093/mnras/stz2928}, 490,
  4786

\bibitem[\protect\citeauthoryear{Grogin et~al.,}{Grogin
  et~al.}{2011}]{Grogin2011}
Grogin N.~A.,  et~al., 2011, \mn@doi [ApJS] {10.1088/0067-0049/197/2/35}, 197,
  35

\bibitem[\protect\citeauthoryear{Hardwick, Cortese, Obreschkow, Catinella  \&
  Cook}{Hardwick et~al.}{2022}]{Hardwick2022}
Hardwick J.~A.,  Cortese L.,  Obreschkow D.,  Catinella B.,   Cook R. H.~W.,
  2022, \mn@doi [MNRAS] {10.1093/mnras/stab3261}, 509, 3751

\bibitem[\protect\citeauthoryear{Hernquist \& Mihos}{Hernquist \&
  Mihos}{1995}]{Hernquist1995}
Hernquist L.,  Mihos J.~C.,  1995, \mn@doi [ApJ] {10.1086/175940}, 448, 41

\bibitem[\protect\citeauthoryear{Hirschmann, Dolag, Saro, Bachmann, Borgani  \&
  Burkert}{Hirschmann et~al.}{2014}]{Hirschmann2014}
Hirschmann M.,  Dolag K.,  Saro A.,  Bachmann L.,  Borgani S.,   Burkert A.,
  2014, \mn@doi [MNRAS] {10.1093/mnras/stu1023}, 442, 2304

\bibitem[\protect\citeauthoryear{Hopkins, Wellons, Angl{\'{e}}s-Alc{\'{a}}zar,
  Faucher-Gigu{\`{e}}re  \& Grudi{\'{c}}}{Hopkins et~al.}{2022}]{Hopkins2022}
Hopkins P.~F.,  Wellons S.,  Angl{\'{e}}s-Alc{\'{a}}zar D.,
  Faucher-Gigu{\`{e}}re C.-A.,   Grudi{\'{c}} M.~Y.,  2022, \mn@doi [MNRAS]
  {10.1093/mnras/stab3458}, 510, 630

\bibitem[\protect\citeauthoryear{Huang et~al.,}{Huang et~al.}{2017}]{Huang2017}
Huang K.-H.,  et~al., 2017, \mn@doi [ApJ] {10.3847/1538-4357/aa62a6}, 838, 6

\bibitem[\protect\citeauthoryear{Huertas-Company et~al.,}{Huertas-Company
  et~al.}{2019}]{Huertas-Company2019}
Huertas-Company M.,  et~al., 2019, \mn@doi [MNRAS] {10.1093/mnras/stz2191},
  489, 1859

\bibitem[\protect\citeauthoryear{Ibarra-Medel, Avila-Reese, S{\'{a}}nchez,
  Gonz{\'{a}}lez-Samaniego  \& Rodr{\'{i}}guez-Puebla}{Ibarra-Medel
  et~al.}{2019}]{Ibarra-Medel2019}
Ibarra-Medel H.~J.,  Avila-Reese V.,  S{\'{a}}nchez S.~F.,
  Gonz{\'{a}}lez-Samaniego A.,   Rodr{\'{i}}guez-Puebla A.,  2019, \mn@doi
  [MNRAS] {10.1093/mnras/sty3256}, 483, 4525

\bibitem[\protect\citeauthoryear{Irodotou, Thomas, Henriques, Sargent  \&
  Hislop}{Irodotou et~al.}{2019}]{Irodotou2019}
Irodotou D.,  Thomas P.~A.,  Henriques B.~M.,  Sargent M.~T.,   Hislop J.~M.,
  2019, \mn@doi [MNRAS] {10.1093/mnras/stz2365}, 489, 3609

\bibitem[\protect\citeauthoryear{Jiang et~al.,}{Jiang et~al.}{2019}]{Jiang2019}
Jiang F.,  et~al., 2019, \mn@doi [MNRAS] {10.1093/mnras/stz1952}, 488, 4801

\bibitem[\protect\citeauthoryear{Kassin, Devriendt, Fall, de Jong, Allgood  \&
  Primack}{Kassin et~al.}{2012}]{Kassin2012}
Kassin S.~A.,  Devriendt J.,  Fall S.~M.,  de Jong R.~S.,  Allgood B.,
  Primack J.~R.,  2012, \mn@doi [MNRAS] {10.1111/j.1365-2966.2012.21219.x},
  424, 502

\bibitem[\protect\citeauthoryear{Koekemoer et~al.,}{Koekemoer
  et~al.}{2011}]{Koekemoer2011}
Koekemoer A.~M.,  et~al., 2011, \mn@doi [ApJS] {10.1088/0067-0049/197/2/36},
  197, 36

\bibitem[\protect\citeauthoryear{Kormendy \& Ho}{Kormendy \&
  Ho}{2013}]{Kormendy2013}
Kormendy J.,  Ho L.~C.,  2013, \mn@doi [ARA&A]
  {10.1146/annurev-astro-082708-101811}, 51, 511

\bibitem[\protect\citeauthoryear{Kulier, Galaz, Padilla  \& Trayford}{Kulier
  et~al.}{2020}]{Kulier2020}
Kulier A.,  Galaz G.,  Padilla N.~D.,   Trayford J.~W.,  2020, \mn@doi [MNRAS]
  {10.1093/mnras/staa1798}, 496, 3996

\bibitem[\protect\citeauthoryear{Lacey \& Ostriker}{Lacey \&
  Ostriker}{1985}]{Lacey1985}
Lacey C.~G.,  Ostriker J.~P.,  1985, \mn@doi [ApJ] {10.1086/163729}, 299, 633

\bibitem[\protect\citeauthoryear{Lagos, Theuns, Stevens, Cortese, Padilla,
  Davis, Contreras  \& Croton}{Lagos et~al.}{2017}]{Lagos2017}
Lagos C. d.~P.,  Theuns T.,  Stevens A. R.~H.,  Cortese L.,  Padilla N.~D.,
  Davis T.~A.,  Contreras S.,   Croton D.,  2017, \mn@doi [MNRAS]
  {10.1093/mnras/stw2610}, 464, 3850

\bibitem[\protect\citeauthoryear{Lagos et~al.,}{Lagos et~al.}{2018}]{Lagos2018}
Lagos C. d.~P.,  et~al., 2018, \mn@doi [MNRAS] {10.1093/mnras/stx2667}, 473,
  4956

\bibitem[\protect\citeauthoryear{Lapi et~al.,}{Lapi et~al.}{2018a}]{Lapi2018a}
Lapi A.,  et~al., 2018a, \mn@doi [ApJ] {10.3847/1538-4357/aab6af}, 857, 22

\bibitem[\protect\citeauthoryear{Lapi, Salucci  \& Danese}{Lapi
  et~al.}{2018b}]{Lapi2018}
Lapi A.,  Salucci P.,   Danese L.,  2018b, \mn@doi [ApJ]
  {10.3847/1538-4357/aabf35}, 859, 2

\bibitem[\protect\citeauthoryear{Lee \& Yi}{Lee \& Yi}{2013}]{Lee2013}
Lee J.,  Yi S.~K.,  2013, \mn@doi [ApJ] {10.1088/0004-637X/766/1/38}, 766, 38

\bibitem[\protect\citeauthoryear{Lewis, Challinor  \& Lasenby}{Lewis
  et~al.}{2000}]{Lewis2000}
Lewis A.,  Challinor A.,   Lasenby A.,  2000, \mn@doi [ApJ] {10.1086/309179},
  538, 473

\bibitem[\protect\citeauthoryear{Li et~al.,}{Li et~al.}{2020}]{Li2020}
Li Y.,  et~al., 2020, \mn@doi [ApJ] {10.3847/1538-4357/ab8f8d}, 895, 102

\bibitem[\protect\citeauthoryear{Lu et~al.,}{Lu et~al.}{2022}]{Lu2022}
Lu S.,  et~al., 2022, \mn@doi [MNRAS] {10.1093/mnras/stab3169}, 509, 2707

\bibitem[\protect\citeauthoryear{Ludlow, Fall, Schaye  \& Obreschkow}{Ludlow
  et~al.}{2021}]{Ludlow2021}
Ludlow A.~D.,  Fall S.~M.,  Schaye J.,   Obreschkow D.,  2021, \mn@doi [MNRAS]
  {10.1093/mnras/stab2770}, 508, 5114

\bibitem[\protect\citeauthoryear{Maccio, Dutton, {Van Den Bosch}, Moore, Potter
   \& Stadel}{Maccio et~al.}{2007}]{Maccio2007}
Maccio A.~V.,  Dutton A.~A.,  {Van Den Bosch} F.~C.,  Moore B.,  Potter D.,
  Stadel J.,  2007, \mn@doi [MNRAS] {10.1111/j.1365-2966.2007.11720.x}, 378, 55

\bibitem[\protect\citeauthoryear{{Mancera Pi{\~{n}}a}, Posti, Fraternali, Adams
   \& Oosterloo}{{Mancera Pi{\~{n}}a} et~al.}{2021a}]{ManceraPina2021}
{Mancera Pi{\~{n}}a} P.~E.,  Posti L.,  Fraternali F.,  Adams E. A.~K.,
  Oosterloo T.,  2021a, \mn@doi [A&A] {10.1051/0004-6361/202039340}, 647, A76

\bibitem[\protect\citeauthoryear{{Mancera Pi{\~{n}}a}, Posti, Pezzulli,
  Fraternali, Fall, Oosterloo  \& Adams}{{Mancera Pi{\~{n}}a}
  et~al.}{2021b}]{ManceraPina2021a}
{Mancera Pi{\~{n}}a} P.~E.,  Posti L.,  Pezzulli G.,  Fraternali F.,  Fall
  S.~M.,  Oosterloo T.,   Adams E. A.~K.,  2021b, \mn@doi [A&A]
  {10.1051/0004-6361/202141574}, 651, L15

\bibitem[\protect\citeauthoryear{Marinacci et~al.,}{Marinacci
  et~al.}{2018}]{Marinacci2018}
Marinacci F.,  et~al., 2018, \mn@doi [MNRAS] {10.1093/mnras/sty2206}, 5139,
  5113

\bibitem[\protect\citeauthoryear{Mo, Mao  \& White}{Mo et~al.}{1998}]{Mo1998}
Mo H.~J.,  Mao S.,   White S. D.~M.,  1998, \mn@doi [MNRAS]
  {10.1046/j.1365-8711.1998.01227.x}, 295, 319

\bibitem[\protect\citeauthoryear{Naab \& Ostriker}{Naab \&
  Ostriker}{2017}]{Naab2017}
Naab T.,  Ostriker J.~P.,  2017, \mn@doi [ARA&A]
  {10.1146/annurev-astro-081913-040019}, 55, 59

\bibitem[\protect\citeauthoryear{Naiman et~al.,}{Naiman
  et~al.}{2018}]{Naiman2018}
Naiman J.~P.,  et~al., 2018, \mn@doi [MNRAS] {10.1093/mnras/sty618}, 477, 1206

\bibitem[\protect\citeauthoryear{Navarro \& Steinmetz}{Navarro \&
  Steinmetz}{1997}]{Navarro1997}
Navarro J.~F.,  Steinmetz M.,  1997, \mn@doi [ApJ] {10.1086/303763}, 478, 13

\bibitem[\protect\citeauthoryear{Navarro, Frenk  \& White}{Navarro
  et~al.}{1995}]{Navarro1995}
Navarro J.~F.,  Frenk C.~S.,   White S. D.~M.,  1995, MNRAS, 275, 56

\bibitem[\protect\citeauthoryear{Nelson et~al.,}{Nelson
  et~al.}{2015}]{Nelson2015}
Nelson D.,  et~al., 2015, \mn@doi [Astron. Comput.]
  {10.1016/j.ascom.2015.09.003}, 13, 12

\bibitem[\protect\citeauthoryear{Nelson et~al.,}{Nelson
  et~al.}{2018}]{Nelson2018}
Nelson D.,  et~al., 2018, \mn@doi [MNRAS] {10.1093/mnras/stx3040}, 475, 624

\bibitem[\protect\citeauthoryear{Nelson et~al.,}{Nelson
  et~al.}{2019a}]{Nelson2019}
Nelson D.,  et~al., 2019a, \mn@doi [Comput. Astrophys. Cosmol.]
  {10.1186/s40668-019-0028-x}, 6, 2

\bibitem[\protect\citeauthoryear{Nelson et~al.,}{Nelson
  et~al.}{2019b}]{Nelson2019a}
Nelson D.,  et~al., 2019b, \mn@doi [MNRAS] {10.1093/mnras/stz2306}, 490, 3234

\bibitem[\protect\citeauthoryear{Obreja, Stinson, Dutton, Macci{\`{o}}, Wang
  \& Kang}{Obreja et~al.}{2016}]{Obreja2016}
Obreja A.,  Stinson G.~S.,  Dutton A.~A.,  Macci{\`{o}} A.~V.,  Wang L.,   Kang
  X.,  2016, \mn@doi [MNRAS] {10.1093/mnras/stw690}, 459, 467

\bibitem[\protect\citeauthoryear{Obreja et~al.,}{Obreja
  et~al.}{2019}]{Obreja2019}
Obreja A.,  et~al., 2019, \mn@doi [MNRAS] {10.1093/mnras/stz1563}, 487, 4424

\bibitem[\protect\citeauthoryear{Obreschkow \& Glazebrook}{Obreschkow \&
  Glazebrook}{2014}]{Obreschkow2014}
Obreschkow D.,  Glazebrook K.,  2014, \mn@doi [ApJ]
  {10.1088/0004-637X/784/1/26}, 784, 26

\bibitem[\protect\citeauthoryear{Pakmor \& Springel}{Pakmor \&
  Springel}{2013}]{Pakmor2013}
Pakmor R.,  Springel V.,  2013, \mn@doi [MNRAS] {10.1093/mnras/stt428}, 432,
  176

\bibitem[\protect\citeauthoryear{Pakmor, Springel, Bauer, Mocz, Munoz, Ohlmann,
  Schaal  \& Zhu}{Pakmor et~al.}{2016}]{Pakmor2016}
Pakmor R.,  Springel V.,  Bauer A.,  Mocz P.,  Munoz D.~J.,  Ohlmann S.~T.,
  Schaal K.,   Zhu C.,  2016, \mn@doi [MNRAS] {10.1093/mnras/stv2380}, 455,
  1134

\bibitem[\protect\citeauthoryear{Pedrosa \& Tissera}{Pedrosa \&
  Tissera}{2015}]{Pedrosa2015}
Pedrosa S.~E.,  Tissera P.~B.,  2015, \mn@doi [A&A]
  {10.1051/0004-6361/201526440}, 584, A43

\bibitem[\protect\citeauthoryear{Peebles}{Peebles}{1969}]{Peebles1969}
Peebles P. J.~E.,  1969, \mn@doi [ApJ] {10.1086/149876}, 155, 393

\bibitem[\protect\citeauthoryear{P{\'{e}}rez-Monta{\~{n}}o \& {Cervantes
  Sodi}}{P{\'{e}}rez-Monta{\~{n}}o \& {Cervantes
  Sodi}}{2019}]{Perez-Montano2019}
P{\'{e}}rez-Monta{\~{n}}o L.~E.,  {Cervantes Sodi} B.,  2019, \mn@doi [MNRAS]
  {10.1093/mnras/stz2847}, 490, 3772

\bibitem[\protect\citeauthoryear{Pillepich et~al.,}{Pillepich
  et~al.}{2018a}]{Pillepich2018}
Pillepich A.,  et~al., 2018a, \mn@doi [MNRAS] {10.1093/mnras/stx2656}, 473,
  4077

\bibitem[\protect\citeauthoryear{Pillepich et~al.,}{Pillepich
  et~al.}{2018b}]{Pillepich2018a}
Pillepich A.,  et~al., 2018b, \mn@doi [MNRAS] {10.1093/mnras/stx3112}, 475, 648

\bibitem[\protect\citeauthoryear{Pillepich et~al.,}{Pillepich
  et~al.}{2019}]{Pillepich2019}
Pillepich A.,  et~al., 2019, \mn@doi [MNRAS] {10.1093/mnras/stz2338}, 490, 3196

\bibitem[\protect\citeauthoryear{{Planck Collaboration XIII}}{{Planck
  Collaboration XIII}}{2016}]{Planck2016}
{Planck Collaboration XIII} 2016, \mn@doi [A&A] {10.1051/0004-6361/201525830},
  594, A13

\bibitem[\protect\citeauthoryear{Posti \& Fall}{Posti \&
  Fall}{2021}]{Posti2021}
Posti L.,  Fall S.~M.,  2021, \mn@doi [A&A] {10.1051/0004-6361/202040256}, 649,
  A119

\bibitem[\protect\citeauthoryear{Posti, Pezzulli, Fraternali  \& {Di
  Teodoro}}{Posti et~al.}{2018a}]{Posti2018}
Posti L.,  Pezzulli G.,  Fraternali F.,   {Di Teodoro} E.~M.,  2018a, \mn@doi
  [MNRAS] {10.1093/mnras/stx3168}, 475, 232

\bibitem[\protect\citeauthoryear{Posti, Fraternali, {Di Teodoro}  \&
  Pezzulli}{Posti et~al.}{2018b}]{Posti2018a}
Posti L.,  Fraternali F.,  {Di Teodoro} E.~M.,   Pezzulli G.,  2018b, \mn@doi
  [A&A] {10.1051/0004-6361/201833091}, 612, L6

\bibitem[\protect\citeauthoryear{Posti, Fraternali  \& Marasco}{Posti
  et~al.}{2019a}]{Posti2019}
Posti L.,  Fraternali F.,   Marasco A.,  2019a, \mn@doi [A&A]
  {10.1051/0004-6361/201935553}, 626, A56

\bibitem[\protect\citeauthoryear{Posti, Marasco, Fraternali  \& Famaey}{Posti
  et~al.}{2019b}]{Posti2019a}
Posti L.,  Marasco A.,  Fraternali F.,   Famaey B.,  2019b, \mn@doi [A&A]
  {10.1051/0004-6361/201935982}, 629, A59

\bibitem[\protect\citeauthoryear{Rodriguez-Gomez et~al.,}{Rodriguez-Gomez
  et~al.}{2016}]{Rodriguez-Gomez2016}
Rodriguez-Gomez V.,  et~al., 2016, \mn@doi [MNRAS] {10.1093/mnras/stw456}, 458,
  2371

\bibitem[\protect\citeauthoryear{Rodriguez-Gomez et~al.,}{Rodriguez-Gomez
  et~al.}{2017}]{Rodriguez-Gomez2017}
Rodriguez-Gomez V.,  et~al., 2017, \mn@doi [MNRAS] {10.1093/mnras/stx305}, 467,
  3083

\bibitem[\protect\citeauthoryear{Rodriguez-Gomez et~al.,}{Rodriguez-Gomez
  et~al.}{2019}]{Rodriguez-Gomez2019}
Rodriguez-Gomez V.,  et~al., 2019, \mn@doi [MNRAS] {10.1093/mnras/sty3345},
  483, 4140

\bibitem[\protect\citeauthoryear{Rodr{\'{i}}guez-Puebla, Avila-Reese, Yang,
  Foucaud, Drory  \& Jing}{Rodr{\'{i}}guez-Puebla
  et~al.}{2015}]{Rodriguez-Puebla2015}
Rodr{\'{i}}guez-Puebla A.,  Avila-Reese V.,  Yang X.,  Foucaud S.,  Drory N.,
  Jing Y.~P.,  2015, \mn@doi [ApJ] {10.1088/0004-637X/799/2/130}, 799, 130

\bibitem[\protect\citeauthoryear{Romanowsky \& Fall}{Romanowsky \&
  Fall}{2012}]{Romanowsky2012}
Romanowsky A.~J.,  Fall S.~M.,  2012, \mn@doi [ApJS]
  {10.1088/0067-0049/203/2/17}, 203, 17

\bibitem[\protect\citeauthoryear{Sales, Navarro, Schaye, Vecchia, Springel  \&
  Booth}{Sales et~al.}{2010}]{Sales2010}
Sales L.~V.,  Navarro J.~F.,  Schaye J.,  Vecchia C.~D.,  Springel V.,   Booth
  C.~M.,  2010, \mn@doi [MNRAS] {10.1111/j.1365-2966.2010.17391.x}, 409, 1541

\bibitem[\protect\citeauthoryear{Sales, Navarro, Theuns, Schaye, White, Frenk,
  Crain  \& {Dalla Vecchia}}{Sales et~al.}{2012}]{Sales2012}
Sales L.~V.,  Navarro J.~F.,  Theuns T.,  Schaye J.,  White S. D.~M.,  Frenk
  C.~S.,  Crain R.~A.,   {Dalla Vecchia} C.,  2012, \mn@doi [MNRAS]
  {10.1111/j.1365-2966.2012.20975.x}, 423, 1544

\bibitem[\protect\citeauthoryear{Salinas \& Galaz}{Salinas \&
  Galaz}{2021}]{Salinas2021}
Salinas V.~H.,  Galaz G.,  2021, \mn@doi [ApJ] {10.3847/1538-4357/ac043d}, 915,
  125

\bibitem[\protect\citeauthoryear{Salpeter}{Salpeter}{1955}]{Salpeter1955}
Salpeter E.~E.,  1955, \mn@doi [ApJ] {10.1086/145971}, 121, 161

\bibitem[\protect\citeauthoryear{Scannapieco, White, Springel  \&
  Tissera}{Scannapieco et~al.}{2009}]{Scannapieco2009}
Scannapieco C.,  White S. D.~M.,  Springel V.,   Tissera P.~B.,  2009, \mn@doi
  [MNRAS] {10.1111/j.1365-2966.2009.14764.x}, 396, 696

\bibitem[\protect\citeauthoryear{Scannapieco et~al.,}{Scannapieco
  et~al.}{2012}]{Scannapieco2012}
Scannapieco C.,  et~al., 2012, \mn@doi [MNRAS]
  {10.1111/j.1365-2966.2012.20993.x}, 423, 1726

\bibitem[\protect\citeauthoryear{Schaye et~al.,}{Schaye
  et~al.}{2015}]{Schaye2015}
Schaye J.,  et~al., 2015, \mn@doi [MNRAS] {10.1093/mnras/stu2058}, 446, 521

\bibitem[\protect\citeauthoryear{Shakura \& Sunyaev}{Shakura \&
  Sunyaev}{1973}]{Shakura1973}
Shakura N.~I.,  Sunyaev R.~A.,  1973, \mn@doi [A&A]
  {10.1017/s007418090010035x}, 24, 337

\bibitem[\protect\citeauthoryear{Shi, Lapi, Mancuso, Wang  \& Danese}{Shi
  et~al.}{2017}]{Shi2017}
Shi J.,  Lapi A.,  Mancuso C.,  Wang H.,   Danese L.,  2017, \mn@doi [ApJ]
  {10.3847/1538-4357/aa7893}, 843, 105

\bibitem[\protect\citeauthoryear{Sijacki, Springel, {Di Matteo}  \&
  Hernquist}{Sijacki et~al.}{2007}]{Sijacki2007}
Sijacki D.,  Springel V.,  {Di Matteo} T.,   Hernquist L.,  2007, \mn@doi
  [MNRAS] {10.1111/j.1365-2966.2007.12153.x}, 380, 877

\bibitem[\protect\citeauthoryear{Snyder et~al.,}{Snyder
  et~al.}{2015}]{Snyder2015}
Snyder G.~F.,  et~al., 2015, \mn@doi [MNRAS] {10.1093/mnras/stv2078}, 454, 1886

\bibitem[\protect\citeauthoryear{Soko{\l}owska, Capelo, Fall, Mayer, Shen  \&
  Bonoli}{Soko{\l}owska et~al.}{2017}]{Sokolowska2017}
Soko{\l}owska A.,  Capelo P.~R.,  Fall S.~M.,  Mayer L.,  Shen S.,   Bonoli S.,
   2017, \mn@doi [ApJ] {10.3847/1538-4357/835/2/289}, 835, 289

\bibitem[\protect\citeauthoryear{Somerville \& Dav{\'{e}}}{Somerville \&
  Dav{\'{e}}}{2015}]{Somerville2015}
Somerville R.~S.,  Dav{\'{e}} R.,  2015, \mn@doi [ARA&A]
  {10.1146/annurev-astro-082812-140951}, 53, 51

\bibitem[\protect\citeauthoryear{Sommer‐Larsen, Gelato  \&
  Vedel}{Sommer‐Larsen et~al.}{1999}]{SommerLarsen1999}
Sommer‐Larsen J.,  Gelato S.,   Vedel H.,  1999, \mn@doi [ApJ]
  {10.1086/307374}, 519, 501

\bibitem[\protect\citeauthoryear{Springel}{Springel}{2005}]{Springel2005a}
Springel V.,  2005, \mn@doi [MNRAS] {10.1111/j.1365-2966.2005.09655.x}, 364,
  1105

\bibitem[\protect\citeauthoryear{Springel}{Springel}{2010}]{Springel2010}
Springel V.,  2010, \mn@doi [MNRAS] {10.1111/j.1365-2966.2009.15715.x}, 401,
  791

\bibitem[\protect\citeauthoryear{Springel \& Hernquist}{Springel \&
  Hernquist}{2003}]{Springel2003}
Springel V.,  Hernquist L.,  2003, \mn@doi [MNRAS]
  {10.1046/j.1365-8711.2003.06206.x}, 339, 289

\bibitem[\protect\citeauthoryear{Springel \& Hernquist}{Springel \&
  Hernquist}{2005}]{Springel2005}
Springel V.,  Hernquist L.,  2005, \mn@doi [ApJ] {10.1086/429486}, 622, L9

\bibitem[\protect\citeauthoryear{Springel, White, Tormen  \&
  Kauffmann}{Springel et~al.}{2001}]{Springel2001}
Springel V.,  White S. D.~M.,  Tormen G.,   Kauffmann G.,  2001, \mn@doi
  [MNRAS] {10.1046/j.1365-8711.2001.04912.x}, 328, 726

\bibitem[\protect\citeauthoryear{Springel, {Di Matteo}  \& Hernquist}{Springel
  et~al.}{2005}]{Springel2005b}
Springel V.,  {Di Matteo} T.,   Hernquist L.,  2005, \mn@doi [MNRAS]
  {10.1111/j.1365-2966.2005.09238.x}, 361, 776

\bibitem[\protect\citeauthoryear{Springel et~al.,}{Springel
  et~al.}{2018}]{Springel2018}
Springel V.,  et~al., 2018, \mn@doi [MNRAS] {10.1093/mnras/stx3304}, 475, 676

\bibitem[\protect\citeauthoryear{Starkenburg, Sales, Genel, Manzano-King,
  Canalizo  \& Hernquist}{Starkenburg et~al.}{2019}]{Starkenburg2019}
Starkenburg T.~K.,  Sales L.~V.,  Genel S.,  Manzano-King C.,  Canalizo G.,
  Hernquist L.,  2019, \mn@doi [ApJ] {10.3847/1538-4357/ab2128}, 878, 143

\bibitem[\protect\citeauthoryear{Stevens, Croton  \& Mutch}{Stevens
  et~al.}{2016}]{Stevens2016}
Stevens A. R.~H.,  Croton D.~J.,   Mutch S.~J.,  2016, \mn@doi [MNRAS]
  {10.1093/mnras/stw1332}, 461, 859

\bibitem[\protect\citeauthoryear{Tabor, Merrifield, Arag{\'{o}}n-Salamanca,
  Fraser-McKelvie, Peterken, Smethurst, Drory  \& Lane}{Tabor
  et~al.}{2019}]{Tabor2019}
Tabor M.,  Merrifield M.,  Arag{\'{o}}n-Salamanca A.,  Fraser-McKelvie A.,
  Peterken T.,  Smethurst R.,  Drory N.,   Lane R.~R.,  2019, \mn@doi [MNRAS]
  {10.1093/mnras/stz431}, 485, 1546

\bibitem[\protect\citeauthoryear{Tacchella et~al.,}{Tacchella
  et~al.}{2019}]{Tacchella2019}
Tacchella S.,  et~al., 2019, \mn@doi [MNRAS] {10.1093/mnras/stz1657}, 487, 5416

\bibitem[\protect\citeauthoryear{Teklu, Remus, Dolag, Beck, Burkert, Schmidt,
  Schulze  \& Steinborn}{Teklu et~al.}{2015}]{Teklu2015}
Teklu A.~F.,  Remus R.-S.,  Dolag K.,  Beck A.~M.,  Burkert A.,  Schmidt A.~S.,
   Schulze F.,   Steinborn L.~K.,  2015, \mn@doi [ApJ]
  {10.1088/0004-637X/812/1/29}, 812, 29

\bibitem[\protect\citeauthoryear{Thacker \& Couchman}{Thacker \&
  Couchman}{2001}]{Thacker2001}
Thacker R.~J.,  Couchman H. M.~P.,  2001, \mn@doi [ApJ] {10.1086/321739}, 555,
  L17

\bibitem[\protect\citeauthoryear{Torrey, Vogelsberger, Sijacki, Springel  \&
  Hernquist}{Torrey et~al.}{2012}]{Torrey2012}
Torrey P.,  Vogelsberger M.,  Sijacki D.,  Springel V.,   Hernquist L.,  2012,
  \mn@doi [MNRAS] {10.1111/j.1365-2966.2012.22082.x}, 427, 2224

\bibitem[\protect\citeauthoryear{Torrey, Vogelsberger, Genel, Sijacki, Springel
   \& Hernquist}{Torrey et~al.}{2014}]{Torrey2014}
Torrey P.,  Vogelsberger M.,  Genel S.,  Sijacki D.,  Springel V.,   Hernquist
  L.,  2014, \mn@doi [MNRAS] {10.1093/mnras/stt2295}, 438, 1985

\bibitem[\protect\citeauthoryear{{\"{U}}bler, Naab, Oser, Aumer, Sales  \&
  White}{{\"{U}}bler et~al.}{2014}]{Ubler2014}
{\"{U}}bler H.,  Naab T.,  Oser L.,  Aumer M.,  Sales L.~V.,   White S. D.~M.,
  2014, \mn@doi [MNRAS] {10.1093/mnras/stu1275}, 443, 2092

\bibitem[\protect\citeauthoryear{Varma et~al.,}{Varma et~al.}{2022}]{Varma2022}
Varma S.,  et~al., 2022, \mn@doi [MNRAS] {10.1093/mnras/stab3149}, 509, 2654

\bibitem[\protect\citeauthoryear{Vogelsberger, Genel, Sijacki, Torrey, Springel
   \& Hernquist}{Vogelsberger et~al.}{2013}]{Vogelsberger2013}
Vogelsberger M.,  Genel S.,  Sijacki D.,  Torrey P.,  Springel V.,   Hernquist
  L.,  2013, \mn@doi [MNRAS] {10.1093/mnras/stt1789}, 436, 3031

\bibitem[\protect\citeauthoryear{Vogelsberger et~al.,}{Vogelsberger
  et~al.}{2014a}]{Vogelsberger2014a}
Vogelsberger M.,  et~al., 2014a, \mn@doi [MNRAS] {10.1093/mnras/stu1536}, 444,
  1518

\bibitem[\protect\citeauthoryear{Vogelsberger et~al.,}{Vogelsberger
  et~al.}{2014b}]{Vogelsberger2014}
Vogelsberger M.,  et~al., 2014b, \mn@doi [Nature] {10.1038/nature13316}, 509,
  177

\bibitem[\protect\citeauthoryear{Vogelsberger, Marinacci, Torrey  \&
  Puchwein}{Vogelsberger et~al.}{2020}]{Vogelsberger2020}
Vogelsberger M.,  Marinacci F.,  Torrey P.,   Puchwein E.,  2020, \mn@doi [Nat.
  Rev. Phys.] {10.1038/s42254-019-0127-2}, 2, 42

\bibitem[\protect\citeauthoryear{Weinberger et~al.,}{Weinberger
  et~al.}{2017}]{Weinberger2017}
Weinberger R.,  et~al., 2017, \mn@doi [MNRAS] {10.1093/mnras/stw2944}, 465,
  3291

\bibitem[\protect\citeauthoryear{White}{White}{1984}]{White1984}
White S. D.~M.,  1984, \mn@doi [ApJ] {10.1086/162573}, 286, 38

\bibitem[\protect\citeauthoryear{Yuan \& Narayan}{Yuan \&
  Narayan}{2014}]{Yuan2014}
Yuan F.,  Narayan R.,  2014, \mn@doi [ARA&A]
  {10.1146/annurev-astro-082812-141003}, 52, 529

\bibitem[\protect\citeauthoryear{Zavala et~al.,}{Zavala
  et~al.}{2016}]{Zavala2016}
Zavala J.,  et~al., 2016, \mn@doi [MNRAS] {10.1093/mnras/stw1286}, 460, 4466

\bibitem[\protect\citeauthoryear{Zjupa \& Springel}{Zjupa \&
  Springel}{2017}]{Zjupa2017}
Zjupa J.,  Springel V.,  2017, \mn@doi [MNRAS] {10.1093/mnras/stw2945}, 466,
  1625

\bibitem[\protect\citeauthoryear{Zoldan, {De Lucia}, Xie, Fontanot  \&
  Hirschmann}{Zoldan et~al.}{2018}]{Zoldan2018}
Zoldan A.,  {De Lucia} G.,  Xie L.,  Fontanot F.,   Hirschmann M.,  2018,
  \mn@doi [MNRAS] {10.1093/mnras/sty2343}, 481, 1376

\bibitem[\protect\citeauthoryear{van~den Bosch}{van~den
  Bosch}{1998}]{VandenBosch1998}
van~den Bosch F.~C.,  1998, \mn@doi [ApJ] {10.1086/306354}, 507, 601

\bibitem[\protect\citeauthoryear{van~der Wel et~al.,}{van~der Wel
  et~al.}{2014}]{vanderWel2014}
van~der Wel A.,  et~al., 2014, \mn@doi [ApJ] {10.1088/0004-637X/788/1/28}, 788,
  28

\makeatother
\end{thebibliography}

\end{document}